\documentclass[11pt]{article}
\pdfoutput=1

\usepackage{jheppub}
\makeatletter
\gdef\@fpheader{}
\makeatother

\usepackage{mathtools}
\usepackage{amsmath, amssymb, amsthm, amsbsy, bbm}
\usepackage{dsfont}
\newtheorem{theorem}{Theorem}
\DeclareMathOperator{\tr}{Tr}

\usepackage[percent]{overpic}
\usepackage[table,xcdraw]{xcolor}
\usepackage[utf8]{inputenc}
\usepackage{enumitem}
\usepackage{hyperref}
\usepackage{graphicx}
\usepackage{caption}
\usepackage{subcaption}
\usepackage{tablefootnote}

\usepackage[normalem]{ulem}
\usepackage{etoolbox}
\usepackage{cancel}
\usepackage{todonotes}

\theoremstyle{remark}

\theoremstyle{definition}

\newcommand{\Tr}{\mathrm{Tr}}

\newcommand{\bra}[1]{\langle #1 \vert}
\newcommand{\ket}[1]{\vert #1 \rangle}

\def\sl#1{\setbox0=\hbox{$#1$}           
   \dimen0=\wd0                                 
   \setbox1=\hbox{/} \dimen1=\wd1               
   \ifdim\dimen0>\dimen1                        
      \rlap{\hbox to \dimen0{\hfil/\hfil}}      
      #1                                        
   \else                                        
      \rlap{\hbox to \dimen1{\hfil$#1$\hfil}}   
      /                                         
   \fi}    
   
\makeatletter
\renewcommand\d[1]{\mspace{2mu}\mathrm{d}#1\@ifnextchar\d{\mspace{-3mu}}{}\mspace{4mu}}
\makeatother

\newtoggle{editing}
\togglefalse{editing}

\iftoggle{editing}{%
	\setlength{\marginparwidth}{2cm}
	\usepackage[draft]{showlabels}
	\newcommand{\jamie}[1]{\textbf{\color{red}[#1  - Jamie]}}
	\newcommand{\AK}[1]{\textbf{\color{blue}[AK: #1]}}

}{%
	\newcommand{\jamie}[1]{\textbf{\color{red}[]}}
    \newcommand{\AK}[1]{\textbf{\color{blue}[]}}
	
}

\usepackage{subfiles}

\title{Random Matrix Theory for Complexity Growth and Black Hole Interiors}
\author{Arjun Kar,}
\author{Lampros Lamprou,}
\author{Moshe Rozali,}
\author{James Sully}

\affiliation{Department of Physics and Astronomy, University of British Columbia,\\
6224 Agricultural Road, Vancouver, BC V6T 1Z1, Canada}

\emailAdd{arjunkar@phas.ubc.ca}
\emailAdd{llamprou@mit.edu}
\emailAdd{rozali@phas.ubc.ca}
\emailAdd{sully@phas.ubc.ca}

\abstract{ We study a precise and computationally tractable notion of operator complexity in holographic quantum theories, including the ensemble dual of Jackiw-Teitelboim gravity and two-dimensional holographic conformal field theories. This is a refined, ``microcanonical'' version of K-complexity that applies to theories with infinite or continuous spectra (including quantum field theories), and in the holographic theories we study exhibits exponential growth for a scrambling time, followed by linear growth until saturation at a time exponential in the entropy ---a behavior that is characteristic of chaos. 
We show that the linear growth regime implies a universal random matrix description of the operator dynamics after scrambling. 
Our main tool for establishing this connection is a ``complexity renormalization group'' framework we develop that allows us to study the effective operator dynamics for different timescales  by ``integrating out'' large K-complexities. 
In the dual gravity setting, we comment on the empirical match between our version of K-complexity and the maximal volume proposal, and speculate on a connection between the universal random matrix theory dynamics of operator growth after scrambling and the spatial translation symmetry of smooth black hole interiors.
}

\begin{document}

\maketitle
\section{Post-scrambling operator growth and the black hole interior}

The physics of black holes is intimately tied to quantum chaos. Eigenstate thermalization, maximal scrambling, and level repulsion in the high energy spectrum all play a role in explaining universal properties of semi-classical black holes, e.g. their near horizon geometry and the Page curve for the entropy of their Hawking radiation. In this work, we aim to view through the quantum chaos lens yet another one of their universal features: the geometry of their interiors.

\begin{figure}
    \centering
    \includegraphics[width=0.44\textwidth]{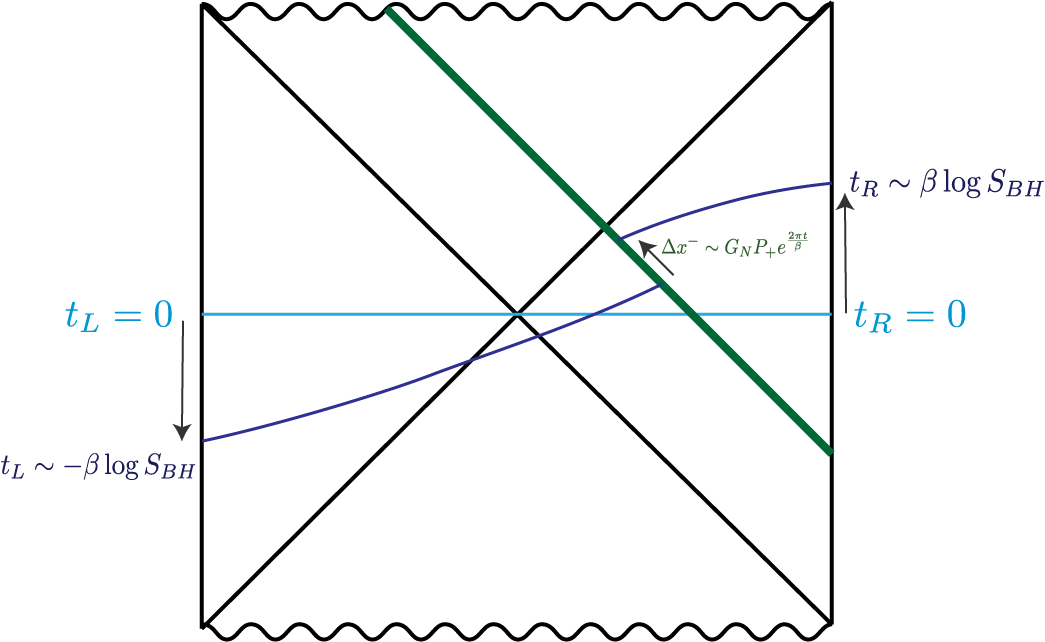}
    \includegraphics[width=0.55\textwidth]{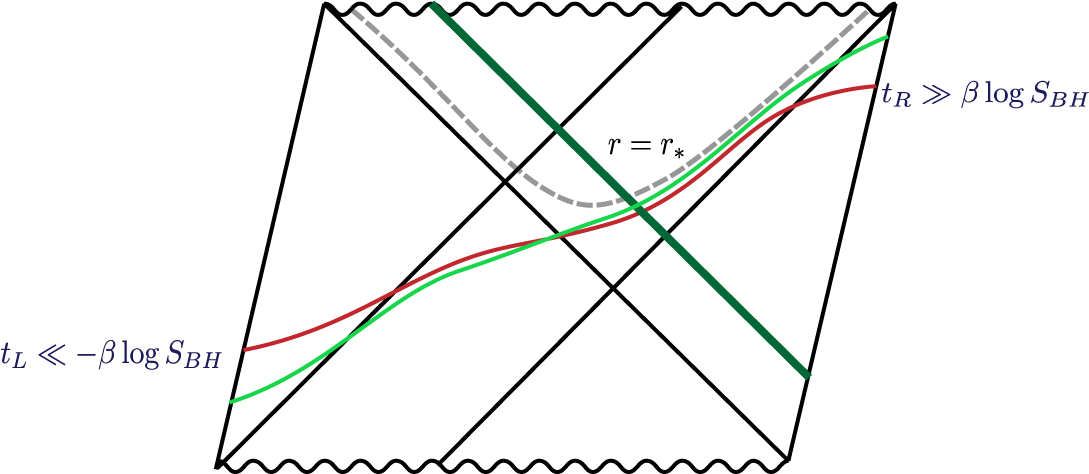}\\[0.5cm]
    \includegraphics[width=0.45\textwidth]{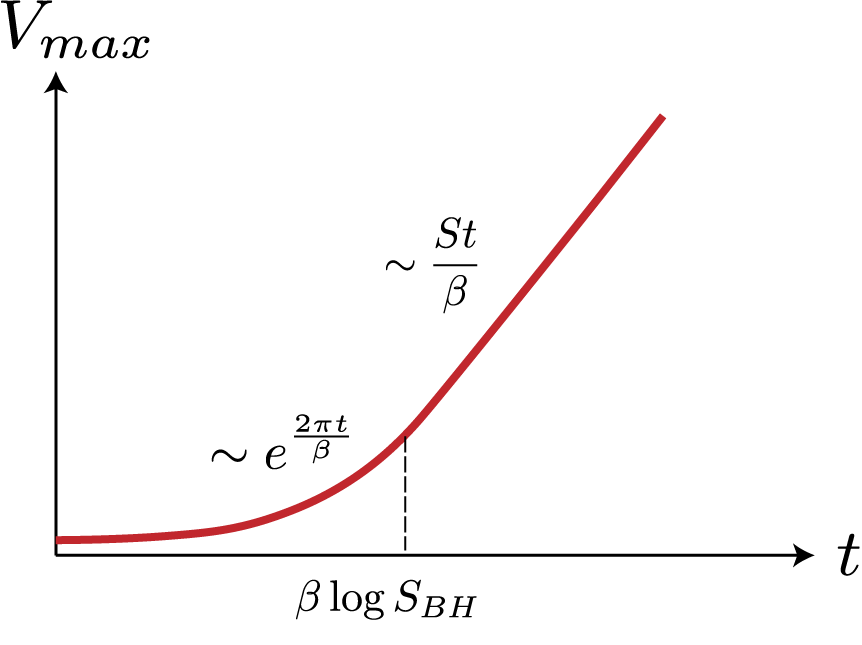}
    \caption{\small{\textbf{(Top)} The two figures illustrate the behavior of maximal volume slices in an eternal AdS black hole spacetime with an infalling excitation, at different boundary times $t_R=-t_L=t$.  \textbf{(Left)} As the particle approaches the horizon, its exponential blueshift leads to backreaction on the volume due to the shockwave effect. \textbf{(Right)} As we approach the scrambling time, the backreaction becomes $O(1)$ and the appropriate Penrose diagram contains an extended right black hole interior. Maximal volume slices accumulate near an interior time-slice $r=r_* \sim O(R_{sch})$ over a segment that is linearly increasing in $t$, as a result of the interior isometry. \textbf{(Bottom)} Dependence of maximal volume as a function of time.}}
    \label{fig:blackhole}
\end{figure}

The horizon of an Anti-de Sitter (AdS) black hole is a reflection of the maximally chaotic dynamics of the dual conformal field theory (CFT). Figure~\ref{fig:blackhole} illustrates the meaning of this statement. A freely falling particle exhibits a universal exponentially growing \emph{radial momentum} in boundary time,\footnote{This statement is only meaningful provided a choice of gauge. The canonical bulk slicing for these discussions consists of maximal volume Cauchy slices anchored at boundary times $t$.} as the particle approaches the black hole. The growing momentum results in gravitational backreaction which, in turn, imprints this exponential growth on the time-dependence of the volume of the maximal volume slices, and by extension on the particle's distance from the boundary along these slices. This characteristic behavior of momentum has been compellingly linked to the dual operator ``size'' which increases at the same exponential rate due to scrambling \cite{Magan:2018nmu,Qi:2018bje,Roberts:2018mnp,Magan:2020iac,Susskind:2020gnl}; similarly, the corresponding maximal volume growth is expected to reflect the evolution of \emph{operator complexity} \cite{Magan:2018nmu,Zhao:2020gxq,Haehl:2021sib,Haehl:2021prg,Haehl:2021dto}---assuming an appropriate definition of the latter. This early-time chaotic behavior ceases around $t\sim\beta \log S_{BH}$ when the CFT operator's size saturates at its maximal $O(S_{BH})$ value and the bulk particle is swallowed up by the \emph{shifted} horizon of the \emph{backreacted} geometry. The phenomena of quantum scrambling and of classical black hole absorption are thus identified by the AdS/CFT correspondence \cite{Sekino:2008he,Shenker:2013pqa}.

At later times $\beta \log S_{BH}\ll t \ll \beta e^{S_{BH}}$, the particle lies in the interior of this larger black hole. Radial momentum along maximal volume slices approaches a constant, in keeping with the saturation of operator size. In turn, the maximal volumes as well as the particle's distance from the boundary transition to a linear growth in boundary $t$, owing in part to the (spatial) Schwarzschild translation symmetry of the interior geometry \cite{Susskind:2014moa}. In this work, we discuss a precise complexity measure that is computationally tractable in holographic quantum theories and exhibits a similar exponential-to-linear growth (Figure~\ref{fig:seq-wvfn-kcomp}), with the right rates to be considered a probe of this interior geometry. We then connect its linear growth after scrambling to a universal random matrix theory (RMT) description of chaotic operator dynamics at large complexities. As is described in detail below, this is a new application of random matrix theory where the random matrix describes the adjoint action of the Hamiltonian on simple operators, $[H,O]$---a superoperator\footnote{i.e. an operator acting on the Hilbert space of operators.} called the \emph{Liouvillian}---and is distinct from the more conventional application of RMT for characterizing spectral properties of the Hamiltonian.

Our starting point is a notion of operator complexity discussed in a series of recent interesting works \cite{Parker:2018yvk,Barbon:2019wsy,Rabinovici:2020ryf} called \emph{Krylov complexity} (K-complexity) which is predicated on the simple idea reviewed in Section \ref{sec:krylov}: starting with a ``simple'' operator $O$ and a state $\rho$, we build an increasing-complexity operator ladder. This is achieved by applying nested commutators  with the Hamiltonian $H$ and orthonormalizing the operators produced at every step, using as our inner product correlators in the state $\rho$. The appeal of the resulting K-complexity basis stems from the following three properties.
\begin{enumerate}
    \item The time evolution of an operator is encoded in its wavefunction on the one-dimensional K-complexity chain which propagates according to a local wave equation (Section \ref{sec:k-basis-refined}).
    \item In maximally chaotic systems with a bounded spectrum, the dynamical information in this wave equation exhibits a substantial degree of universality \cite{Barbon:2019wsy} (Section \ref{sec:universal-growth}).
    \item These universal features imply that the K-complexity of $O(t)$, measured by its average position on the K-chain, grows exponentially for a scrambling time, before transitioning to a linear increase which is only disturbed by non-perturbative $O(e^{-S})$ effects that eventually lead to its saturation at $O(e^{2S})$ values \cite{Rabinovici:2020ryf}.
\end{enumerate}

\paragraph{Refined Krylov complexity for non-compact spectra and gravity}
The goal of the first part of this paper, undertaken in Section \ref{sec:k-basis-refined}, is to refine the notion of Krylov complexity so that it may be meaningfully applied to systems with a non-compact spectrum, for example holographic field theories. We emphasize that the new quantum chaotic universality we find, and its correspondence with the interior geometry of the black hole (described below), depend in detail on our prescription for defining the Krylov complexity for systems with non-compact spectrum, which we motivate on physical and mathematical grounds. 

In Section \ref{sec:complexitygravity}, we apply our refined definition to low dimensional gravity theories, including the ensemble dual of Jackiw-Teitelboim (JT) gravity, and show that (even though the spectrum is infinite) our refined K-complexity definition exhibits the same universal K-complexity growth regimes as the previously studied maximally chaotic systems with compact spectrum (Section \ref{sec:k-comp-jt}). While we relegate a thorough analysis of the precise bulk dual of our K-complexity notion to future work ---a task that appears to be within reach given the precise boundary definition--- we note the consistency of our result with a holographic interpretation as the volume of the maximal slice, in accordance with the discussions in \cite{Susskind:2014moa}, or as a measure of distance of the infalling particle from the boundary along the slice. Such a holographic perspective, if confirmed, would promote K-complexity to a precise boundary probe of the black hole interior geometry. 

\paragraph{Linear complexity growth and a new quantum chaotic universality} In the second part of this paper, we convert the Krylov formalism into a powerful tool for studying spectral properties of the Liouvillian and operator wavefunctions: a complexity renormalization group (RG) framework. This formalism will enable us to link the linear K-complexity growth to a universal description of operator dynamics after scrambling. 

Exploiting a direct relation between the Krylov data and an operator-weighted measure of the Liouvillian spectrum discussed in Section \ref{sec:lanczostospectrum}, we develop a method for coarse-graining over the large complexity degrees of freedom. The resulting coarse-grained Liouvillian is described by a random matrix theory, whose potential depends on a selected complexity RG scale $k$. Similarly, the operator wavefunction flows with $k$ so that its evolution with the effective RMT Hamiltonian accurately reproduces the exact time-evolution of the microscopic operator for time-intervals set by the complexity scale $k$. This works even though the underlying RMT may have a very different spectrum of eigenvalues than the original microscopic theory.

Our central result is that the effective RMT potential flows to a \emph{universal ``infinite well'' potential} for complexity scales larger than the entropy, $k \to \infty, S\to \infty$ with $k/S=r\gg 1$ (Section \ref{sec:scalinglimits}). In maximally chaotic systems, for example in the dual of JT gravity, this universal RMT controls the evolution of simple operators after scrambling $t\gg O(\beta \log S)$ and until times of order $t\lesssim O( \beta\, e^{S})$ when, as we show, the fine-grained features of the spectrum become of increasing relevance. 
The uniform matrix probability distribution of the effective Liouvillian underlies the linear K-complexity growth for $O(\beta \log S) \ll t \ll O( \beta\, e^{S})$. Its rapid onset after scrambling is a {\it new example of quantum chaotic universality}. Our complexity RG formalism also allows us to characterize the non-perturbative $O(e^{-2S})$ effects that slow down the linear K-complexity growth and lead to its ultimate saturation at $t\sim e^{2S}$ (Section \ref{sec:finitesize}).

This insight into the physics of linear K-complexity growth, combined with its appearance in the JT ensemble and its interpretation as probe of the behind-the-horizon geometry along maximal volume slices, outlines an intriguing link between smoothness of the black hole interior ---as reflected in its spatial homogeneity--- and the universal RMT dynamics at large complexities. We comment on this connection further in the Discussion (Section~\ref{sec:discussion}) which also includes a summary of our results and directions for future study.

\section{K-complexity growth in chaotic quantum field theories}\label{sec:krylov}

We start by reviewing the definitions required for the discussion of Krylov complexity, and adapt them to systems with a non-compact spectrum.

\subsection{Krylov basis and (a refined version of) K-complexity}\label{sec:k-basis-refined}
Operator complexity is an elusive concept. The most commonly used notion, circuit complexity, is defined as the minimum number of ``simple'' operations required to represent the action of the operator on a state. While being a useful intuitive guide, a mathematical definition of circuit complexity often requires a number of ad hoc choices about the preferred set of gates, their complexity weight and the acceptable error margin in the operator representation, in addition to being technically impractical in non-qubit systems such as continuum quantum field theories.

In recent works \cite{Parker:2018yvk,Barbon:2019wsy,Rabinovici:2020ryf,Magan:2020iac,Jian:2020qpp}, a different definition of operator complexity was proposed, one attuned to the study of complexity generated by time evolution, that depends only on the system's Hamiltonian and a reference quantum state. The idea capitalizes on the intuition that acting with the Hamiltonian on an initially simple operator $O$ via commutators generally help us ascend the complexity ladder. We therefore start with the set of operators:
\begin{equation}
    O, \,\,[H,O], \,\,[H,[H,O]],\,\, \dots \, , \label{nonorthonormalbasis}
\end{equation}
which are loosely associated with increasing complexity and linearly span the subspace of operator space explored by time evolution of $O$. For chaotic systems, this subspace is expected to be dense in the full operator space of the theory. It is convenient to represent operators of our quantum system's operator algebra $\mathcal{A}$ as vectors in the GNS Hilbert space $A\in \mathcal{A} \to |A)\in \mathcal{H}_{GNS}$ and define the \emph{Liouvillian} $\mathcal{L}$ as the operator on $\mathcal{H}_{GNS}$ that acts as:
\begin{equation}
    \mathcal{L}|A) = |[H,A]) \label{liouvillian} \, .
\end{equation}

Each application of the Liouvillian, however, does not unambiguously generate a more complex element but rather produces an operator with support on both higher and lower complexities. The key in constructing a meaningful complexity ladder is to introduce an operator inner product, completing the definition of $\mathcal{H}_{GNS}$, that will allow us to orthonormalize the basis (\ref{nonorthonormalbasis}). This is the point where the second ingredient of the Krylov formalism comes in: the reference state. Since we are interested in studying propagation of excitations in a black hole spacetime, the natural choice is the thermal state \cite{Parker:2018yvk} for which the inner product is simply
\begin{equation}
    (A|B)=\Tr [\rho^{\frac{1}{2}}A^\dagger \rho^{\frac{1}{2}}B] \, , \label{innerproduct}
\end{equation}
where $\rho=e^{-\beta H}$. This can also be thought of as a two-sided correlator in the thermo-field double state $|TFD\rangle$.  With this definition of the inner product, the Liouvillian is a Hermitian operator on $\mathcal{H}_{GNS}$.

The Krylov basis is then constructed by applying the Gram-Schmidt algorithm to the sequence of elements ${\cal L}^n|O)$  starting from $|O)$ and using the inner product (\ref{innerproduct}) in order to remove the overlap of each element ${\cal L}^n |O)$ with all the previous ones ${\cal L}^m |O)$, $0\leq m<n$ \cite{Viswanath1994TheRM,Parker:2018yvk}. The resulting orthonormal basis, called the Krylov basis, consists of the operators:
\begin{equation}
{\cal B}_K= \{ |O_n)= P_n({\cal L}) |O)\}_{n=0}^D \, ,
\end{equation}
where $P_n({\cal L})$ is a polynomial of the Liouvillian of degree $n$ and $D=\text{dim}[{\cal H}_{GNS}]$.

\paragraph{A refinement of the Krylov formalism} The approach outlined above has been used in a number of recent interesting studies \cite{Parker:2018yvk,Barbon:2019wsy,Rabinovici:2020ryf,Magan:2020iac,Jian:2020qpp}. For systems with a finite number of states, this formalism elegantly highlights robust signatures of quantum chaos \cite{Viswanath1994TheRM,Parker:2018yvk}, as we review below. 

In particular, an early exponentially growing K-complexity regime and a subsequent long linear growth regime were identified in these finite systems. This growth is similar to the leading time-dependence of the geometric complexity (volume of maximal slices) in a black hole spacetime with an infalling particle. While this is exciting, the derivation of that K-complexity scaling relies heavily on the system having finitely many states, whereas this is not expected to be the case in a thermal system whose holographic dual is a black hole. This motivates us to discuss the subject in continuum quantum field theories.

For systems with an infinite spectrum, it is usually more involved to define signatures of quantum chaos and demonstrate their universality. Indeed, a na\"ive application of the prescription outlined above to a system with an infinite spectrum obscures the aforementioned quantum chaotic signatures. The relevant complication ---and its remedy--- can be understood by expressing the action of the Liouvillian on operators (\ref{liouvillian}) in the energy projector eigenbasis:
\begin{equation}
    {\cal L} |O) = \sum_{i,j} (E_{i}-E_j)\,\langle E_j| O|E_i\rangle \,\, |E_{i}\rangle \langle E_j| = \int_0^\infty dE \int_{-2E}^{2E} d\omega \, \rho(E, \omega) \,O(E, \omega)  \,\,\omega \,\, |E+\frac{\omega}{2}\rangle \langle E-\frac{\omega}{2}| \, ,\label{liouvillianEnergyBasis}
\end{equation}
where in the second equality we introduced the shorthand $O(E, \omega)=\langle E+\frac{\omega}{2}| O|E-\frac{\omega}{2}\rangle $ and the spectral measure
\begin{equation}
    \rho(E,\omega) = \sum_{i,j} \delta \left(E- \frac{E_i+E_j}{2}\right) \,\delta  \left(\omega- E_i+E_j\right) \, ,\label{rhodefinition}
\end{equation}
with the limits of the $\omega-$integration following from the lower bound of the energy spectrum which we set to 0, without loss of generality. Note that while we are using notation usually adapted when making a continuous approximation to the spectrum, the spectrum is kept discrete.

The key observation is that the Liouvillian ${\cal L}$ probes only the \emph{energy difference}, $\omega$, of the operator wavefunction $O(E,\omega)$ and \emph{does not mix different average energy sectors}. Formally, the average energy observable $\mathcal{E}$ on $\mathcal{H}_{GNS}$ defined by $\mathcal{E}|O) = \tfrac{1}{2} |\{H,O\})$ obeys $[\mathcal{E},{\cal L}]=0$ and thus corresponds to a symmetry of the Liouvillian.  In other words, the average energy $E$ is essentially a conserved charge for $\mathcal{L}$.  The increasing complexity of the operator generated by iterative applications of ${\cal L}$ on $O$ takes place in every fixed $E$ sector separately, a fact we ought to take into account in our definition of the complexity basis and the associated complexity measure, else we risk blurring out chaotic features of the Liouvillian action just as conserved charges for a Hamiltonian can obscure signatures of e.g. level repulsion. The natural way to deal with this is to orthonormalize the Krylov elements of each fixed $E$ sector separately using the ``microcanonical'' inner product:
\begin{align}
    (A|B)_E&= \int_{-2E}^{2E} d\omega\, \rho (E,\omega)\, A^*(E,\omega) \,B(E,\omega) \, , \label{microinnerproduct}
\end{align}
where $A(E,\omega)$ and $B(E,\omega)$ are the operator wavefunctions of $A$ and $B$, respectively, and $\rho({ E},\omega)$ is given by (\ref{rhodefinition}).\footnote{More precisely, we imagine smearing $\rho(E, \omega)$ within a microcanonical window of width $\epsilon \ll 1$ about the average energy of interest $E$ in order to regulate the delta function dependence of (\ref{rhodefinition}):
\begin{align}
    \rho_{\epsilon} (E,\omega)&= \int_{E-\epsilon}^{E+\epsilon} dE'\,\rho(E',\omega) \, .
\end{align}
We will leave this subtlety implicit in our discussion.} Note that the inner product (\ref{innerproduct}) considered in previous works (when the state is chosen to be the thermal density matrix) is related to (\ref{microinnerproduct}) via a Laplace transform:
\begin{align}
    \Tr [\rho^{\frac{1}{2}}A^\dagger \rho^{\frac{1}{2}}B] &= \int_0^\infty dE \, e^{-\beta E} (A|B)_E \, .\label{Laplacetransform}
\end{align}
In the presence of other conserved charges of the Liouvillian, a similar approach ought to be followed: a K-complexity basis should be constructed for each fixed charge sector separately. This separates the quantum chaotic aspects of the theory from those which are simply determined by symmetry.\footnote{A previous attempt at applying Krylov complexity to field theory appeared in \cite{Dymarsky:2021bjq}, where the canonical picture was applied directly and infinite exponential growth was found even for free and integrable theories. The exponential growth found there is completely generic for all infinite-dimensional theories (including JT gravity, see Appendix~\ref{app:canonical-k-jt}) because it reflects the contributions from sectors of larger and larger average energy as we probe the tails of the thermal distribution. Accordingly, the rate of complexity growth is simply set by the temperature. In contrast, in our refined K-complexity formulation the appearance of an exponential regime is directly related to chaos as we will see explicitly in the case of the ensemble dual of JT gravity.  Another application of Krylov techniques appeared in \cite{Dymarsky:2019elm}.}

\paragraph{Krylov basis revisited} Applying the Gram-Schmidt algorithm to the linear operator basis $\{{\cal L}^n |O)\}_{n=0}^D$ using (\ref{microinnerproduct}) as the inner product in every fixed average energy sector, we construct an orthonormal basis for every $E$ which explicitly reads
\begin{align}
    \mathcal{B}^E_K&=\Big\{\,|O_{E,n})= {\cal N}(E)^{-\frac{1}{2}}\int_{-2E}^{2E} d\omega \,\rho(E,\omega)  \,O(E,\omega) \, P_n^E(\omega)\,|E+\frac{\omega}{2}\rangle \langle E-\frac{\omega}{2}| \,\Big\}_{n=0}^{D_E} \label{kbasis}\\
    \text{with: } &(O_{E,n}|O_{E,m})_E =\delta_{nm} \, , \label{orthogonality1}
\end{align}
where $P^E_n(\omega)$ is again a polynomial of degree $n$, $D_E=\text{dim}[\mathcal{H}^E_{GNS}]\sim e^{2S(E)}$ and ${\cal N}(E)$ an overall normalization of the polynomials $P_n^E$ introduced for later convenience:
\begin{equation}
    {\cal N}(E)= \int_{-2E}^{2E} dE \, \rho(E,\omega) \,|O(E,\omega)|^2 \, . \label{normalization}
\end{equation}
Importantly, $P^E_n$ in (\ref{kbasis}) form a sequence of \emph{orthogonal polynomials} with respect to a measure $\mu_E$ determined by the orthonormality condition (\ref{orthogonality1}) with inner product (\ref{microinnerproduct}):
\begin{align}
   &\int_{-2E}^{2E} d\mu_{E} (\omega) P^E_n(\omega) P^E_m(\omega)=\delta_{nm} \label{orthogonality}\\
    \text{where:  }& d\mu_{E}(\omega) ={\cal N}(E)^{-1} \rho(E,\omega) |O(E,\omega)|^2 \,d\omega \, , \label{themeasure}
\end{align}
which by virtue of (\ref{normalization}) is canonically normalized, $\int_{-2E}^{2E} d\mu_E(\omega) = 1$.  We will often call the measure $\mu_E$ in (\ref{themeasure}) the \emph{operator-weighted spectral measure} to emphasize that it is a re-weighting of the underlying spectral measure by the operator wavefunction. Together with the conditions $P^E_0(\omega)=1$ and $P^E_1(\omega)=\omega$, $\mu_E$ uniquely determines the entire polynomial sequence. The orthogonality of these polynomials will play a central role in deriving many of the important results of this paper.

The time-evolved operator $O(t)= e^{iHt}Oe^{-iHt}$ can be expressed in the basis (\ref{kbasis}) as 
\begin{equation}
    |O(t))= \int_0^\infty dE\, {\cal N}(E)^{\frac{1}{2}} \sum_{n=0}^D \phi_{E,n}(t) i^n |O_{E,n}) \, .\label{kwavefunction}
\end{equation}
We will refer to $\phi_{E,n}(t)$ as the \emph{K-wavefunction} and the inclusion of ${\cal N}(E)$ in (\ref{kwavefunction}) is to ensure that $\sum_{n=0}^{D(E)}|\phi_{E,n}(t)|^2= 1$ for every $E$, e.g. the initial operator $O=O(t=0)$ has the K-wavefunction $\phi_{E,n}(0)=\delta_{0n}$, $\forall E\in [0,\infty)$. Similarly, the time-evolved two-point function (\ref{Laplacetransform}) at canonical temperature $\beta$ reads:
\begin{equation}
    \Tr [e^{-\frac{\beta H}{2}}O(0)e^{-\frac{\beta H}{2}} O(t) ] = \int dE e^{-\beta E} (O(0)|O(t))_E = \int_0^\infty dE \,e^{-\beta E}{\cal N}(E)\, \phi_{E,0}(t) \, . \label{2ptKwavefunction}
\end{equation}

\begin{figure}
    \centering
    \includegraphics[width=.45\textwidth]{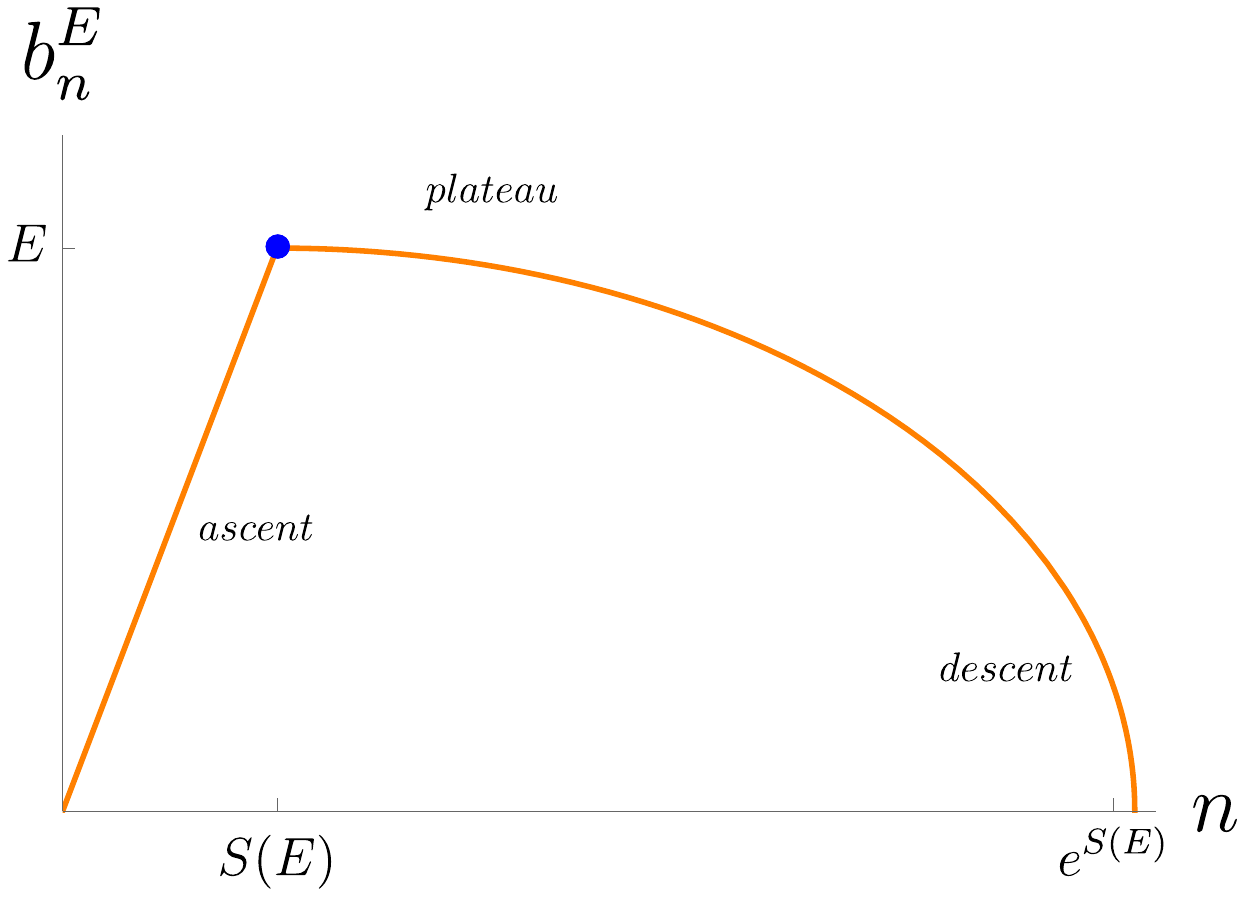}
    \includegraphics[width=.5\textwidth]{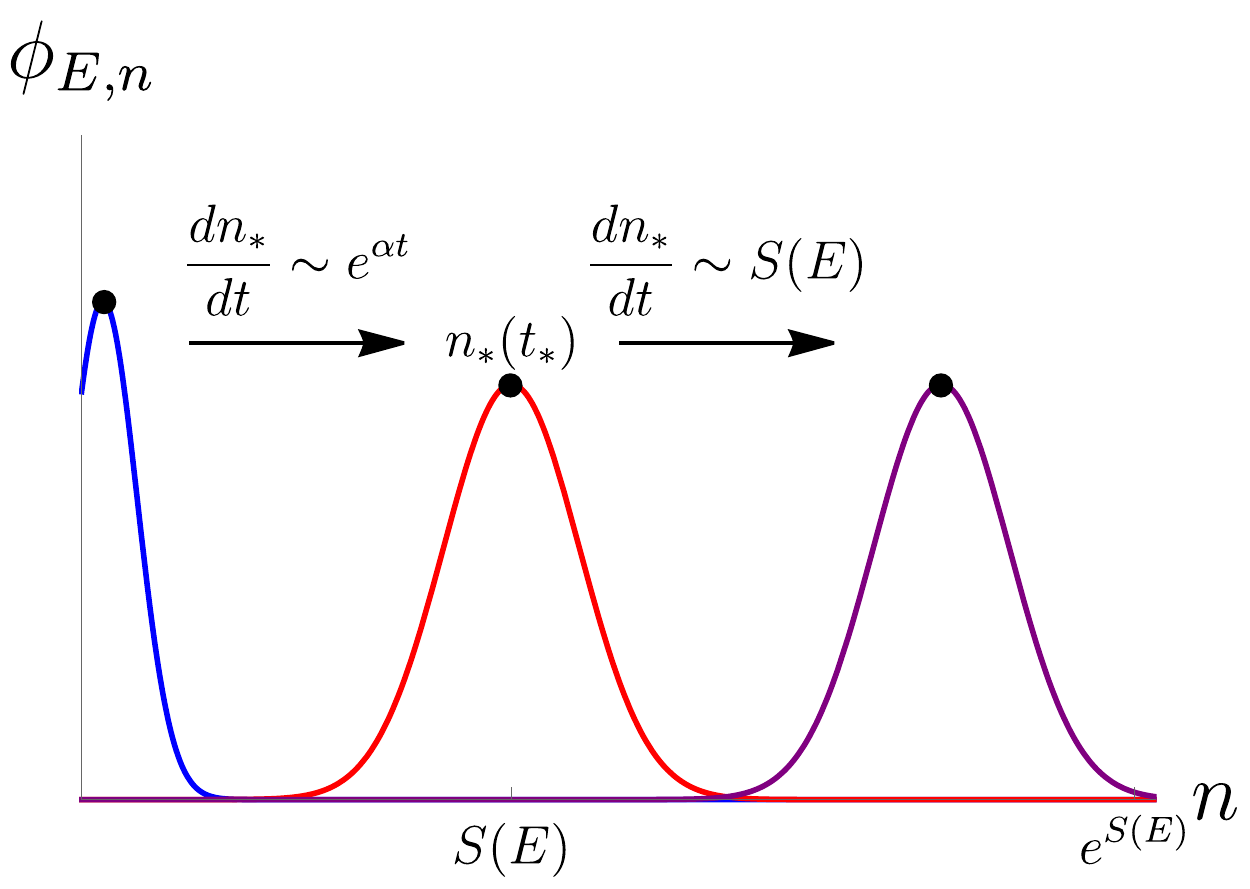}\\
    \includegraphics[width=.45\textwidth]{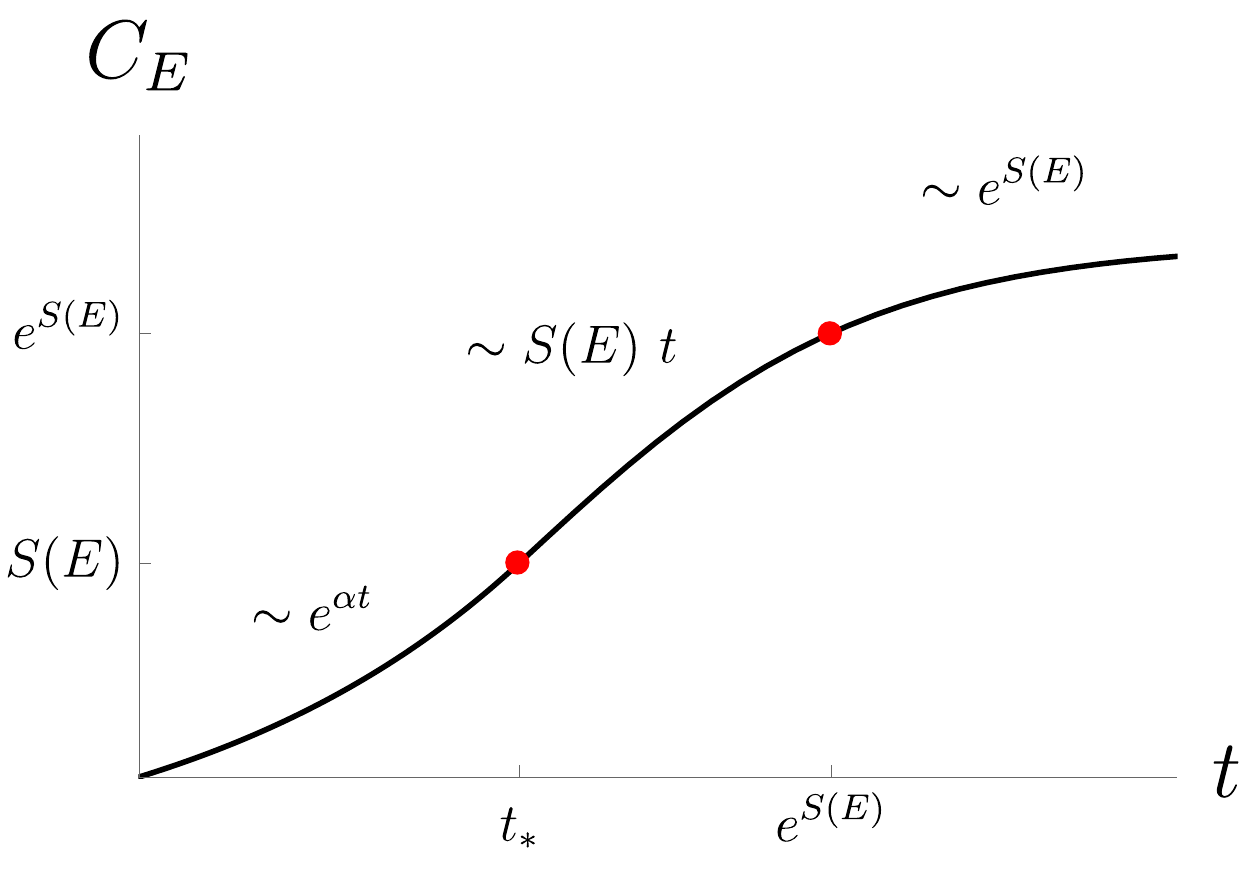}
    \caption{\small{These three schematic plots show our expectations for the refined Krylov complexity growth of a simple local operator in a maximally chaotic quantum system with fixed average energy $E$.  \textbf{(Top left)} The Lanczos sequence.  The Lanczos ascent is the linear region of coefficient index $n < S(E)$ and the Lanczos plateau is the approximately flat region $n > S(E)$.  The Lanczos descent is not clearly separated from the plateau; its slope is initially exponentially small $O(e^{-S(E)})$, and has been greatly exaggerated. \textbf{(Top right)} The Krylov wavefunction $\phi_{E,n}(t)$ before the scrambling time (blue), at the scrambling time $t_*$ (red), and some time after $t_*$ but before $e^{S(E)}$ (purple).  Before $t=t_*$, the peak of the wavefunction $n_*(t)$ decreases in height as the wavepacket spreads.  At the scrambling time, the wavefunction is peaked at $n_*(t_*) = S(E)$, and after this time the peak propagates at a roughly fixed velocity with minimal spreading along the chain as it passes through the long Lanczos plateau region.  \textbf{(Bottom)} The Krylov complexity $C_E$ for a simple local operator.  $C_E$ grows exponentially for a scrambling time as the Krylov wavefunction travels along the Lanczos ascent, and then transitions to a linearly growing regime induced by the Lanczos plateau which lasts for an exponential time.  Toward the end of the plateau, when the descent effects become large, the complexity saturates at an exponential value $e^{S(E)}$. }}
    \label{fig:seq-wvfn-kcomp}
\end{figure}

\paragraph{K-complexity} It is natural to interpret the element $|O_{E,n})$ in the time-evolved operator (\ref{kwavefunction}) as  contributing to the operator complexity by an amount equal to $n$. In any fixed average energy sector $E$, we may define the Krylov complexity $C_E[O(t)]$ of the time evolved operator as
\begin{equation}
    C_E[O(t)]= \sum_{n=0}^{D(E)}\, n |\phi_{E,n}(t)|^2 \, .\label{kcomplexity}
\end{equation}
$C_E$ can be visualized  as the expectation value of the ``position operator'' along the one-dimensional Krylov chain formed by the K-basis elements (\ref{kbasis}) (Figure~\ref{fig:seq-wvfn-kcomp}). 

Of course, more generally, the object of interest is not $C_E(t)$ but rather the complexity of the excitation $O(t)$ about the thermal state $e^{-\beta H}$. Since the different average energy sectors are thermally populated, the \emph{thermal K-complexity of $O(t)$} is defined by the formula:
\begin{equation}
    C_{th}[O(t)]= \frac{\int_0^\infty dE \,e^{-\beta E} {\cal N}(E)\, C_E(t)}{\int_0^\infty dE \,e^{-\beta E} {\cal N}(E)} \, ,\label{kcomplexityave}
\end{equation}
where ${\cal N}(E)$ given by (\ref{normalization}) and we have normalized $C_{th}$ so that $C_{th}[O(0)]=1$. This definition is in direct correspondence to the way the canonical thermal two-point function is related to the fixed $E$ correlator (\ref{2ptKwavefunction}).

\paragraph{Operator evolution on the K-chain and Lanczos coefficients} It is a well-known mathematical fact that a sequence of orthogonal polynomials $\lbrace P^E_n \rbrace$ with measure $\mu_E$ satisfies a second order recurrence relation:
\begin{align}
    \omega P^E_n(\omega)&= b^E_{n+1} P^E_{n+1}(\omega) +b^E_n P^E_{n-1}(\omega) \, , \label{recurrence}\\
    P^E_0(\omega)&=1 \, , \nonumber\\
    P^E_1(\omega)&=\omega \, , \nonumber
\end{align}
with coefficients $b^E_n\geq 0$. The inner product with measure $\mu_E$ and the recurrence relation (\ref{recurrence}) are two different but equivalent definitions of the polynomials $P^E_n$.

By virtue of property (\ref{recurrence}), the Liouvillian generator of operator time evolution in a fixed $E$ sector admits a simple tridiagonal matrix representation in the K-basis (\ref{kbasis}):
\begin{equation}
   \mathcal{L}= \begin{pmatrix}
     0 & b^E_1 & 0 & 0 & \cdots \\
     b^E_1 & 0 & b^E_2 & 0 & \cdots\\
     0 & b^E_2 & 0 & b^E_3 & \cdots \\
     0 & 0 & b^E_3 & 0 & \cdots \\
     \vdots
    \end{pmatrix} 
    \label{tridiagonal}
\end{equation}
and the recurrence coefficients $b^E_n$ in this context are called \emph{Lanczos coefficients}. This allows us to express the time evolution of an operator $O(t)$ in terms of a Schr\"odinger equation for its wavefunction on the one-dimensional Krylov chain (\ref{kwavefunction}):
\begin{equation}
    \dot{\phi}_{E,n}(t)= b^E_{n+1} \phi_{E,n+1}(t) -b^E_n \phi_{E,n-1}(t) \, . \label{kschrodinger}
\end{equation}
This equation is \emph{exact} and valid in any quantum system. All the information about the dynamical properties of the theory has been absorbed in the Lanczos sequence $b^E_n$. As this paper aims to demonstrate, what makes the representation (\ref{tridiagonal}) special is the remarkably simple universal behaviour of Lanczos coefficients in chaotic systems.

\paragraph{The moment method for computing Lanczos coefficients} The Lanczos sequence defined above contains all the dynamical information about our system in the Krylov basis and will be a central object in our discussion. It is, therefore, useful to briefly review the most practical way for obtaining it: the moment method, explained in \cite{Viswanath1994TheRM,Parker:2018yvk}. 

The input is the time-evolved two-point function of the operator in the state of interest. Since in our approach we construct a Krylov chain for every fixed average energy sector separately, the relevant correlator is the inverse Laplace transform of the thermal two-point function, namely:
\begin{equation}
    G_E(t)= (O|O(t))_E= \int_{-2E}^{2E} d\omega\, \rho(E,\omega) |O(E, \omega)|^2 \,e^{i\omega t} \, .
\end{equation}
The frequency distribution of this two-point function is described by a set of moments, defined as:
\begin{equation}
    m^E_{n} = \frac{\left(-i \frac{d}{dt}\right)^n G_E(t)\big|_{t=0} }{G_E(0)} \, .\label{moments}
\end{equation}
It can then be shown that the Lanczos coefficients appearing in the recurrence relation (\ref{recurrence}) and controlling the evolution of the K-wavefunction (\ref{kschrodinger}) can be obtained by the Hankel transform of the moment matrix $M_{ij}= m^E_{i+j}$:
\begin{equation}
    b_1^{2n} b_2^{2n-2} \dots b_n^2 = \det\left[M_{ij}\right]_{0\leq i,j \leq n} \, .\label{Hankeltransform}
\end{equation}
The moment method, while somewhat cumbersome to implement numerically due to stability issues,\footnote{The stability issues are quite severe, with some numerical experiments requiring greater than $O(n)$ decimal places of precision in order to reliably compute $b_n^E$.} is a very useful tool for computing Lanczos coefficients and we will use it in Section \ref{sec:k-comp-jt} to compute them for the ensemble dual to JT gravity.

\subsection{Universal regimes for complexity growth}\label{sec:universal-growth}
We are interested in universal behaviour of the K-complexity in chaotic systems, and how it relates to the dual spacetime geometry in holographic theories. There are two such regimes identified in \cite{Parker:2018yvk} and \cite{Barbon:2019wsy,Rabinovici:2020ryf} in systems with a bounded spectrum. We claim that the same universal behaviour also appears in chaotic Hamiltonians with spectra unbounded from above, albeit for the refined version of K-complexity defined in (\ref{kcomplexityave}) ---as we explicitly show happens in JT gravity. 
We also identify a third, less recognized, universal regime at very late times and discuss its relationship to the un-weighted spectral density. 

\paragraph{``Lanczos ascent'' and maximal scrambling}
The first universal regime of the Lanczos coefficients was identified in \cite{Parker:2018yvk}. It was argued that, for local operators in thermal systems, the sequence $b_n$ cannot grow faster than linearly $b_n \sim \alpha n$ asymptotically as $n\lesssim S \to \infty$, with a coefficient $\alpha \leq \pi T$. This bound was conjectured to be saturated in maximally chaotic theories. For systems with a meaningful Lyapunov exponent $\lambda_L$, it is bounded by (twice) the slope of the Lanczos sequence, $\lambda_L\leq 2 \alpha$. These conjectures were supported by numerical studies in the Sachdev-Ye-Kitaev model and Gaussian unitary ensemble (GUE) random matrix theory in \cite{Rabinovici:2020ryf}. We call this regime of the Lanczos sequence the ``ascent''.

In large $N$ systems, this period is characterized by scrambling and exponential operator growth as reflected in out-of-time-order correlators and a decaying two-point function $\langle O(t)O(0)\rangle$. In holographic maximally scrambling systems, this is the regime where the initial perturbation (corresponding to an insertion of the operator $O$ at the conformal boundary) falls towards the horizon (corresponding to a brief non-universal period of order 1) and becomes exponentially blueshifted, leading to universal classical backreaction in the form of 't Hooft shockwaves that depend only on the near-horizon geometry \cite{Dray:1984ha,Dray:1985yt}.  

The behaviour of the K-complexity in this time regime, defined as the average position on the K-chain via eq.~(\ref{kcomplexityave}) is also universal and exponentially growing, $C_{th}(t) \sim e^{2 \alpha t}$. The relation of the complexity to spacetime geometry is well-understood in this regime: it reflects the exponential rate of the perturbation's approach to the horizon (\cite{Qi:2018bje,Magan:2018nmu,Barbon:2019tuq,Susskind:2020gnl,Jian:2020qpp,Magan:2020iac,Haehl:2021sib} and Section \ref{sec:holographic-k-comp}). One of our goals here is to extend this understanding to later time regimes, ultimately relating the behaviour of the complexity to geometrical features behind the black hole horizon.

\paragraph{``Lanczos plateau'' and linear complexity growth}

The behaviour of the Lanczos sequence after the scrambling regime is discussed in \cite{Barbon:2019wsy}. Even after the initial operator spreads over the entire system, it is still very far from maximally complex with respect to the Liouvillian, and continues to evolve in an interesting way. In large $N$ systems obeying eigenstate thermalization, it was found in \cite{Barbon:2019wsy} that the Lanczos sequence develops a second universal regime, the Lanczos plateau.\footnote{The reader may be conditioned to think about features of the spectral form factor, we will try to prevent that by using the term ``Lanczos plateau''.} The basic intuition behind this fact is as follows: increasing Lanczos coefficients concentrate on matrix elements $\bra{E_i}O\ket{E_j}$ with increasing energy differences $E_i-E_j$. The second universal regime is approached when $E_i-E_j$ is dominated by its maximal value.\footnote{Either because we are in a finite dimensional system, or because we formulate K-complexity in every fixed average energy sector in a quantum field theory, as we argued for in the previous Section.} Once this regime is reached, the behavior of the Lanczos sequence becomes universal. We will elaborate and expand greatly on the physics of this regime in what follows.

Corresponding to  this second universal regime, the K-complexity displays a characteristic linear growth. In what follows we attempt to interpret this universal behaviour in the post-scrambling regime, where it should relate to the spacetime geometry behind the black hole horizon. But first we apply this formalism to systems with a holographic gravitational dual.

\paragraph{``Lanczos descent'' and finite size effects} The Lanczos plateau becomes exact only in the limit $D(E)= e^{2S(E)}\to \infty$, namely when the Krylov space is infinite dimensional and K-complexity has no upper bound. In the systems of interest, however, e.g. holographic quantum theories, the spectral density is finite and, as a result, the same is true for the number of states in a fixed average energy sector. The Krylov complexity chain has a finite size $D(E) \sim e^{2S(E)}$ which signals the existence of a maximal K-complexity $C\sim e^{2S(E)}$ and it means that $b_{n= D(E)}=0$. The results of \cite{Rabinovici:2020ryf} indicate that the approach of the Lanczos sequence to zero typically happens progressively,\footnote{There is no mathematical restriction on how the Lanczos sequence terminates. The progressive approach to zero, as we will explain, depends on physical properties of the system, namely the spectral density at the edge of spectrum.} via the appearance of a non-perturbatively small negative slope $\sim e^{-2S(E)}$ in the Lanczos plateau, which was dubbed the ``descent''. The same numerical studies suggest that the Lanczos descent accelerates as $n$ increases, albeit its precise form is not expected to be universal since it is sensitive to details of the spectrum. This non-perturbative feature of the Lanczos sequence controls the approach to the ``equilibration'' of K-complexity at its maximal value $C \sim e^{S}$ at timescales $t\sim O(e^S)$. In Section \ref{sec:finitesize} we will explain the appearance and shape of the Lanczos descent by computing the effect of these finite size effects on $b_n$ at leading order in $n$.

We encourage readers interested in building further intuition for the relationship between the spectral measure and the Lanczos sequence to consult Appendix \ref{sec:lanczos-dictionary}. There we explain how many characteristic features of a continuum spectral measure are encoded in the Lanczos sequence.

\section{Complexity growth in low dimensional gravity}\label{sec:complexitygravity}

In the following we apply our understanding of the Krylov complexity to holographic duals of low dimensional gravity models, namely JT gravity and two-dimensional topological gravity.
We also discuss the 2d holographic CFT setting which is dual to a static BTZ black hole in 3d.
The calculation for a near-extremal rotating BTZ black hole is effectively captured by the JT analysis.

\subsection{The 2d gravity setup}
Jackiw-Teitelboim gravity has, in the past few years, taught us important lessons about both the nature of semiclassical gravity and the black hole information paradox.
Since it gives us an unprecedented degree of control directly in the bulk, we employ it here as our fundamental example of K-complexity in gravity.
The Euclidean action is
\begin{equation}
    I_\text{JT}[M] = -\frac{S_0}{2\pi} \left[ \frac{1}{2} \int_M \sqrt{g} R + \int_{\partial M} \sqrt{h} K \right] - \left[ \frac{1}{2} \int_M \sqrt{g} \phi (R+2) + \int_{\partial M} \sqrt{h} \phi (K-1) \right] .
\end{equation}
The boundary conditions for the metric and dilaton for the cases where $\partial M$ is a product of circles consist of setting boundary circle lengths $\beta/\epsilon$ and boundary dilaton values $\gamma/\epsilon$, where $\epsilon$ is the usual holographic renormalization parameter which we imagine to be approaching zero.
As we have come to know well, this theory of Euclidean quantum gravity has a UV completion as a double-scaled Hermitian matrix model \cite{Saad:2019lba} with leading order density of states \cite{Stanford:2017thb}
\begin{equation}
    \rho_0(E) = \frac{e^{S_0}\gamma}{2\pi^2} \sinh (2\pi \sqrt{2\gamma E}) .
\label{eq:jt-leading-density}
\end{equation}
The ground state entropy $S_0$ in this theory plays the role of the inverse Newton's constant $G_N^{-1}$, analogous to the central charge $c$ of two-dimensional holographic CFTs or the number of colors $N$ in a holographic gauge theory. The perturbative genus expansion of the double-scaled matrix model exactly matches the genus expansion (in $e^{-S_0}$) of JT gravity, and doubly non-perturbative contributions can be extracted using the matrix model.
However, in addition to $S_0 \gg 1$, we will see that the ``microcanonical entropy'' $2\pi \sqrt{2\gamma E}$ must also be large, and to increase the scale of this quantity we take $S_0 \gg \gamma \gg 1$.\footnote{The boundary dilaton value $\gamma$ is often set to an arbitrary $O(1)$ number since the gravitational analysis of partition functions or correlation functions is more or less insensitive to its precise value \cite{Saad:2019lba,Saad:2019pqd}.  However, in higher dimensions, both the extremal entropy and microcanonical entropy above extremality are controlled by the same parameter, $1/G_N$, motivating our limit $S_0 \gg \gamma \gg 1$ which also appears in e.g. \cite{Ghosh:2019rcj}.  }

The thermal two-point function $\overline{G_\beta (t)}$ has been computed\footnote{We have written $\overline{G_\beta(t)}$ with a bar to distinguish the ensemble-averaged \cite{Saad:2018bqo,Saad:2019lba} two-point function, which is the result of a bulk JT gravity computation, from a microscopic two-point function in some unitary quantum mechanical system.  Also see \cite{Pollack:2020gfa} for a potential mechanism relating the ensemble to a single unitary theory.  } in detail for a free scalar field (dual to an operator $O_\Delta$ of dimension $\Delta$) directly in the bulk \cite{Mertens:2017mtv,Yang:2018gdb,Saad:2019pqd}, and is essentially given by the spectral decomposition
\begin{equation}
    \overline{G_\beta (t)} =  \int dE_1 dE_2 \rho(E_1,E_2) e^{-\frac{\beta}{2}(E_1+E_2)} e^{-it(E_1-E_2)} e^{-S_0} |\bra{E_1} O_\Delta \ket{E_2}|^2 ,
\end{equation}
where $\rho(E_1,E_2)$ is the density pair correlation function in the matrix model, and the operator matrix elements are
\begin{equation}
    |\bra{E_1} O_\Delta \ket{E_2}|^2 = \frac{|\Gamma(\Delta+i(\sqrt{2\gamma E_1}+\sqrt{2\gamma E_2})) \Gamma(\Delta+i(\sqrt{2\gamma E_1}-\sqrt{2\gamma E_2})) |^2  }{2^{2\Delta+1}\Gamma(2\Delta)} , \quad \Delta > 0 .
    \label{eq:JT-scalar-matrix-elts}
\end{equation}
 The bulk calculation which leads to these expressions involves integrating over certain families of geodesics on the disk, the disk with one handle, and the disk with two ``D-brane" boundaries added to the path integral \cite{Saad:2019pqd}.
We have reinstated the factors of $\gamma$, which appears in the boundary condition of the dilaton and also in $\rho_0(E)$, compared to \cite{Saad:2019lba,Saad:2019pqd} where it is set to $\frac{1}{2}$ and 1, respectively.
Though we do not have an exact expression for the density pair correlation function $\rho(E_1,E_2)$, we do have a ``semiclassical" expression which incorporates all leading perturbative and non-perturbative contributions from the matrix integral.
These contributions consist of the disconnected piece, a contact delta function, and the famous sine kernel:
\begin{equation}
    \rho(E_1,E_2) \approx \rho_0(E_1)\rho_0(E_2) + e^{S_0} \delta(E_1-E_2) - \frac{\sin^2 [\pi \rho_0(E_2)(E_1-E_2)]}{\pi^2(E_1-E_2)^2} .
\end{equation}
The third term, the sine kernel, is only valid near $E_1 \approx E_2$, as the exact formula for $\rho(E_1,E_2)$ is symmetric under $E_1 \leftrightarrow E_2$.
Far from $E_1 \approx E_2$, the disconnected piece $\rho_0(E_1)\rho_0(E_2)$ is a good approximation to $\rho(E_1,E_2)$.
For our purposes, namely to reproduce the ascent and plateau of the Lanczos sequence depicted in Figure~\ref{fig:seq-wvfn-kcomp}, we will see that the disconnected approximation to this quantity is sufficient.

Following our definition of K-complexity in Section~\ref{sec:k-basis-refined}, we rewrite the two-point function as
\begin{equation}
    \overline{G_\beta(t)} = \int_0^\infty dE\; e^{-\beta E} {\cal N}(E)  \int_{-2E}^{2E} d\omega \frac{\rho(E,\omega)}{{\cal N}(E)} |O_\Delta(E,\omega)|^2 e^{-i\omega t} ,
\label{eq:jt-thermal-2pt}
\end{equation}
where ${\cal N}(E)$ is given by (\ref{normalization}) and we have changed variables to the total energy $E =\frac{E_1 + E_2}{2}$ and the energy difference $\omega = E_1 - E_2$, and the densities are defined using \eqref{eq:jt-leading-density} as
\begin{equation}
    \rho(E,\omega) = \rho_0\left(E+ \frac{\omega}{2} \right) \rho_0 \left( E-\frac{\omega}{2} \right)
\end{equation}
Our plan now is to compute the Lanczos sequence for a fixed average (equivalently, total) energy sector, using the moment method of Section \ref{sec:k-basis-refined} and then use them to compute the thermal K-complexity via (\ref{kcomplexityave}). The relevant fixed average energy moments in JT gravity are
\begin{equation}
    m_{2n}^E = \frac{1}{{\cal N}(E)}\int_{-2E}^{2E} d\omega \; \sinh (2\pi \sqrt{2\gamma(E+\frac{\omega}{2})}) \sinh(2\pi \sqrt{2\gamma(E-\frac{\omega}{2})}) |\bra{E+\frac{\omega}{2}} O_\Delta \ket{E-\frac{\omega}{2}}|^2 \omega^{2n} , \label{JTmomentexplicit}
\end{equation}
where the constant ${\cal N}(E)$ ensures that $m_0^E = 1$ and the operator wavefunction is (\ref{eq:JT-scalar-matrix-elts}). Explicitly computing (\ref{JTmomentexplicit}) is beyond our abilities; nevertheless it can be computed using a saddle point approximation. The resulting Lanczos coefficients are consistent with the output of the numerical calculation.

The thermal K-complexity of a simple operator in this theory, defined by eq. (\ref{kcomplexityave}) as the thermal average of the K-complexity in different fixed $E$ sectors, is expected to exhibit an early exponential growth for about a scrambling time, reflected in an initial linear ``ascent'' of the Lanczos sequence $b^E_n\sim \alpha n$,  for $1\ll n\lesssim S(E)\sim \sqrt{\gamma E}$ \footnote{We have subtracted the extremal entropy $S_0$ from the entropy $S(E)$ that appears here and in the rest of this Section. All K-complexity time-scales that come out of our analysis below refer to this subtracted entropy, consistent with the observation of \cite{Brown:2018bms} that the extremal black hole microstates do not seem to participate in the quantum computation in the same way as the rest.} with the maximal slope $\alpha=\frac{\pi}{\beta_E}=\frac{\pi}{S'(E)}$. This should be followed by a linear complexity growth for an exponential amount of time $t\ll e^{S(E)}$, as a result of the anticipated long ``Lanczos plateau'' for $S(E) \lesssim n \ll e^{S(E)}$. The following analysis explicitly confirms this expectation, offering further support to the claim that chaotic systems, even those with infinite spectrum, possess a universal Lanczos sequence.

\subsection{Thermal K-complexity in the ensemble dual of JT gravity}
\label{sec:k-comp-jt}
While JT gravity is our main example of a maximally chaotic gravitational system, it will be illuminating to begin with a gravitational theory which is considerably simpler: topological gravity.

\subsection*{Topological gravity}
Topological gravity will serve as a useful and simple demonstration of our moment calculation techniques. The leading order density of states in topological gravity is $\rho_0(E) = \sqrt{E}$. We will see that, at least at the level of the density of states (neglecting the operator wavefunction), this theory has a complexity plateau. The microcanonical moments are
\begin{equation}
    m_{2n}^E = \frac{1}{{\cal N}(E)}\int_{-2E}^{2E} d\omega \frac{\sqrt{4E^2-\omega^2}}{\pi E^2} |O_\Delta(E,\omega)|^2 \omega^{2n} ,
\end{equation}
where ${\cal N}(E)$ is given by (\ref{normalization}) and ensures that $m_0^E = 1$.
The operator matrix elements are not important for a simple demonstration of how we compute moments and K-complexity in gravity, so we neglect them for simplicity.
(Physically, they are of course crucial, and we will return to a model which incorporates precise operator matrix elements in our treatment of JT gravity.)
Thus, for now we will drop the term $|O_\Delta(E,\omega)|^2$.
This leaves us with
\begin{equation}
    m_{2n}^E = \frac{1}{\mathcal{N}(E)} \int_{-2E}^{2E} \frac{\sqrt{4E^2-\omega^2}}{\pi E^2} \omega^{2n} .
\end{equation}
This integral is simple enough to evaluate analytically, but a saddle point approximation is much more illuminating.
The critical point equation for the integrand is quadratic in $\omega$, and has the exact solution
\begin{equation}
    \omega_* = 2E \,\sqrt{\frac{2n}{2n+1}} , \quad m_{2n}^E \approx \frac{1}{\mathcal{N}(E)} \left( \frac{8n E^2}{2n+1} \right)^n \sqrt{\frac{4}{(2n+1) \pi^2 E^2}} .
\end{equation}
The most important feature of these moments, which is very obvious from this saddle point approximation, but would be obscured by the exact integral result, is that the term raised to the power $n$ in $m_{2n}^E$ is $O(n^0)$.
This is also clear from the location of the saddle, which is a constant fraction of $E$ immediately at $n=1$.
This means the fraction $\omega_*/2E$ is never small, and the moments $m_{2n}^E$ in topological gravity are sensitive immediately to the hard edges of the microcanonical spectrum at $|\omega| = 2E$.
 
The implications of this statement for the Lanczos sequence are as follows.
The leading divergence as $n\to \infty$ of the moments is $m_{2n}^E \sim e^{2n\log E}$, and this corresponds to a sequence which is essentially constant
\begin{equation}
    b_n^E \approx E . 
\end{equation}

The infinite Lanczos plateau we have obtained in topological gravity using the microcanonical method should not be taken very seriously, as we dropped the operator wavefunction in the calculation.
However, the saddle point and moment growth regime analysis will appear in a more complicated form in JT gravity, where we hope to extract the universal form for the Lanczos sequence in maximally chaotic theories discussed in \cite{Parker:2018yvk,Barbon:2019wsy}.

\subsection*{JT gravity}

We now turn to the more interesting example of JT gravity, where we will extract the Lanczos ascent and plateau analytically.
We also comment on a potential source of the Lanczos descent in this theory.

\paragraph{The Lanczos ascent} We first focus on the small moments $1\ll n\ll S(E)\sim \sqrt{\gamma E}$ which control the early part of the Lanczos sequence $b_n^E$. The saddle point of (\ref{JTmomentexplicit}) is located at small energy differences $\omega\ll E$ for which the density can be approximated by a single exponential 
\begin{equation}
\rho(E,\omega) \to \exp \left(2\pi \sqrt{2\gamma (E+\frac{\omega}{2})} + 2\pi \sqrt{2\gamma(E-\frac{\omega}{2})}\right)\approx \exp\left(4\pi\sqrt{2\gamma E}\left(1-O(\frac{\omega^2}{E^2})\right)\right).
\end{equation}
The matrix elements of a free scalar field coupled to JT gravity are in turn given by \eqref{eq:JT-scalar-matrix-elts}. For $\omega \ll E$ they read:
\begin{align}
    |\bra{E_1} O_\Delta \ket{E_2}|^2 &= \frac{|\Gamma(\Delta+i(\sqrt{2\gamma E_1}+\sqrt{2\gamma E_2})) \Gamma(\Delta+i(\sqrt{2\gamma E_1}-\sqrt{2\gamma E_2})) |^2  }{2^{2\Delta+1}\Gamma(2\Delta)} , \quad \Delta > 0 \nonumber \\
    &= \frac{|\Gamma(\Delta+i2\sqrt{2\gamma E} +O(\omega^2/E^{3/2})) \Gamma(\Delta+\frac{i}{2}\sqrt{\frac{2\gamma}{E}}\omega +O(\omega^2/E^{3/2})) |^2  }{2^{2\Delta+1}\Gamma(2\Delta)} \label{wavefunctionsmallomega}
\end{align}
For integer $\Delta > 0$, there is a useful simplification of these elements which employs the identity
\begin{equation}
    |\Gamma(\Delta + i r)|^2 = \frac{\pi r}{\sinh(\pi r)} \prod_{k=1}^{\Delta-1} (r^2 + k^2)  . \label{gammaidentity}
\end{equation}
There is no real loss in restricting to integer $\Delta$, as the quantitative behavior we obtain will be general.
Using \eqref{gammaidentity} in \eqref{wavefunctionsmallomega}, the key observation is that the product of Gamma functions may be approximated by
\begin{equation}
    \frac{1}{\sinh(2\pi\sqrt{2\gamma E}) \sinh (\frac{\pi}{2}\sqrt{\frac{2\gamma}{E}} \omega) } \approx \,\exp \left(-S(E) -\frac{\beta_E\, \omega}{2}  \right) ,
\label{eq:operator-decay}
\end{equation}
where we expressed the exponent in terms of the microcanonical inverse temperature
\begin{equation} 
\beta_E= \pi\sqrt{\frac{2\gamma}{E}} = \frac{\partial S(E)}{\partial E} , \label{microtemperature}
\end{equation}
where $S(E)\approx 2\pi \sqrt{2\gamma E}$ the microcanonical entropy of JT gravity above extremality, at leading order in $E$. For average energies $E\sim O(\gamma)$ the microcanonical temperature in (\ref{eq:operator-decay}) is $\beta_E\sim O(1)$. The approximation employed above is valid in the parametric regime:
\begin{equation}
     \beta_E^{-1}\ll \omega \ll E .
\label{eq:linear-regime}
\end{equation}

The moments $m_{2n}^E$ may be evaluated by saddle point approximation for which the critical point equation is (to leading order in $\omega/E$)
\begin{equation}
    \frac{2n}{\omega_*} - \beta_E = 0 , 
\end{equation}
where we have leading contributions only from $\omega^{2n}$ and the exponential decay \eqref{eq:operator-decay}.
The solution and resulting moments are (simplifying both of the Gamma functions into exponentials)
\begin{equation}
    \omega_* = \frac{2n}{\beta_E} , \quad m_{2n}^E \approx \frac{1}{{\cal N}(E)} \left(  \frac{2n}{\beta_E} \right)^{2n} e^{-2n + O(n^{1/4})} .
\label{eq:linear-saddle}
\end{equation}
The saddle point location indeed satisfies the condition \eqref{eq:linear-regime} for our approximation to be valid as long as $n$ is not too small, $n \gg 1$, so the computation is consistent.
This growth of $m_{2n}^E$ corresponds to a Lanczos sequence which is growing linearly with a particular slope:
\begin{equation}
    b_n^E \sim \frac{\pi}{\beta_E} n , \quad n \gg 1 .
\end{equation}
This is indeed the linear ascent expected for a scrambling quantum system at K-complexities $1\ll n\lesssim S$. Moreover, its slope, $\alpha=\frac{\pi}{\beta_E}$ saturates its proposed upper bound $\alpha\leq \frac{\lambda_L}{2}$ with $\lambda_L$ the Lyapunov exponent ---when such a notion is meaningful--- since in the case of JT, $\lambda_L= \frac{2\pi}{\beta_E}$. 
It is also clear from this derivation why we needed $n \gg 1$ to describe the linear ascent: for the hyperbolic sine function to be approximated as an exponential, we needed the saddle point location $\omega_*$ to be large enough so that $\beta_E \omega_* \gg 1$.
This means that we do not have control over a region of moment index $n$ where the saddle point location is comparable to or smaller than $\beta_E^{-1}$.
However, this region is $O(1)$ in terms of the entropy $S(E)$ because something like $\beta_E \omega_* = 10$ would be sufficient for our approximation to hold to very high accuracy, and so this regime is not parametrically large.

\paragraph{Some remarks} The first important point about the early regime $1\ll n \ll S(E)$ is that it represents a set of moments which are more or less blind to the fact that there exists an upper bound to the microcanonical spectrum.
This blindness arises directly from the fact that the operator wavefunction is exponentially decreasing as $\omega$ increases away from zero, and this exponential decay is only truncated by the universal square root behavior $\sqrt{E\pm \frac{\omega}{2} }$ once we begin to probe the spectral edges $|\omega| \approx 2E$ very closely.
For $n \ll \sqrt{\gamma E}$, the saddle point is very well separated from the edges, and sees only the exponential decay controlling the bulk of the microcanonical spectrum $|\omega| \ll E$. In effect, the Lanczos coefficients $b_n$ capture the effective width of the spectral window within which Krylov elements $O_n$ are supported, since $b_n \propto \omega_*(n)$.

A second point worth emphasizing is that the operator wavefunction in the Liouvillian measure (\ref{JTmomentexplicit}) is crucial for obtaining the Lanczos ascent, responsible for the maximal exponential growth of K-complexity that is characteristic of scrambling. Had we instead neglected to include $|\langle E_1|O|E_2\rangle|^2$ in (\ref{JTmomentexplicit}), the same saddle point computation would have given as moments that scale as $m_{2n}^E \sim e^{n\log (nE^{3/2}\gamma^{-1/2}) + o(n)}$, and this translates to an early Lanczos sequence which grows roughly as
\begin{equation}
    b_n^E \sim \sqrt{\gamma^{-1/2} E^{3/2}\,\, n}, \quad n \ll \sqrt{\gamma E} .
\end{equation}
This early square root regime would be the behaviour of the Lanczos sequence associated to the spectral form factor. 
Indeed, this square root regime is essentially capturing the $n \sim 1$ regime we discussed briefly above, where the approximation $\omega_* \gg \beta_E^{-1}$ is invalid.
Since this regime is describing only an $O(1)$ length of the sequence, we will just consider it as a transient effect, and from now on we will pretend the linear regime extends all the way down to $n=1$.

The importance of $|O_\Delta(E,\omega)|^2$ may be somewhat surprising at first sight, as we are quite used to studying questions about black holes, such as the information paradox, using purely spectral information \cite{Cotler:2016fpe,Saad:2018bqo}. In contrast, the early growth of K-complexity appears to rely on the interplay between the black hole and the infalling excitation.
From  a quantum chaos point of view, the central role of the operator wavefunction points at a connection between the scrambling phenomenon and the Eigenstate Thermalization Hypothesis (ETH), i.e. an ansatz for the behavior of simple operator wavefunctions in the energy basis of chaotic systems -- a connection previously explored in \cite{Murthy:2019fgs}. Maximal scrambling, defined as the exponential growth of the operator complexity explored in this paper, can be translated to a precise statement about the operator matrix elements $|O_\Delta(E,\omega)|^2$ between high energy eigenstates: for $\beta_E^{-1}\ll \omega\ll E$, they decay exponentially in the energy difference $\omega$ with an exponent $\beta_E/2$ ---which is a \emph{lower bound} on this decay rate, saturated in maximal scrambling situations. Equivalently, the microcanonical two-point function of $O$ in maximal scrambling theories has a pole at at $t=i\pi \beta_E $, in the $\gamma\to \infty$ limit.

This is, of course, not a generic property of the microcanonical two-point function. It sensitively depends on the fact that the energy basis matrix elements of our simple, bounded operator $O$ decay as slowly as allowed by unitarity as a function of the energy difference $\omega$ ---a property expected in systems that thermalize. The appearance of this linear Lanczos ascent is, therefore, a signature of the chaotic nature of our theory. That is in contrast to the ``canonical'' Lanczos sequence studied in \cite{Dymarsky:2021bjq}  which seems to be uncorrelated to chaos (see also Appendix~\ref{app:canonical-k-jt}).

\paragraph{The Lanczos plateau} The precise moment at which the previous $\omega \ll E$ approximation and saddle point location breaks down is set by $\omega_*/2E \approx 1$, which occurs at
\begin{equation}
    n \sim S(E)\sim \sqrt{\gamma E} .
\end{equation}
After this point, the form of the moments $m_{2n}^E$ will be substantially modified from what we found above.

At $n \gg S(E)$, we know that the leading divergence in $m_{2n}^E$ must transition to roughly $e^{2n \log E}$.
This transition can be seen mechanically by noting that in this late regime, we have a saddle point location $\omega_* \approx 2E$, and here we are only sensitive to the region of the microcanonical spectrum where the approximations $\sinh(2\pi\sqrt{2\gamma(E\pm \frac{\omega}{2})}) \approx 2\pi \sqrt{2\gamma(E\pm \frac{\omega}{2})}$ are valid, and from that point onward we just reproduce the topological gravity moments:
\begin{equation}
    b_n^E \sim E , \quad n \gg S(E) .
\end{equation}
Notice that in this post-scrambling regime, the operator wavefunction is irrelevant and the behaviour of the Lanczos sequence is instead controlled by the square root density shape $\sqrt{E \pm \frac{\omega}{2}}$. Once again, the Lanczos coefficients capture the length of the spectral interval effectively seen by the orthogonal polynomials ---in this case the entire bandwidth of the fixed average energy sector of the Liouvillian.

All told, once we include the operator matrix elements, we have the following Lanczos sequence in JT gravity:
\begin{equation}
b_n^E \sim
\begin{cases} 
     \frac{\pi}{\beta_E} n , & n < S(E)  ,\\
     E , & n > S(E) . 
\end{cases}
\label{eq:microcanonical-jt-seq}
\end{equation}
This sequence is continuous so it is safe to assume that the parametric linear regime \eqref{eq:linear-regime} and the late constant regime are capturing all of the relevant physics aside from transient effects near $n=1$ and $n = \sqrt{\gamma E}$.

\paragraph{The Lanczos descent}
The understanding of the Lanczos ascent and plateau that we have outlined above did not rely on any non-perturbative contributions to the JT density of states.
In fact, it was sufficient to approximate the density pair correlation function with the leading order disconnected result.
Though we will not address it in this work, it is an important problem to understand whether the Lanczos descent can be extracted in the dual matrix model by incorporating higher terms in the gravitational genus expansion.

We can make the following qualitative points.
For the purpose of black hole information recovery, the connected terms in the density pair correlation function (captured at leading order by the sine kernel) are very important.
So, we might na\"ively expect that these terms also yield the Lanczos descent.
However, by the time we reach the Lanczos plateau, the saddle point of the moment integral is already far from $\omega = 0$, so the Lanczos descent is unlikely to arise from connected contributions which are only relevant in the vicinity of $\omega = 0$.
On the other hand, there are non-perturbative (in $e^{-S_0}$) contributions to the \textit{disconnected} terms which induce wiggles in the one-point function of the density.
Gravitationally, these are analogous to D-brane effects just like the sine kernel \cite{Saad:2019lba}.
The saddle point calculation of $m_{2n}^E$ can be substantially affected by these wiggles when there are only a small number of them which are relevant, as the true saddle point location gets ``stuck" on the individual eigenvalue peaks near the edge of the spectrum $\omega \approx 2E$.
This mechanism may provide a microscopic understanding of the Lanczos descent in JT gravity, but we leave a more detailed treatment for future work. 

In Section \ref{sec:finitesize} we will substantiate this intuition and describe how finite density effects control the shape of the Lanczos descent in the complexity renormalization group formalism we develop in Section \ref{sec:rmt}. We will demonstrate the validity of our effective prescription in a toy example where the microscopic calculation is tractable, but the analogous computation in JT gravity will not be pursued.

\paragraph{Thermal K-complexity} The K-complexity of the excitation $O$ about the thermal state $e^{-\beta H}$ can now be computed using the definitions (\ref{kcomplexity}) and (\ref{kcomplexityave}), and reads overall
\begin{align}
    C_{th}[O(t)]&= \frac{\int_0^\infty dE \, e^{-\beta E} {\cal N}(E) \sum_{n=0}^{D(E)} n |\phi_{E,n}(t)|^2}{\int_0^\infty dE \, e^{-\beta E} {\cal N}(E)} \, , 
\end{align}
where each $\phi_{E,n}(t)$ is the K-wavefunction (\ref{kwavefunction}) of $O(t)$ which evolves via the local K-chain equation (\ref{kschrodinger}) and is normalized so that $\sum_{n=0}^{D(E)}  |\phi_{E,n}(t)|^2 =1$, $\forall E,t$. It is straightforward to show that as a result of the Lanczos coefficients we obtained above (\ref{eq:microcanonical-jt-seq}) we have 
\begin{equation}
    C_E[O(t)]=  \sum_{n=0}^{D(E)} n |\phi_{E,n}(t)|^2 \sim \begin{cases} 
     e^{\frac{2\pi}{\beta_E} t} , & t \lesssim \beta_E \log S(E)  ,\\
     E t , & \beta_E \log S(E)\lesssim t\ll e^{S(E)} . 
\end{cases}
\label{CEJT}
\end{equation}
For time-scales short compared to $e^{S}$, $C_E \ll e^S$. On the other hand, the factor ${\cal N}(E)$ is equal to:
\begin{align}
    {\cal N}(E) &= \int_{-2E}^{2E} d\omega \; \sinh (2\pi \sqrt{2\gamma(E+\frac{\omega}{2})}) \sinh(2\pi \sqrt{2\gamma(E-\frac{\omega}{2})}) |\bra{E+\frac{\omega}{2}} O_\Delta \ket{E-\frac{\omega}{2}}|^2\nonumber\\
    &\sim \exp [S(E)]\, f(E) \, ,
\end{align}
where in the second step we used the fact the squared operator matrix elements decay with the average energy $E$ as $e^{-S(E)}$ as can be seen by the explicit formulas (\ref{wavefunctionsmallomega}) and (\ref{gammaidentity}) but can also be argued to hold in general chaotic theories, due to the ETH. $f(E)$ is a function containing all the subleading corrections that do not grow exponentially with $E$.

By virtue of this, we are justified to estimate the integrals in the thermal K-complexity formula above using a saddle point approximation which yields the final result:
\begin{align}
    C_{th}[O(t)]&\sim  \begin{cases} 
     e^{\frac{2\pi}{\beta_{E_*}} t} , & t \lesssim \beta_{E_* }\log S(E_*)  ,\\
     E_* t , & \beta_{E_*} \log S(E_*)\lesssim t\ll e^{S(E_*)} \, ,
\end{cases} \label{JTthermalKcomplexity}\\
\text{with: } &\frac{d}{dE}(S(E)-\beta E)|_{E=E_*}\approx 0 \, . \label{saddlepointKth}
\end{align}
At the saddle point (\ref{saddlepointKth}) the microcanonical temperature coincides with the canonical one, $\beta_{E_*}=\beta$, and $E_*$ is simply the expectation value of the energy in the thermal state $E_*=\Tr [e^{-\beta H} H]$.   

\subsection{Holographic 2d CFTs}

Note that there were just two key features of the JT theory (more precisely, the dual matrix model) which led to the pattern of complexity growth written above.
These key features were (1) an exponential density of states with a sharp edge and (2) an operator wavefunction which was exponentially suppressed in the energy difference $\omega$.
We note that these features are known to be present in 2d holographic CFTs quantized on a circle due to (respectively) (1) the Cardy formula for the density of states at high energy and (2) a subtle universality that is fundamentally due to Virasoro representation theory controlling the operator product expansion (OPE) coefficients \cite{Collier:2019weq}.
Let us be a bit more quantitative by recalling the relevant formulas and sketching their derivation, following closely the presentation of \cite{Collier:2019weq}.

\subsection*{Density of states}
We begin with the density of states.
In any 2d CFT quantized on a circle $S^1$, the density of states at high energies is determined by modular invariance, which is an equivalence of thermal partition functions
\begin{equation}
    Z(\beta) = Z(4\pi^2/\beta), \quad Z(\beta) \equiv \tr e^{-\beta H} .
\end{equation}
This equivalence arises from the fact that the path integral representation of $Z(\beta)$, a torus $T^2$ with cycles of length $2\pi$ and $\beta$, is invariant under a large diffeomorphism (the so-called modular S transformation) which exchanges these cycles.
As $\beta \to \infty$, the left hand side $Z(\beta)$ receives a contribution only from the vacuum state $\ket{0}$ which has a Casimir energy $\bra{0}H\ket{0} = -c/12$ on $S^1$, where $c$ is the central charge.
This yields a partition function 
\begin{equation} 
Z(\beta \to \infty) = e^{\beta c / 12} .
\end{equation}
By modular invariance, we can find the behavior as $\beta \to 0$ by replacing $\beta \to 4\pi^2/\beta$, which yields
\begin{equation}
    Z(\beta \to 0) = e^{\pi^2 c / 3\beta} .
\end{equation}
The corresponding high energy density of Virasoro primary states is simply \cite{Cardy:1986ie}
\begin{equation}
    \rho(E) = \exp \left( 2\pi \sqrt{\frac{c}{3}E} . \right) , \quad Z(\beta) = \int_0^\infty dE e^{-\beta E} \rho(E) ,
\end{equation}
where the latter Laplace transform formula is meant to hold only at leading order as $\beta \to 0$.
The derivation we have presented here only strictly applies to large energies $E \to \infty$, but under some standard holographic assumptions \cite{Hartman:2014oaa} the square root shape of the exponent in $\rho(E)$ can be shown to hold down to $E=0$ (the $M=0$ BTZ black hole, in this normalization).

A refinement of the Cardy density can be obtained by introducing the modular S kernel which acts on Virasoro characters $\chi_h(\tau)$ via
\begin{equation}
    \chi_h(-1/\tau) = \int dh' \chi_{h'}(\tau) \mathbb{S}_{h' h}[0] ,
\end{equation}
where $\tau$ is the modular parameter of the torus.
The modular S kernel comes with an operator argument (the identity in the case of the partition function or characters), as it is the generalization of the modular S matrix to torus one-point functions.
We will not be concerned with its precise form.
We need only know that it interchanges cycles on a torus with a single operator insertion.
Using the precise expression for $\mathbb{S}$ (see \cite{Collier:2019weq} for details), the refined Cardy formula which produces the image under modular S of the entire vacuum character is roughly
\begin{equation}
    \rho(E) \propto \sinh \left( 2\pi \sqrt{\frac{c}{3} E} \right) ,
\end{equation}
which is very reminiscent of the leading JT density of states \eqref{eq:jt-leading-density}.
However, we caution that this entropy has a different interpretation: here we are discussing a static BTZ black hole.
If we were instead discussing a near-extremal rotating BTZ black hole, the 2d CFT discussion of K-complexity for those states would proceed more or less identically to the JT derivation given in Section~\ref{sec:k-comp-jt}.
For a static BTZ black hole, we must continue to utilize the results of \cite{Collier:2019weq}.

\subsection*{Operator wavefunction}
Turning now to the operator wavefunction, we recall the results of \cite{Collier:2019weq} concerning the determination of OPE coefficients via Virasoro representation theory, which also governs OPE coefficients in Liouville theory (and in that context lead to, for example, the DOZZ formula).
Just as the modular S transformation sends the exponentiated Casimir energy of the ground state to the Cardy form of the high temperature partition function, there is a corresponding relationship between the universal identity OPE coefficient $C_{h h 0} = 1$ and a function $C_0(h, h_1, h_2)$
where $h_1, h_2 \sim c$ but $h$ is fixed as $c \to \infty$.
The form of $C_0$ is obtained by enforcing modular covariance of the torus two-point function of $O_{h}$, which is precisely the Euclidean correlation function that would (after continuation to Lorentzian time) determine the Lanczos sequence and K-complexity for a simple operator in a thermal 2d CFT on $S^1$.

By the state-operator correspondence, the torus two-point function is the partition function on a two-holed torus $T^2(h,h)$ with appropriate boundary conditions (those of $O_h$) at the holes.
Then general strategy for deriving $C_0$ is to write the two-holed torus partition function $Z[T^2(h,h)]$ in two channels which are related by a certain modular covariance transformation.
A channel in this context means a decomposition of the manifold into a set of three-holed spheres which are glued together: each such sphere, with boundary conditions of operators $h_1$, $h_2$, and $h_3$ produces an OPE coefficient $C_{h_1h_2h_3}$ along with some kinematic information in the form of conformal blocks.
While the total two-point function is independent of the channel, the individual conformal blocks (each associated with a Virasoro primary) make different contributions in each channel.
Generically, a single block (or set of blocks) in one channel is a linear combination of blocks in another channel, and the linear combination in question is determined fully by conformal invariance.
Finally, $C_0$ is obtained by noticing that in one channel, the identity block dominates, and this combined with the modular covariance relation determines the OPE coefficient behavior of the high energy states appearing in the other channel.

In more detail, the two relevant ways to decompose $T^2(h,h)$ into two three-holed spheres are
\begin{equation}
\begin{split}
    Z[T^2(h,h)] & = \sum_{j,\ell} Z[S^2(h,h,h_j)] \times Z[S^2(h_j,h_\ell,h_\ell] \\
    & = \sum_{j,m} Z[S^2(h,h_j,h_m)] \times Z[S^2(h,h_j,h_m)] ,
\end{split}
\end{equation}
where \cite{Collier:2019weq} refer to the first equality as the ``OPE" channel and the second as the ``necklace" channel.\footnote{Our notation is actually a bit ambiguous.  We must also specify that in the OPE channel, the sum over states $h_\ell$ joins two boundaries of the second three-holed sphere into the longitude of the resulting one-holed torus, as opposed to the meridian.  See Figure 6 in \cite{Collier:2019weq} for more detail.}
The modular covariance transformation relating a pair of blocks $(h_1',h_2')$ in the OPE channel to a pair $(h_1,h_2)$ in the necklace channel is given by
\begin{equation}
    \mathbb{K}_{h_1,h_2;h_1',h_2'}[h] = \mathbb{S}_{h_1h_1'}[h_2'] \times \mathbb{F}_{h_2h_2'} \left[ \begin{smallmatrix}
 h & h_1 \\
 h & h_1 \\
\end{smallmatrix} \right] ,
\end{equation}
where $\mathbb{S}$ is again the modular S kernel (now with a nontrivial boundary condition) and $\mathbb{F}$ is the so-called fusion kernel which interchanges the equator and prime meridian of a four-holed sphere with boundary conditions specified in the brackets.
The main fact which we would like to emphasize here is that the form of $C_0$ depends only on this product of $\mathbb{S}$ and $\mathbb{F}$, which are determined entirely by Virasoro and modular representation theory, along with some technical assumptions about identity dominance which are expected to hold in holographic theories.
So, just as in the Cardy density case, the OPE coefficient behavior we need is determined in 2d holographic CFTs largely by the enhanced conformal symmetry algebra.
When the OPE channel is dominated by identity exchange, the necklace channel blocks which contribute are simply
\begin{equation}
    \mathbb{K}_{h_1,h_2;0,0}[h] = \mathbb{S}_{h_10}[0] \times \mathbb{F}_{h_20} \left[ \begin{smallmatrix}
 h & h_1 \\
 h & h_1 \\
\end{smallmatrix} \right] \propto C_0(h,h_1,h_2) ,
\end{equation}
where $h_1$ and $h_2$ are heavy (scaling with $c$).
Using known \cite{Collier:2018exn} asymptotic forms for $\mathbb{S}_{h_10}$ and $\mathbb{F}_{h_20}$, \cite{Collier:2019weq} find that the OPE coefficient of two heavy operators with a light probe operator of conformal weight $h$ (fixed in the $c \to \infty$ limit) is\footnote{Equation (6.3) of \cite{Collier:2019weq}.}
\begin{equation}
    C_0 \sim \frac{\Gamma(h+2i\delta) \Gamma(h-2i\delta)}{\Gamma(2h)} ,
\end{equation}
where $\delta$ determines a finite difference in energies between the two heavy operators at large $c$.
We also have the heuristic relationship
\begin{equation}
    |\bra{E+\frac{\omega}{2}} O_h \ket{E-\frac{\omega}{2}}|^2 \sim |C_0|^2 , \quad \omega = 4p\delta ,
\end{equation}
where $p$ and $\delta$ are kept finite to ensure $h_1,h_2,c \to \infty$ with $h_1/c$ and $h_2/c$ fixed.
The coefficient of the diverging average energy is $E/c \sim p^2$, so the decay of the wavefunction is
\begin{equation}
    |C_0|^2 \sim \exp \left( - \sqrt{\frac{c}{E}}\, \omega \right) \, ,
\end{equation}
which is analogous to the JT result \eqref{eq:operator-decay} with the microcanonical entropy scale parameter here being the central charge instead of the dilaton boundary value.
We therefore see that a light probe operator two-point function in a holographic 2d CFT has an operator wavefunction in the static BTZ black hole sector which is almost identical to the matrix elements for a scalar in the JT theory \eqref{eq:JT-scalar-matrix-elts}.

As we saw in the JT case, the two features of (1) an exponential density of states with a hard square root edge and (2) an exponentially decaying operator wavefunction as the energy difference is increased are sufficient to recover the universal Lanczos ascent and plateau regimes.
Therefore, since we have found these same features in 2d holographic CFTs (in particular, in the sector which is dual to static BTZ black holes), our analysis of K-complexity extends naturally to this setting.
\subsection*{Higher dimensions}
In higher dimensions, the existence of an exponential density of states may be demonstrated holographically by a bulk Euclidean black hole saddle point calculation, but the form of a generic operator wavefunction is (to our knowledge) not completely under control.
Indeed, the derivation of an exponentially decaying wavefunction appears to require a large degree of control over quantum gravitational geometry fluctuations.
In the JT theory and 3d pure gravity, these fluctuations are localized on the boundary (the so-called boundary gravitons), and so there is enough control in these theories to find enough of the wavefunctions.
In four or more dimensions, an analogous derivation seems out of reach at present.

\subsection{K-complexity and the black hole interior}
\label{sec:holographic-k-comp} 
A number of works have suggested that certain measures of quantum complexity may have a semiclassical bulk dual \cite{Stanford:2014jda,Susskind:2014moa,Brown:2015lvg,Susskind:2020gnl}. Notions of state complexity \cite{Jefferson:2017sdb,Chapman:2017rqy,Hashimoto:2017fga,Kim:2017qrq,Chapman:2018hou,Hackl:2018ptj,Khan:2018rzm,Balasubramanian:2018hsu,Erdmenger:2020sup} or unitary complexity of time evolution \cite{Susskind:2018fmx,Magan:2018nmu,Caputa:2018kdj,Balasubramanian:2019wgd,Bouland:2019pvu,Brandao:2019sgy,Flory:2020eot,Balasubramanian:2021mxo} have been intriguingly linked to measures of ``size'' of the bulk universe at the given boundary time, e.g. the volume of the bulk maximal volume slice \cite{Susskind:2014moa,Belin:2018bpg,Brown:2018bms} anchored at that time, or the gravitational action in the corresponding Wheeler-de Witt diamond \cite{Brown:2015lvg}. In parallel developments, measures of operator size \cite{Qi:2018bje} have been compellingly argued to track the evolution of the \emph{radial momentum} of the bulk particle, where the radial component is defined with respect to the maximal volume slicing of the bulk spacetime. The two approaches were connected in \cite{Magan:2018nmu,Barbon:2019tuq,Barbon:2020olv,Barbon:2020uux,Magan:2020iac,Susskind:2020gnl,Haehl:2021prg,Haehl:2021sib,Haehl:2021dto} where it was observed that the boundary rate of maximal volume increase in the universe with an infalling excitation equals the particle's radial momentum, suggesting that operator size and circuit complexity of the state excited by the operator are related by the formula
\begin{equation}
    \frac{dC[O]}{dt}= 2\pi S[O] \label{sizecomplexity}
\end{equation}

In this work, we are exploring a notion of complexity associated with an excitation of a thermal state which is distinct from circuit complexity. K-complexity has the comparative advantage that its definition involves no intractable global optimization and, as this Section demonstrated, its computation is feasible in realistic holographic systems. The natural next step is, therefore, a detailed analysis of the representation of $C_{th}$ on the gravitational side of the duality. 

While we leave this investigation for future work, in light of the fact that the circuit complexity/maximal volume correspondence is supported by empirical evidence, namely the match of the leading time-dependence of the two quantities, we are inclined to perform the same comparison here. The volumes of the maximal slices in the spacetime of an AdS wormhole with an infalling excitation as a function of the (two-side symmetric) boundary time were computed in previous works, see e.g. \cite{Susskind:2020gnl}. We quote here the outcome for $\Delta V(t)$, i.e. the difference in the maximal volume between the perturbed black hole and the original one as a function of boundary time:
\begin{equation}
    \Delta V(t) = \begin{cases}
         \frac{2\pi E_{inf} L_{AdS}^2}{R_{h}} \sinh \left(\frac{R_h t}{L_{AdS}^2}\right)\, ,\quad &t\lesssim \frac{\beta}{2\pi} \log \frac{ S}{\delta S} \\
         \frac{S}{\beta}t \, ,\quad & \frac{\beta}{2\pi} \log \frac{ S}{\delta S} \lesssim t\ll \beta e^S
    \end{cases}
    \label{maxvolume}
\end{equation}
where $S$, $R_h$ are the entropy and horizon radius of the original black hole respectively, $\delta S$ is the increase of the entropy cause by the absorption of the extra particle and $E_{inf}$ is the energy of the latter. From black hole thermodynamics we know that $\frac{R_h}{L_{AdS}^2} =\frac{2\pi}{\beta}$ and $\frac{S}{\beta}=E_{BH}$, hence the behavior (\ref{maxvolume}) is identical, to leading order in $t$, with (\ref{JTthermalKcomplexity}), providing a non-trivial match of the two quantities. 

It is important to note, however, that the same time-evolution is exhibited by a number of related geometric quantities in the eternal black hole spacetime, such as measures of the distance of the particle from the asymptotic boundary along maximal volume slices, e.g. the volume seen by the particle. More careful analysis of the behavior of K-complexity is needed to disambiguate these candidate duals. Nevertheless, the physics underlying the time-dependence of all the quantities in this example is the same, hence it is interesting to contemplate the consequences of such a potential holographic relation.

The two characteristic volume growth regimes above result from different gravitational physics. The early exponential growth is a consequence of the 't Hooft shockwave-like backreaction of the infalling particle on the maximal volume due to its exponential blueshift near the horizon \cite{Dray:1984ha,Dray:1985yt,Sfetsos:1994xa,Cai:1999dz,Shenker:2013pqa} (Figure~\ref{fig:blackhole}). The exponential blueshift, in turn, follows from the near horizon Poincar\'e symmetry because Schwarzschild time acts like a Rindler boost near the horizon, while the particle's momentum $P_-$ is approximately null, and therefore $e^{iH t}P_- e^{-iHt} \sim e^{\frac{2\pi}{\beta}t}P_-$ for the appropriate normalization of $H$ and $t$.

More interestingly, the late linear regime is a direct probe of the backreacted black hole interior. The now absorbed particle has increased the mass of the black hole which has already equilibrated and its interior has expanded on one side. The maximal volume slices traverse a sizable part of this newly acquired interior in a very specific way: they exponentially accumulate near a fixed interior time-slice $r=r_* \sim R_h$ for a Schwarzschild length proportional to the boundary time $t$ (Figure~\ref{fig:blackhole}). Something that is not often emphasized in the literature is that the linear dependence of the proper volume of this interior part on $t$ is a consequence of the $t$ translation symmetry of the AdS-Schwarzschild interior metric, i.e. the spatial homogeneity of the $r=const.$ timeslices. Both characteristic regimes of (\ref{maxvolume}) are thus tied to the Schwarzschild translation symmetry of the black hole background on each side of the horizon. 

Allowing ourselves to interpret our K-complexity as a measure of this maximal volume or of the particle's distance from the boundary based merely on the empirical matching of (\ref{JTthermalKcomplexity}) and (\ref{maxvolume}) for now, we may conclude that the physics underlying the linear complexity increase of K-complexity after scrambling is related to the smooth black hole interior geometry ---as the latter is reflected by its spatial homogeneity. We shall see in the next Section that the plateau in the Lanczos coefficients that is responsible for the linear K-complexity regime, in turn, implies a parametrically large time (or complexity) regime in which the Liouvillian acts effectively as a random Hermitian matrix of fixed bandwidth drawn from the flat matrix distribution, in every fixed average energy sector. An exciting reading of this connection is that this universal RMT is a holographic glimpse of the smooth interior. If this speculation is correct, we should be able to diagnose deviations from smoothness, for example the scattering off of some excitation thrown in from the second exterior of an AdS wormhole, using the K-complexity analysis about the relevant excited state.

\section{Random matrix theory for chaotic operator dynamics}\label{sec:rmt}
In the previous Section, we presented evidence for the existence of a universal linear K-complexity growth regime in chaotic systems and in low dimensional gravity, after scrambling and for an $O(e^S)$ amount of time. The goal of this Section is to link the universal linear complexity growth to a precise dynamical property of chaotic systems. We will show that linear complexity growth implies a random matrix effective description of the Liouvillian in any fixed average energy sector, with a \emph{flat} matrix potential at large complexities.

\subsection{From Lanczos coefficients to spectral data}
\label{sec:lanczostospectrum}

A central objective of this paper is to identify the spectral properties that are responsible for the universality of linear K-complexity growth in large $N$ chaotic theories after scrambling.\footnote{In spin systems, the relationship between spectral properties of the thermal two-point function and the Lanczos sequence has been explored in a variety of models \cite{Magnus1987,Lubinsky1987ASO,Viswanath1994}.} We devise two strategies for extracting spectral signatures of this complexity regime, which we summarize below and explore in detail in Section \ref{sec:scalinglimits} and \ref{sec:environment} respectively.

\subsection*{Approach 1: Effective spectrum of ``zoomed in'' Liouvillian}
We will first approach the problem by zooming in on the Krylov chain at different scales and studying the resulting effective spectrum of the Liouvillian. An obvious, albeit na\"ive, way to do this is to explicitly truncate the K-complexity chain at some site $k$, which defines what we will call the \emph{system Liouvillian} ${\cal L}_{sys}^{(k)}$ containing the $0\leq i,j \leq k$ matrix elements of the tridiagonal matrix ${\cal L}$ (\ref{tridiagonal}). Its spectrum can be studied by considering an arbitrary function $f({\cal L}_{sys}^{(k)})$ and computing its trace in the corresponding truncated GNS subspace 
\begin{equation}
 \frac{1}{k}\text{Tr}^{(k)}_{GNS,E} [f({\cal L}_{sys}^{(k)})] = \int_{-2E}^{2E} d\omega\, w_k(\omega) f(\omega) .
\end{equation}
where $w_k(\omega)$ the effective spectral density of interest.

\begin{figure}
    \centering
    \includegraphics[width=0.9\textwidth]{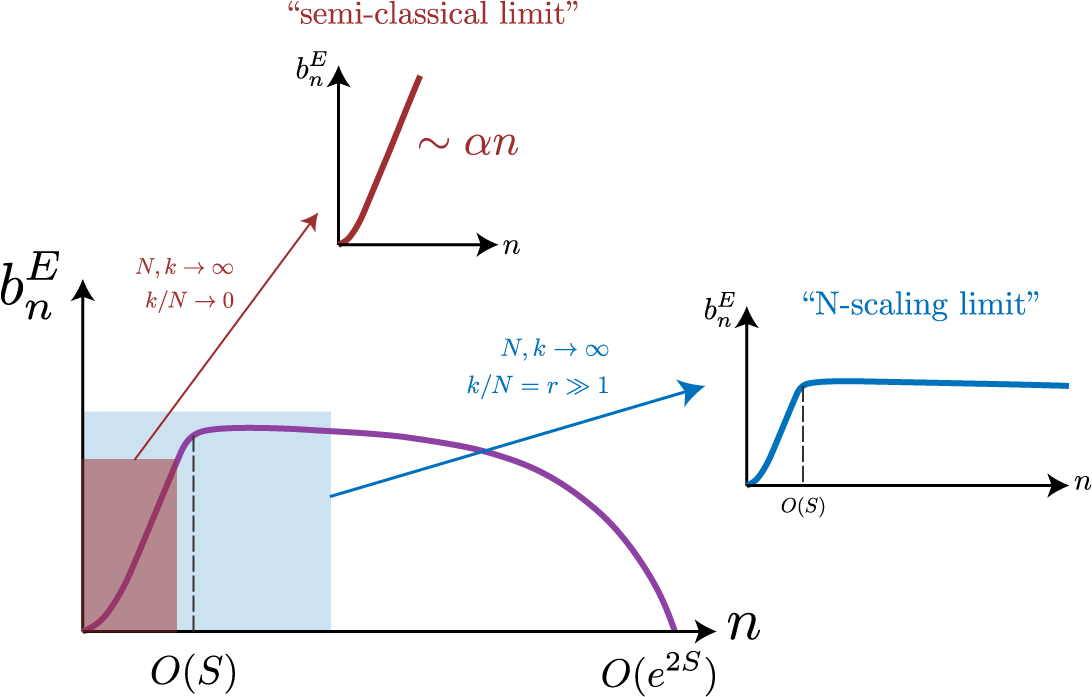}
    \caption{\small{Illustration of a Lanczos sequence in a chaotic system, exhibiting all universal regimes: an early linear ascent, a parametrically long plateau and a slow ultimate descent controlled by $O(e^{-2S})$ effects. An effective Liouvillian ${\cal L}_{sys}^{(k)}$ for the low K-complexity operator subspace, i.e the ``system'', can be constructed by truncating the sequence at a site $k$ and taking appropriate scaling limits of $k,N\to \infty$. The resulting infinitely long, ``zoomed in'' Krylov chains define an effective Liouvillian at different complexity RG scales. The two limits discussed in Section~\ref{sec:scalinglimits} are shown. In the semiclassical limit, the scrambling time is infinite and the asymptotic linear ascent describes thermal dissipation and exponential K-complexity growth ---thus outside-the-horizon physics. The $N-$scaling limit contains an infinite plateau after scrambling, as a result of which the effective Liouvillian acts like a random matrix drawn from the ``infinite well'' ensemble (\ref{infinitewell1}) at $t\gg t_{scr}$. The latter is reflected in linear K-complexity growth and holographically in behind-the-horiozon physics. }}
    \label{fig:scalinglimits}
\end{figure}

The right procedure, however, needs to be more subtle: A na\"ive truncation to the first $k$ sites is no better approximation to the properties of the theory than the use of the first $k$ orders of the Taylor expansion of the two-point function. A remedy that is available in systems with a large number $N$ of local degrees of freedom (and thus with entropy that scales as $S\sim N$) is to instead study the spectrum ${\cal L}_{sys}^{(k)}$ in different scaling limits $k\to \infty$, $N\to \infty$ (Section \ref{sec:scalinglimits} and Figure~\ref{fig:scalinglimits}).

Each scaling limit defines an effective system with an infinitely long K-chain, while at the same time zooming in the fraction of the original Krylov space we are interested in. As we discuss in the next Subsection we can understand this as a description of the original system at different \emph{complexity RG scales}. As we demonstrate, at every scale the effective Liouvillian can be understood as a \emph{random matrix}  (in every fixed average energy sector). We then show that for $k\to \infty$, $N\to \infty$, $k/N=r \gg 1$, when we maintain a sizable part of the Lanczos plateau, the matrix probability of the Liouvillian approaches the Haar measure over matrices with a spectrum confined in $\omega\in [-2E, 2E]$, what we dub the ``infinite well'' random matrix ensemble. This is the first dynamical signature of the linear complexity growth we identify in this work.

\subsection*{Approach 2: Operator resolvent for high-complexity subspace}

Another interesting way to look at the physics of the Lanczos plateau is to study its implications for the operator-weighted spectral measure $d\mu_E(\omega)$ (\ref{themeasure}). This measure combines the spectral information of the Liouvillian with the operator matrix elements and controls the two-point function, since $G_E(t)=(O|O(t))_E = \int d\mu_E(\omega) e^{-i\omega t}$. As we explain below, the high-complexity Lanczos data can be ``coarse-grained'' to obtain an effective measure. The central result of our analysis is that the operator-weighted spectral measure at high complexities approaches the Wigner semicircle when the theory posseses a parametrically long Lanczos plateau.  

Let us make this idea more precise. An efficient way to study the connection between $d\mu_E$ and the Lanczos sequence is to introduce the ``operator resolvent''

\begin{equation}
    R_0(z)= (O_0|\frac{1}{z-{\cal L}} |O_0)_E=\int  \frac{d\mu_{E}(\omega)}{z-\omega} . \label{resolvent} 
    \end{equation}
We will drop the subscript $E$ on all quantities, since we will only study the physics in a fixed average energy sector. The spectral measure $d\mu$ is encoded in the analytic structure of $R_0(z)$, $z\in \mathbb{C}$: Poles mark the location of isolated eigenvalues while branch cuts develop along continuous parts of the spectrum and describe the local spectral density:
\begin{equation}
    \lim_{\epsilon \to 0} \left( R_0(\omega+i\epsilon)-R_0(\omega-i\epsilon)\right) = \pi i \frac{d\mu(\omega)}{d\omega} \, . \label{branchcut} 
\end{equation}

The operator resolvent can be related to the Lanczos coefficients, as explained in detail in \cite{Haydock1980}. 
By virtue of the tridiagonal form of ${\cal L}$, $R_0(z)$ admits an elegant representation in terms of a continued fraction whose elements are the Lanczos coefficients $b_n$ \cite{Grosso1985},
\begin{equation}
    R_0(z) = \frac{1}{z-\frac{b_1^2}{z-\frac{b_2^2}{z-\ddots}}} . \label{continuedfrac}
\end{equation}

Our strategy for studying the effect of the Lanczos plateau on this measure will be the following. We define the operator resolvent $R_{k}(z)$ for the environment Liouvillian ${\cal L}_{env}^{(k)}$, containing the contribution to (\ref{continuedfrac}) from the Lanczos elements $b_{n\geq k}$, i.e. the large complexity Krylov subspace (Section \ref{sec:environment}). By virtue of (\ref{continuedfrac}), the microscopic $R_0(z)$ can be expressed as a finite continued fraction terminated by $R_k(z)$
   \begin{equation}
    R_0(z) = \frac{1}{z-\frac{b_1^2}{z-\frac{b_2^2}{z-\frac{\ddots}{z-b^2_{k-1}\,R_k(z)}}}} \, . \label{continuedfrac2}
\end{equation}
In the language of \cite{Viswanath1994TheRM}, $R_k(z)$ is a ``terminator function'' and it allows us to ``integrate out'' the contribution of high-complexity data to obtain a coarse-grained operator-weighted measure $d\mu_E(\omega)$.

The lesson there will be that, in chaotic theories, the existence of a parametrically long Lanczos plateau implies that the environment operator-weighted measure $d\mu^{(k)}(\omega)$\footnote{That is, the measure with respect to which \begin{equation}
    R_k(z)=\int\frac{d\mu^{(k)}(\omega)}{z-\omega} \, .
\end{equation}} for $k,N\to \infty$ with $k/N=r\gg 1$ is \emph{universal} and approaches a Gaussian unitary ensemble density. This corresponds to the terminator function
\begin{equation}
    R_k(z)= z-\sqrt{z^2-4E^2} \quad z\in \mathbb{C}\setminus [-2E,2E].
\end{equation}

By exploiting this universality of $R_k(z)$ in (\ref{continuedfrac2}), the high-complexity part of the Krylov subspace can be ``integrated out'', reducing the computation of the effective measure controlling chaotic operator evolution for $t\lesssim O(\beta e^S)$ to that of an $O(S)$, instead an $O(e^S)$, number of Lanczos coefficients.

It is worth noting that the two approaches outlined above are not independent. The scaling limit utilized by the first approach is implicitly a choice for the environment resolvent, $R_k(z)$, used to terminate the continued fraction in the second. They do, however, refer to distinct physical signatures of the Lanczos plateau and rely on different mathematical tools so it is worth presenting both perspectives.

\subsection{Complexity as an RG scale}
Before we dive into the computations summarized above, we find it useful to articulate the conceptual underpinning of our strategy in this Section. It is illuminating to approach the ideas of this Section from a Renormalization Group point of view. In our attempt to describe the physical world, the algebra of observables is rarely bestowed on us in its entirety, owing to our finite extent in space and time and our computational capabilities. The obstacle is best described as stemming from operator complexity, a practical constraint that only permits our study of quantum systems through the lens of a small subset of simple observables, relegating most of the operator algebra to a complex, inaccessible environment. Similarly, any finite amount of time evolution will, by definition, only explore a finite part of the complexity space. Operator dynamics within any chosen time-band are hence not sensitive to the microscopic details of the Liouvillian spectrum and operator wavefunction, but instead to an \emph{effective} coarse-grained  distribution over frequencies associated to the restriction of ${\cal L}$ to the appropriate complexity subspace.

In the K-complexity framework utilized in this paper, these ideas can be precisely articulated by splitting up the K-complexity chain in two parts: the interval containing sites $n=0$ to $n=k-1$ which we will refer to as the \emph{system}, and the \emph{environment} containing the high-complexity elements of the Krylov sequence, at sites $n\geq k$. The Liouvllian splits accordingly:
\begin{equation}
    \mathcal{L}=\begin{pmatrix}
    \mathcal{L}^{(k)}_{sys} & \begin{pmatrix} 0 &  & \cdots & 0\\ \vdots \\ 0 \\ b_{k} & 0 &\cdots & 0 \end{pmatrix} \\
    \begin{pmatrix} 0 & \cdots &0 & b_{k}\\ \vdots & & & 0\\ & & &\vdots \\ 0 & \cdots & &0 \end{pmatrix} & \mathcal{L}^{(k)}_{env}
\end{pmatrix} \, .\label{Ldecomposition}
\end{equation}
 We would like to think of the location $k$ of the split as defining a \emph{complexity RG scale}. This separation between system and environment is neither fundamental nor sharp and the properties of the two components of the decomposition may depend on the complexity scale of the selected divide. 
 
As alluded to in the previous Subsection, following this split we may proceed in two possible ways: directly study ${\cal L}_{sys}^{(k)}$ in an appropriate scaling limit $k\to \infty$, $N\to \infty$, or, compute the operator resolvent $R_k(z)$ of ${\cal L}_{env}^{(k)}$ in order to integrate out the high-complexity contributions to the operator evolution using the appropriate resolvent  in (\ref{continuedfrac2}). The transformation of the spectrum as we vary the scale $k$ defines a complexity RG flow whose properties we study in detail. 

\begin{figure}
    \centering
    \includegraphics[width=0.60\textwidth]{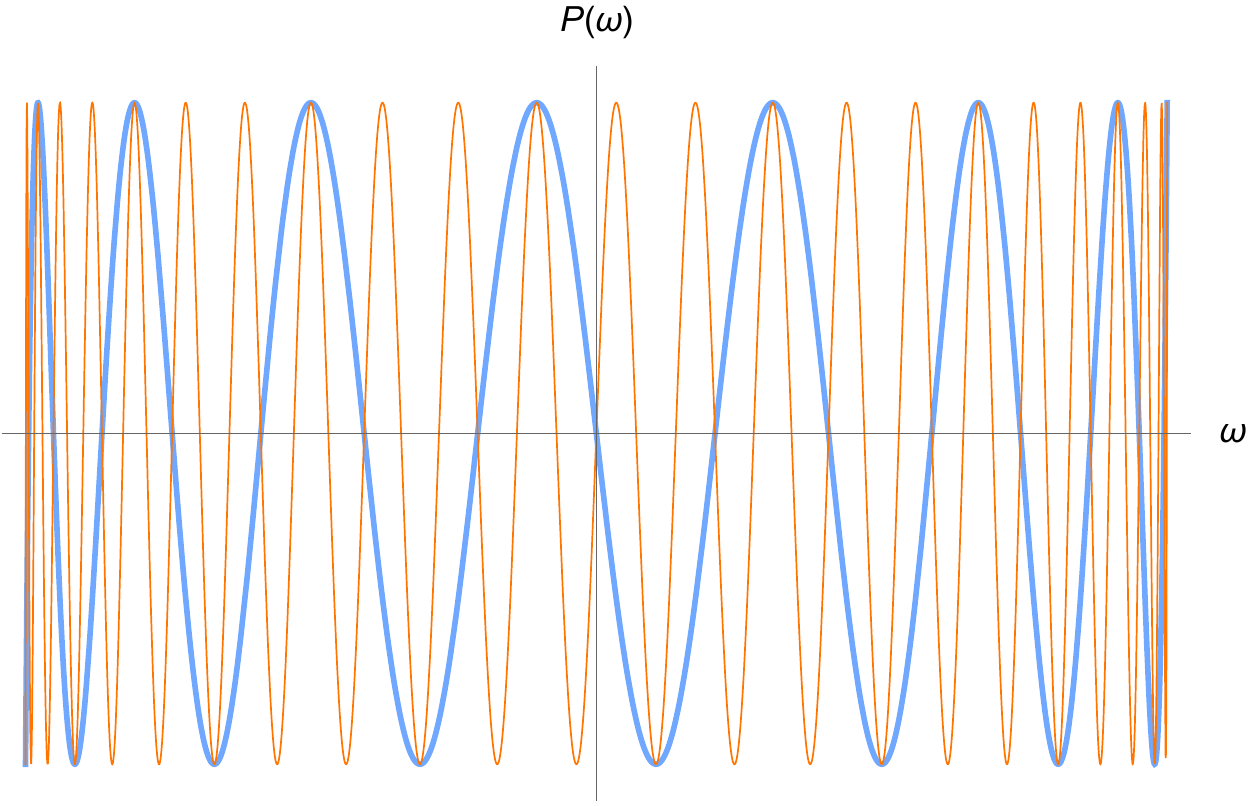}
    \caption{\small{Illustration of the behavior of the orthogonal polynomial smearing functions $P^E_n(\omega)$ that define the Krylov elements with definite K-complexity values (\ref{kbasis}) in every fixed average energy sector. Here we show an example of $n=15$ (blue) and $n=45$ (orange).}}
    \label{fig:chebyshev}
\end{figure}

\paragraph{An intuitive look at the complexity RG flow}
The matrix representation of the Krylov operators $O_n$ in the Hamiltonian eigenbasis and a fixed average energy sector $E$ is 
\begin{equation}
    [O_n]^E_{ij}= P^E_n(E_i-E_j) \langle E_i |O |E_j\rangle 
\end{equation}
The weighting functions $P_n$ are the orthogonal polynomials (\ref{kbasis}) that depend only on the location along the anti-diagonal of the matrix, i.e. the energy difference between the bra and ket states. As we increase the complexity index $n$, the ``wavelength'' of $P_n$ becomes shorter, thus probing matrix elements between eigenstates with increasingly fine energy splittings ---Figure~\ref{fig:chebyshev} illustrates this fact. The role of our complexity cutoff is to remove the very ``short wavelength modes'' and in effect coarse-grain nearby energy eigenstates. The truncation of the system along the K-complexity axis, therefore, places a \emph{constraint on our resolution of matrix elements between states with small energy differences}. The complexity RG flow we develop in this work describes the dependence of operator dynamics on this choice of resolution.

It is worth emphasizing that this is a natural choice of cutoff in general quantum systems. By the time-energy uncertainty principle, studying the properties of states with extremely small energy differences $\delta E$ requires very lengthy experiments $\delta t \sim 1/\delta E$. As a result, any observer with finite lifetime is subject to such a limitation. In the more specific context of holographic systems, the high energy part of the spectrum is densely populated by black hole microstates. From a semi-classical gravity point of view, these microstates are indistinguishable within any fixed charge sector. The coarse-graining that is implicit in the dual gravitational description of the CFT dynamics, therefore, places a similar resolution constraint.
This relationship between the black hole entropy, time-energy uncertainty, and limits on experiment complexity is an old idea \cite{Balasubramanian:2006iw}, and the complexity RG flow we introduce in this work provides a precise notion of how the cutoff on complexity may be implemented.

\subsection{Complexity RG flow and random matrix theory}
\subsubsection{Scaling limits of chaotic Liouvillian spectrum} 
\label{sec:scalinglimits}

We begin with the analysis of the ``system'' Liouvillian ${\cal L}_{sys}^{(k)}$, controlling the dynamics of the low-complexity $n\leq k$  Krylov subspace, in different scaling limits $k\to \infty, N\to \infty$. We will show that this coarse-grained Liouvillian  (in a fixed average energy sector) can be described by a random matrix drawn from a Hermitian ensemble whose potential flows to the ``infinite well'' (\ref{infinitewell1}), in the complexity regime $S\ll k \lesssim e^S$.  This is an instance of chaotic universality in operator dynamics. This universal RMT description breaks down at $k \sim e^S$ when microscopic eigenvalues become important, as we will explicitly see.

To begin, consider an arbitrary function of the system Liouvillian, $f({\cal L}_{sys}^{(k)})$.\footnote{More precisely, we should consider a polynomial of degree $m$, where $m\lesssim k$ and $k$ is large. In this case we have that the trace in the cutoff GNS Hilbert space at fixed average energy $E$ is approximately given by
\begin{equation}
    \frac{1}{k}\text{Tr}^{(k)}_{GNS,E}[f_m({\cal L}_{sys}^{(k)})] \approx \frac{1}{k}\text{Tr}^{(k)}_{GNS,E} [f_m({\cal L})]\, .
\end{equation}} 
One can then show that
\begin{equation}
    \frac{1}{k}\text{Tr}^{(k)}_{GNS,E}[f(\mathcal{L}_{sys}^{(k)})] \approx\frac{1}{k}\sum_{n=0}^k (O_n| f(\mathcal{L}) |O_n)_E = \int d\omega\,\left(\frac{1}{k}\frac{ d\mu_{E,N}(\omega)}{d\omega} \,  \sum_{n=0}^{k} {P^E_n}(\omega)^2\right)\, f(\omega) \, . \label{cutoffgnstrace} 
\end{equation}
It follows that the spectral density of the ``system'' Liouvillian is equal to:
\begin{equation}
    w_{E,k}(\omega)\,\,d\omega= \frac{1}{k} \sum_{n=0}^{k} {P^E_n}(\omega)^2\,d\mu_{E,N}(\omega) \, . \label{cutoffLRMTdensity}
\end{equation}
The subscript $N$ was introduced above as a reminder of the dependence of the measure (and the corresponding polynomials ${P^E_n}$ that are orthogonal with respect to it) on the number of degrees of freedom. This dependence is both explicit and implicit. On the one hand, the measure (\ref{themeasure}) has a functional dependence on the parameter $N$ through the spectral density and the operator wavefunction, as can be seen explicitly in the JT example (\ref{JTmomentexplicit}) where the role of $N^2$ is played by the dilaton boundary condition $\gamma$. On the other hand, the spectrum in a given fixed average $E$ sector is discrete, with an eigenvalue gap controlled by $D(E)^{-1}\sim e^{-S(E)}\sim e^{-N}$. As we take $D(E)$ larger, the level spacing decreases, until we obtain a continuous approximation at $D=\infty$. We will denote this continuous approximation to the measure as $d\mu^{cont}_{E,N}$. The interplay between these two types of $N$ dependence will be important in the analysis to follow.

\subsection*{Complexity random matrix theory}
As explained in Section \ref{sec:lanczostospectrum}, the truncated Liouvillian is of little interest unless we consider an $N\to \infty$, $k\to \infty$ scaling limit. The reader may observe that in such a limit, eq. (\ref{cutoffLRMTdensity}) is identical to the general random matrix formula (\ref{spectraldensity2}) for the equilibrium measure of a Hermitian matrix ensemble. This invites us to interpret the effective Liouvillian we obtain after our complexity coarse-graining as a random matrix, whose probability measure is determined by properties of the Krylov chain 
\begin{equation}
    \lim_{k,N\to \infty} {\cal L}_{sys}^{(k)} \quad \rightarrow \quad \text{Random Matrix.}
\end{equation}
There are a number of ways to approach the $N\to \infty$, $k\to \infty$ limits which maintain different fractions of the Krylov chain and thus probe physics at different complexity RG scales. The different scaling limits (Figure~\ref{fig:scalinglimits}) lead to distinct RMT potentials for the corresponding Liouvillian. This is the subject we turn to next.

\begin{figure}
    \centering
    \includegraphics[width=.45\textwidth]{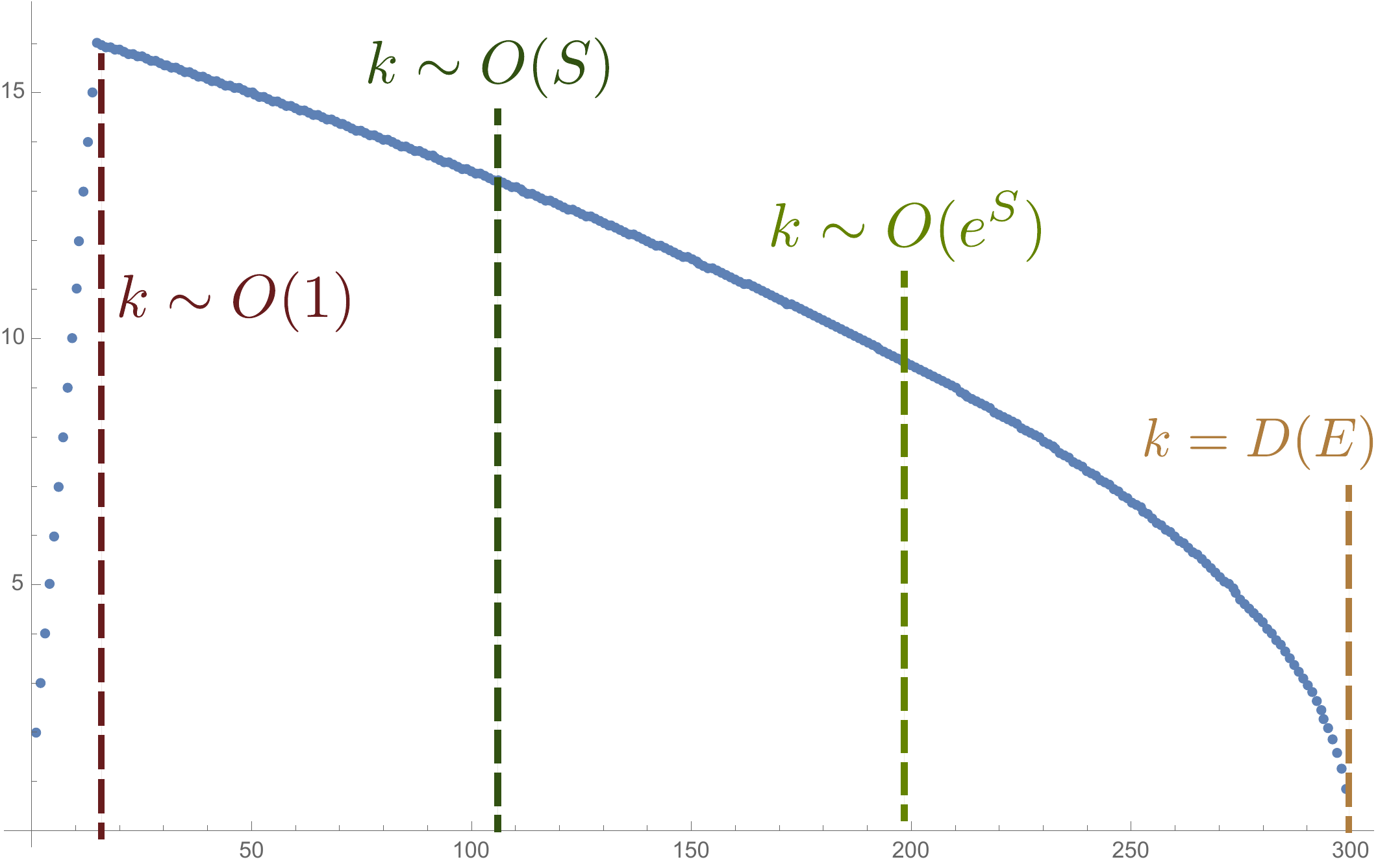}
     \includegraphics[width=.45\textwidth]{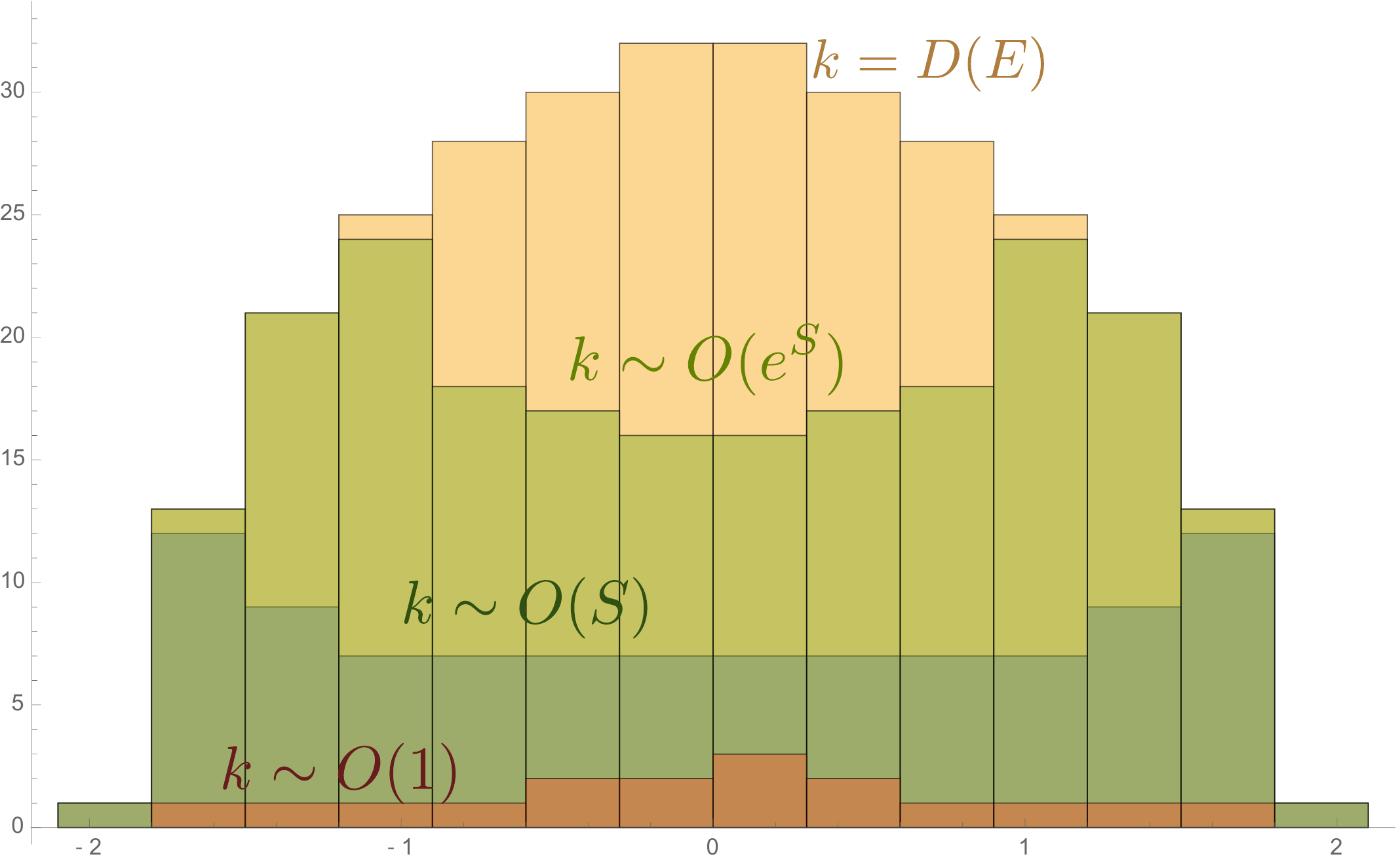}
    \caption{\small{\textbf{(Left)} An example of a Lanczos sequence $b_n$ as a function of $n$, where $S$ is the microcanonical entropy and $D(E)\sim e^{2S(E)}$ the total size of the Krylov space. Our Lanczos sequence contains all the characteristic regimes discussed in Section \ref{sec:universal-growth}: A fast early linear ascent and a parametrically long plateau, corrected by a non-perturbatively small $O(e^{-2S})$ descent rate. \textbf{(Right)} A histogram of the spectrum of ${\cal L}^{(k)}_{sys}$ ---the Liouvillian projected on a complexity subspace set by the complexity RG scale $k$. $k=D(E)$ is the spectrum of the exact microscopic ${\cal L}$, $\rho(E,\omega)$ (\ref{rhodefinition}). When $k\sim O(1)$ the spectrum is generally not universal but has an exponential tail in systems that thermalize (the exponential is somewhat obscured by the discretization in the figure but it can be confirmed by direct diagonalization of the corresponding ${\cal L}_{sys}^{(k)}$). When $k\sim O(S)$ the spectrum develops a ``belly'' that approaches the inverse semi-circle distribution (\ref{densityflow}). This is a signature of the universal ``infinite well'' random matrix ensemble controlling the dynamics in the post-scrambling regime (\ref{infinitewell1}) which underlies the linear growth of K-complexity. For $k\sim O(e^{2S})$ the effective spectrum matches the microscopic one in the regions where the inverse semicircle would exceed $\rho(E,\omega)$, as described around eq. (\ref{finitesize}) ---in the example this happens near the edge of the spectrum. In the main text we extract this RG flow of the Liouvillian spectrum analytically by taking appropriate scaling limits of the spectrum of ${\cal L}_{sys}^{(k)}$.}}
    \label{fig:RGexample}
\end{figure}

Since the analysis of this Section is somewhat subtle and technical, we demonstrate in Figure~\ref{fig:RGexample} the effective Liouvillian spectrum at different complexity RG scales for a discrete system, where the computation can be performed numerically. We advise the reader to study the figure before diving into the technical analysis that follows.

\paragraph{The ``semiclassical'' limit} The simplest way to obtain a random matrix spectral density from (\ref{cutoffLRMTdensity}) is to first take $N\to \infty$ and then $k\to \infty$. In this limit, the measure becomes continuous, since the level spacing decreases exponentially fast in $N$, and the parameter $N$ that appears explicitly in $d\mu_{E,N}^{cont}$ is taken to infinity as well:
\begin{equation}
    d\mu_{E,N} \to  d\mu^{cont}_{E,N=\infty} \, .
\end{equation}
We thus obtain the RMT equilibrium measure from (\ref{cutoffLRMTdensity}): 
    \begin{equation}
        w_{s.c.}(\omega) \, d\omega= \lim_{k\to \infty} \left(\lim_{N\to \infty} w_{E,k}(\omega)  \right)\,d\omega = \lim_{k\to \infty} \frac{1}{k} \sum_{n=0}^{k} {P^E_n}(\omega)^2\,d\mu^{cont}_{E,N=\infty}(\omega) \, . \label{semiclassicallimit}
    \end{equation}
We will call this the \emph{semiclassical limit}. Since the scrambling time of our system is $O(\log N)$ it becomes infinite in this limit and the Lanczos plateau, which begins at a K-chain location $k_c \sim O(N)$, is pushed asymptotically far  and is thus invisible. 
The quantity $k_c$ more generally represents the rough location on the K-chain where the Lanczos ascent turns over to the Lanczos plateau, and in the semiclassical limit this location diverges.
The semiclassical limit effectively zooms in the early, pre-scrambling area of the the K-chain and the Lanczos coefficients grow asymptotically as $b^E_{n\to \infty} \sim \frac{\pi\, n}{\beta_E} $ in a maximally chaotic theory (or are otherwise upper-bounded by this growth) as discussed in Section \ref{sec:universal-growth} and reference \cite{Parker:2018yvk}, and as was demonstrated in detail for JT gravity in Section \ref{sec:k-comp-jt}. As discussed in the previous Sections, the linear growth corresponds to a universal exponential tail at large frequencies $w_{s.c}(\omega)\sim e^{-\alpha \omega}$, for $\omega \gg \beta_E^{-1}$, where $\alpha \geq \beta_E/\pi$. (The entire equilibrium density $w_{s.c}(\omega)$ is not necessarily completely universal in a chaotic theory.)

\paragraph{The $N-$scaling limit} Next, we can take the large $N$ limit while maintaining an $O(rN)$ piece of the Lanczos plateau. This is achieved by the scaling limit $N\to \infty$, $k\to \infty$ with the ratio $\frac{k}{N}= \frac{2k}{\log D} = r>1$ kept fixed. Since the discreteness of the measure $d\mu_{E,N}$ is controlled by the frequency spacing of order $D^{-1}\sim e^{-N}$, $d\mu_{E,N}$ approaches the continuous measure $d\mu_{E,N}^{cont}$ parametrically faster than $N$, which also appears explicitly in the measure and the upper limit $k=r N$ of the polynomial sum in (\ref{cutoffLRMTdensity}). We can, therefore, again use the continuous approximation to $d\mu_{E,N}$ in (\ref{cutoffLRMTdensity}) but take the $N$ appearing explicitly in the measure to $\infty$ \emph{together} with the upper limit of the orthogonal polynomial sum. The eigenvalue density of $\mathcal{L}_{sys}^{(k)}$ obtained in this limit is the RMT equilibrium measure
    \begin{equation}
       w_{r}(\omega) d\omega =\lim_{N\to \infty}w_{E, k=rN} d\omega = \lim_{N\to \infty} \frac{1}{rN}\sum_{n=0}^{rN} P^E_{n}(\omega)^2 d\mu_{E,N}^{cont}(\omega) \, .\label{scalinglimit}
    \end{equation} 
It is convenient to separate the contribution coming from $P_n$ with $n>k_c = r_c N$ in (\ref{scalinglimit}) coming from the Lanczos plateau which is now included in the part of the Krylov subspace we keep. We can then utilize the asymptotic formula (\ref{P2measure}) of Theorem \ref{thm1} discussed in Appendix \ref{app:asymptotics} for the asymptotic behavior of orthogonal polynomials with Lanczos coefficients that asymptote to a constant. We find:
    \begin{align}
       w_r(\omega) d\omega &= \lim_{N\to \infty} \left( \frac{1}{rN}\sum_{n=0}^{k_c} P_{n}^2(\omega) d\mu_{E,N}^{cont}(\omega) + \frac{1}{rN}\sum_{n=k_c}^{rN} P_{n}^2(\omega) d\mu_{E,N}^{cont}(\omega) \right) \nonumber\\
       &=\frac{r_c}{r} \lim_{k_c \to \infty} \frac{1}{k_c}\sum_{n=0}^{k_c} P_{n}^2(\omega) d\mu_{E,N}^{cont}(\omega) + \frac{r-r_c}{\pi \,r} \frac{d\omega}{\sqrt{4E^2-\omega^2}} \nonumber\\
       &\approx \frac{r_c}{r} w_{s.c.}(\omega)\, d\omega + \frac{r-r_c}{\pi \,r} \frac{d\omega}{\sqrt{4E^2-\omega^2}}. \label{Nscalinglimit}
    \end{align}
where in the last step we used the fact that the contribution from early part of the Krylov chain (before we reach the Lanczos plateau) will be approximately equal to the semiclassical limit of the spectral density (\ref{semiclassicallimit}). 

Notice that our final formula (\ref{Nscalinglimit}) for the $N-$scaling limit of the Liouvillian spectral density depends on two parameters $r$ and $r_c$. The former is our complexity RG scale, controlling which fraction of the K-complexity chain we keep. Changing $r$ is a Renormalization Group transformation and induces a flow of the spectral density $w_r(\omega)$. The latter, $r_c$, is a \emph{physical} parameter demarcating the transition point between the early Lanczos ascent and the Lanczos plateau.\footnote{While there need not be a sharp transition between the two regimes, there will be an approximate scale at which the transition happens and we presume a transition scale can almost always be suitably well-defined.} As we dial our complexity RG scale, the spectral density interpolates between the two asymptotic expressions:
\begin{equation}
    w_r(\omega) = \begin{cases}
    w_{s.c.}(\omega)\,\, ,\,\,\, \,\,\,\, r\to r_c\\
    \frac{1}{\pi \sqrt{4E^2-\omega^2}} \,\, , \,\,\,\,\,\,\, r\gg r_c \, .
    \end{cases} \label{densityflow}
\end{equation}

\paragraph{Lesson 1: Universal RMT at large complexities} Equation (\ref{densityflow}) is one of the main results of this paper. The parametrically long Lanczos plateau implies a universal inverse semi-circle\footnote{A perhaps related feature appears in analysis of ETH \cite{Richter:2020bkf}.} eigenvalue distribution for the effective Liouvillian, for complexity RG scales $k_c\ll k\lesssim e^S$.
The significance of this result is more clearly illuminated by its random matrix interpretation: The inverse semi-circle is the equilibrium measure for the RMT with an ``infinite well'' potential
\begin{equation}
 V(\omega)=\begin{cases}
 0,\,\, \text{for } \omega \in [-2E,2E]\\
 \infty\,\, \text{for }\omega \notin [-2E,2E] \, .
 \end{cases} \label{infinitewell1}
\end{equation} 
At complexity RG scales $k_c\ll k \lesssim e^S$, the Liouvillian is to a good approximation a Hermitian random matrix drawn from the \emph{uniform distribution}, constrained only by the fact that its spectrum must be confined within a fixed interval (since we are working in a fixed average energy sector which has a finite bandwidth as explained in Section \ref{sec:k-basis-refined}). Only a small fraction of its eigenvalues, those distributed according to $w_{s.c}$, contain information about the particular system the Liouvillian and operator wavefunction describe. 

In chaotic systems, where $k_c \sim S$ due to the rapid rise of the Lanczos sequence to the plateau, $w_{s.c}$ will describe a minimal fraction of the sequence. And since the Lanczos plateau is parametrically long ($\sim e^S$), this fraction becomes negligible for a large range of complexities $k\gg S$. In this regime, the Liouvillian is, thus, a universal random matrix. This universality underlies the linear K-complexity growth regime discussed in the Section \ref{sec:universal-growth}. 
\paragraph{Lesson 2: Effective operator wavefunction}\label{sec:eff-operator-wfn}
So far we have highlighted the universal aspects of our low-complexity effective field theory and how it explains the universal linear growth of complexity after scrambling. But we can also use our effective field theory to calculate non-universal quantities---in particular, the time-dependent two-point function of the simple operator we used to construct our Krylov basis. In order to do this, we need to understand the \emph{effective operator wavefunction} in our cut-off theory.\footnote{It would be interesting to compare this notion to the effective field theory pursued in \cite{Altland:2021rqn}. }

One state in our microscopic theory that we would like to map into the effective theory is given by the initial $n=0$ state in \eqref{kbasis}:
\begin{equation}
        |O_{E})
    = {\cal N}(E)^{-\frac{1}{2}}\int_{-2E}^{2E} d\omega \,\rho(E,\omega)  \,O(E,\omega) \,  |E,\omega) \, .
    \label{eq:n0-state}
\end{equation}
The state in our effective theory must result in the same measure over frequencies as \eqref{eq:n0-state}, but we want to write it as the product of a new effective operator wavefunction and the new effective spectral density \eqref{cutoffLRMTdensity}, $w_{E,k}(\omega)$:
\begin{equation}
    w_{E,k}(\omega)\,\,d\omega= \frac{1}{k} \sum_{n=0}^{k} {P^E_n}(\omega)^2\,d\mu_{E,N}(\omega) \, . \label{eq:cutoffLRMTdensity-repeat}
\end{equation}
The difference between the effective theory spectral measure above and the original operator-weighted measure is just the sum over the first $k$ orthogonal polynomials. Thus, we can write a state that gives the original operator-weighted spectral measure in the form
\begin{equation}
        |\tilde O^{(k)}_{E})
    = {\cal N}(E)^{-\frac{1}{2}}\int_{-2E}^{2E} d\omega \,w_{E,k}(\omega)  \,\tilde O^{(k)}(E,\omega) \,|E,\omega)\, .
    \label{eq:n0-effective-state}
\end{equation}
with 
\begin{equation}
    \tilde O^{(k)}(E,\omega) = \left(\frac{1}{k} \sum_{n=0}^{k} {P^E_n}(\omega)^2\right)^{-1/2} \, .
    \label{eq:n0-effective-wfn}
\end{equation}
Note that this operator wavefunction \emph{is not} universal at large $k$, unlike the spectral measure (\ref{eq:cutoffLRMTdensity-repeat}).
In order to express this effective state in the frequency eigenbasis of the effective Liouvillian ${\cal L}^{(k)}_{sys}$, we need to find weights $W_i$ 
\begin{equation}
        |\tilde O^{(k)}_{E})
    = \sum_i \sqrt{W_i} \,|E,\omega_i)\, .
    \label{eq:n0-effective-state-finite}
\end{equation}
where $\omega_i$ are the eigenfrequencies of ${\cal L}_{sys}^{(k)}$ so that the effective time evolved two-point function matches the microscopic one $G_E(t)$ for the relevant time interval set by $k$:
\begin{equation}
    G_E(t)= (\tilde{O}_E^{(k)}| e^{i{\cal L}_{sys}^{(k)}t}|\tilde{O}_E^{(k)}) 
\end{equation}
 This amounts to solving the problem 
\begin{equation}
    \sum_{i=1}^k W_i f(\omega_i) = \int_{-2E}^{2E} \, d\mu_{E}(\omega) \, f(\omega)
\end{equation}
for polynomial functions $f(\omega)$ of order less than $k$.

Phrased thusly, the problem and its solution are known as Gaussian quadrature. The well-known solution is 
\begin{equation}
    W_i = \frac{a_n}{a_{n-1} {P^E_n}^{\, '}(\omega_i){P^E_{n-1}}(\omega_i)} \, ,
\end{equation}
where $a_n$ is the leading coefficient of $P^E_n$. The eigenfrequencies $\omega_i$ of ${\cal L}^{(k)}_{sys}$ are simply the roots of $P^E_n(\omega)$.

\subsubsection{Finite size effects, ``Lanczos descent'' and complexity growth slow-down}
\label{sec:finitesize}
\paragraph{The $e^{2N}-$scaling limit} Lastly, we can attempt to preserve an $O(1)$ fraction of the entire K-chain as we take the $N\to \infty$ limit. This corresponds to the case $N\to \infty$, $k\to \infty$ with the ratio $e^{-2N}k = \frac{k}{D}=t$, $t\in [0,1]$ kept fixed. The discrete measure $d\mu_{E,N}$ in (\ref{cutoffLRMTdensity}) can no longer be approximated by the continuous limit and the spectral density is sensitive to the precise microscopic spectrum of $\mathcal{L}$ described by the spectral density $\rho(E, \omega)$ (\ref{rhodefinition})
\begin{equation}
    \tilde{w}_t(\omega) d\omega = \lim_{N\to \infty} w_{E,k=tD} (\omega) d\omega= \lim_{N\to \infty} \frac{1}{tD} \sum_{n=0}^{tD} P_n^2(\omega)\,d\mu_{E,N}(\omega) \, . \label{e2Nscalingdensity}
\end{equation}
Expression (\ref{e2Nscalingdensity}), however, is formal because, unlike the previous two cases, we have not explained how the $N\to \infty$ limit is meant to be taken or whether it is even defined. In order to compute this equilibrium measure we have to employ a different strategy. 

What is new in the case where we maintain an $O(1)$ fraction of the Lanczos sequence is that the diagonalization of ${\cal L}_{sys}^{(k)}$ will start yielding subsets of the precise eigenvalues of the microscopic ${\cal L}$ (Figure~\ref{fig:RGexample}). Part of the effective spectrum $w_t(\omega)$ is thus expected to coincide with the exact $\rho(E,\omega)$ over progressively larger regions as we increase $t$.  To describe this behavior, we first observe that, by continuity and, crucially, due to the parametrically long plateau, we expect that the limit $\lim_{t\to 0} \tilde{w}_t$ of (\ref{e2Nscalingdensity}) should match with the limit $\lim_{r\to \infty} w_r $ of (\ref{Nscalinglimit}). We thus expect:
\begin{equation}
    \tilde{w}_{t\to 0}(\omega) \to \frac{1}{\pi\sqrt{4E^2-\omega^2}} \, . \label{tto0limit}
\end{equation}
As mentioned in Appendix \ref{app:rmtreview}, the spectal density (\ref{tto0limit}) is the equilibrium measure for the matrix ensemble with the ``infinite well'' potential (\ref{infinitewell1}). This extremizes the ``energy functional'':
\begin{equation}
    I[w] = -2\int d\omega_1 d\omega_2\, w(\omega_1) w(\omega_2) \log|\omega_2 - \omega_1| \, . \label{logarithmicenergy}
\end{equation}
Reference \cite{Kuijlaars/10.1137/S089547989935527X} proved that the equilibrium measure of the spectrum of $\mathcal{L}^{(k=tD)}_{sys}$  for finite $t=\frac{k}{D}$ is the \emph{unique} measure $\tilde{w}_t\, d\omega$ that extremizes (\ref{logarithmicenergy}) subject to the constraint:
\begin{equation}
    0\leq t \tilde{w}_t(\omega) \leq \rho(E,\omega) \,\, ,\,\,\,\,\,\,\,\, \int d\omega \,\tilde{w}_t = 1
    \label{finitesize}
\end{equation}
where $\rho(E,\omega)$ the microscopic spectral density of the Liouvillian in a fixed average energy sector (\ref{rhodefinition}).

\begin{figure}
    \centering
    \includegraphics[width=0.8\textwidth]{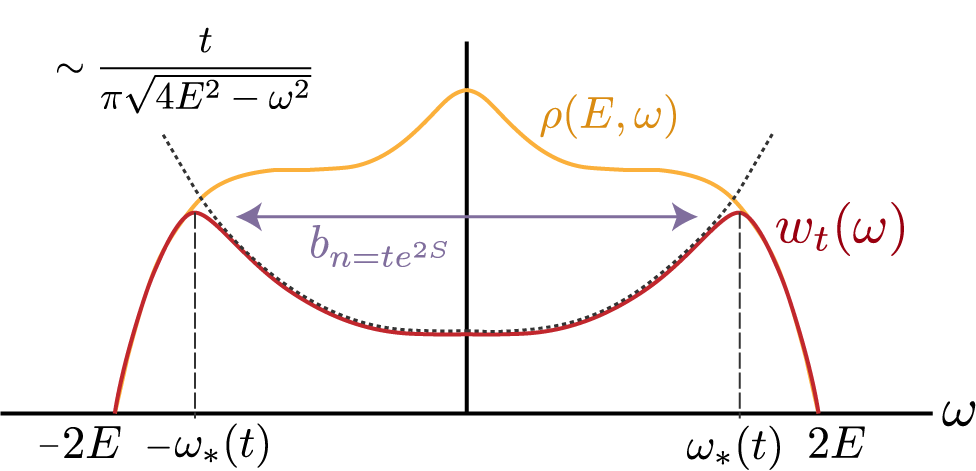}
    \caption{\small{An example of a microscopic Liouvillian spectral density $\rho(E,\omega)$, together with the spectrum $w_t(\omega)$ of the effective Liouvillian ${\cal L}_{sys}^{(k)}$ when $k$ is an $t=O(1)$ fraction of the whole Krylov sequence $D(E)$. The effective measure obtained by extremizing (\ref{logarithmicenergy}) subject to (\ref{finitesize}) consists roughly of two regions: (a) regions where $\rho(E,\omega)\geq \frac{t}{\pi\sqrt{4E^2-\omega^2}}$ and $w_t(\omega)$ is approximated by the universal inverse-semicircle form, and (b) regions where $\rho(E,\omega)\leq \frac{t}{\pi\sqrt{4E^2-\omega^2}}$ and $w_t(\omega)$ matches the precise microscopic spectrum $\rho(E,\omega)$. The overall width $\omega_*(t)$ of the former region sets the magnitude of the Lanczos coefficients $b_n$ in this late part of the Lanczos sequence, at leading order in $n$. As we increase $t$, more microscopic eigenvalues are being accessed and $\omega_*(t)$ decreases ---an $O(e^{-2S})$ effect that controls the shape of the ``Lanczos descent'' and the slow-down of K-complexity growth as it approaches saturation. }}
    \label{fig:descentRG}
\end{figure}

The intuition behind it is simple (Figure~\ref{fig:descentRG}): the eigenvalues of the truncated Liouvillian $\mathcal{L}^{(k=tD)}_{sys}$, $t\in [0,1]$ will roughly follow the ``fixed point'' equilibrium distribution (\ref{tto0limit}) as long as the local density does not exceed the actual density of the microscopic spectrum, divided by $t$ to account for the fact that the number of eigenvalues of $\mathcal{L}^{(k=tD)}_{sys}$ is a $t$ fraction of the microscopic number $D$. In turn, when $\frac{t}{\pi \sqrt{4E^2-\omega^2}}$ locally exceeds $\rho(E,\omega)$ the local density starts reflecting the precise spectrum $t^{-1}\rho(E,\omega)$.

\paragraph{Lesson 3: The shape of the Lanczos descent} The analysis of the $e^{2N}-$scaling limit above shows that the actual spectrum of the Liouvillian becomes gradually available once we recover an $O(1)$ fraction of the full Lanczos sequence. This starts roughly from the regions where the eigenvalue density is smaller than the inverse semicircle ``equilibrium'' distribution, which is typically found near the edge of the spectrum $|\omega|\sim 2E$. For any given (sufficiently small) fraction $t$ of the Krylov chain, the boundaries $\omega_* (t)$ of the region in which $w_t(\omega)=t^{-1} \rho(E,\omega)$ are effectively determined by the equation (Figure~\ref{fig:descentRG}):\footnote{We say effectively because, strictly speaking, to properly compute $\omega_*(t)$ we need to first obtain $w_t$ by extremizing the potential (\ref{logarithmicenergy}) subject to the constraint (\ref{finitesize}). This will change the inverse semicircle measure in the neighborhood of $\omega_*(t)$ in order to smoothly connect to $t^{-1}\rho(E,\omega)$ (Figure~\ref{fig:descentRG}). This subtlety, however, is not crucial for estimating the dependence of $\omega_*$ on the fraction $t$.}
\begin{equation}
    \rho(E,\omega_*(t)) = \frac{t}{\pi \sqrt{4E^2-\omega^2_*(t)}} \, .  \label{microscopicspecregions}
\end{equation}

Any interval in which the microscopic spectrum has been detected at some complexity scale $k=tD$, gets frozen; the same eigenvalues will continue to be part of the effective Liouvillian spectrum for any $t'>t$. Moreover, these regions no longer contribute to the orthogonalization of the new Krylov elements because the behavior of the orthogonal polynomials $P_{n>tD}$ in them is fixed once and for all: all their roots in these intervals coincide with the precise microscopic eigenvalues of ${\cal L}$. This means that the boundaries $\omega_*(t)$ are effectively hard walls of the spectrum, with the largest $\omega_*^{max}(t)$ setting the overall width of the interval within which the higher degree polynomials differ and thus need to be orthogonalized.

As we observed in Section \ref{sec:k-comp-jt}, the Lanczos coefficients $b_n$ are a measure of the width of the spectral interval effectively seen by the orthogonal polynomials of the corresponding degree $P_n$. Combined with the observation of the previous paragraph, this means that the leading order in $n$ form of $b_n$ at $n\sim O(e^{2S})$ is expected to be (Figure~\ref{fig:descentRG}):
\begin{equation}
    b_n \propto \omega_*^{max}(ne^{-2S(E)}) \, . \label{lanczosdescent}
\end{equation}
We argue that the prescription (\ref{lanczosdescent}) reproduces the shape of the non-perturbatively slow ``descent'' of the Lanczos coefficients observed in \cite{Rabinovici:2020ryf} caused by the finite spectral density of the system and which results in a slow-down of the linear K-complexity growth as it approaches saturation at values $e^{2S}$.

As a concrete illustration, consider the toy example of a Liouvillian whose spectrum is a Wigner semi-circle $\rho(\omega) = \sqrt{1-\omega^2}$. Then eq.~(\ref{microscopicspecregions}) and (\ref{lanczosdescent}) predict that the large Lanczos coefficients behave as:
\begin{equation}
    b_{n=te^{2S}}^{GUE}\propto \sqrt{1-t} \, . \label{GUEdescent}
\end{equation}
This is an example of a Lanczos descent. We can then directly compare the prediction (\ref{GUEdescent}) to the exact value of the ensemble averaged coefficients $b_n^{GUE,exact}$ in the finite size GUE which can be computed analytically \cite{rmtnotes}:
\begin{equation}
    b_{n=tN}^{GUE\,exact}= \frac{\sqrt{2}}{\sqrt{N}}\frac{\Gamma(N-n+\frac{1}{2})}{\Gamma(N-n)} \overset{N, n \to \infty}{\underset{n/N=t}\longrightarrow} \sqrt{2}\sqrt{1-t} \, .
\end{equation}
The prescription (\ref{microscopicspecregions}), therefore, successfully captures the non-perturbative descent, up to a numerical coefficient which ought to be fixed by matching with the value of $b_n$ at the Lanczos plateau. Notice that the Lanczos descent computed by (\ref{microscopicspecregions}) and (\ref{lanczosdescent}) is an example of a non-perturbative effect that leads to the progressive deceleration of the K-complexity growth as we approach its maximal $O(e^{2S})$ value and, in the JT dual, may furnish an effective prescription for computing the D-brane effects discussed in Section \ref{sec:k-comp-jt}.

\subsubsection{Universal GUE contribution from large complexity subspace}
\label{sec:environment}

An alternative way to glimpse at the same chaos universality found above is to directly study the operator-weighted spectral measure of the Liouvillian on the large complexity Krylov subspace, i.e the ``environment'' ${\cal L}_{env}^{(k)}$. This can be used compute the operator resolvent $R_k$ that terminates the continued fraction (\ref{continuedfrac2}) in order to integrate out high complexities. We will show that in chaotic quantum systems the operator-weighted measure of the environment is universally described by the Gaussian unitary ensemble in the $k,S\to \infty$, $k/S=r\gg 1$ limit.

\paragraph{Krylov basis re-revisited} The Krylov modes for the environment live on the $n\geq k$ part of the original K-chain. The lowest complexity vector that serves as the starting point in this Krylov construction is $|O_k)_E=P^E_k(\mathcal{L})|O)_E$. The rest of the environment basis $\mathcal{B}^k_{env}$ is composed by the high-complexity elements of $\mathcal{H}_{GNS}$ which can again be expressed as polynomials of $\mathcal{L}^{(k)}_{env}$ acting on the new starting vector $|O_k)_E$  
\begin{equation}
   \mathcal{B}^k_{env}=\big\{\,|O_{n+k})_E \equiv P_n^{(k)}(\mathcal{L}^{(k)}_{env})|O_k)_E\,\big\}_{n=0}^{D(E)-k}  \, .
\end{equation}
$P^{(k)}_n$ are now polynomial solutions to a different recurrence relation,\footnote{We suppress the superscript $E$ of the polynomials $P_n^{(k)}$ denoting the fixed average energy sector the Krylov basis is being constructed in order to minimize clutter.} i.e. the one generated by acting with $\mathcal{L}^{(k)}_{env}$:
\begin{align}
    \omega P_n^{(k)}(\omega)&= b_{n+k+1} P_{n+1}^{(k)}(\omega) +b_{n+k}P_{n-1}^{(k)}(\omega) \, , \label{associatedpol}\\
    P_0^{(k)}(\omega)&=1 \, , \nonumber\\
    P_1^{(k)}(\omega)&=\omega  \, .\nonumber
\end{align}
This new family of polynomials is closely related to the original $P_n$'s and is known in the mathematical literature as the \emph{associated polynomials of degree $k$} \cite{VANASSCHE1991237}. $P^{(k)}_n$ are also orthogonal with respect to a new measure $\mu^{(k)}$.

The $k$-dependence of the basis $\mathcal{B}^k_{env}$ can be understood physically as an ``RG transformation'' of the environment as we flow towards higher complexities. The explicit form of this transformation is provided by the theory of orthogonal polynomials and it reads:
\begin{align}
    P_n^{(k)}(\omega)&= b_k \left( P_{n+k}(\omega) Q_{k-1}(\omega) - P_{k-1}(\omega) Q_{n+k}(\omega)\right) \label{RGmap}\\
    \text{where: } Q_n(\omega)&= \int d\mu(x) \frac{P_n(x)}{\omega-x}= (O|\frac{1}{\omega-\mathcal{L}}|O_n) \, . \label{Qdefinition}
\end{align}
$Q_n$ is referred to as the Stieltjes transform of $P_n$ (\ref{stieltjes}).

\paragraph{RG flow of the environment measure} Varying our complexity RG scale $k$ induces a flow of the measure of the environement $d\mu^{(k)}$ for the associated polynomials (\ref{associatedpol}). From the mathematics of orthogonal polynomials we learn that this operator resolvent $R_k$ for the environment satisfies:
\begin{align}
   R_k(\omega)= \int \frac{d\mu^{(k)}(x)}{\omega-x}= \frac{1}{b_k}\frac{Q_{k}(\omega)}{Q_{k-1}(\omega)} \, , \label{associatemeasure}
\end{align}
where $Q_n$ are the functions of the second kind (\ref{Qdefinition}). The operator-weighted spectral measure of $\mathcal{L}_{env}^{(k)}$ at an arbitrary point along the RG trajectory is, therefore, given by:
\begin{equation}
    \frac{d\mu^{(k)}(\omega)}{d\omega}=\lim_{\epsilon\to 0}\frac{1}{2b_k}\left( \frac{Q_{k}(\omega+i\epsilon)}{Q_{k-1}(\omega+i\epsilon)}-\frac{Q_{k}(\omega-i\epsilon)}{Q_{k-1}(\omega-i\epsilon)} \right) \, . \label{spectralflow}
\end{equation}

\paragraph{Gaussian unitary ensemble fixed point} In the limit $k,S\to \infty$, $k/S=r\gg 1$ the Lanczos plateau $b_k=b$, $\forall k>k_c\sim O(S)$ becomes infinitely long. This implies a \emph{fixed point} in the complexity RG transformation (\ref{RGmap}): The elements of $\mathcal{B}^k_{env}$ and $\mathcal{B}^{k'}_{env}$, $\forall k',k>k_c\sim O(S)$ obey the same recurrence relation (\ref{associatedpol}). By the one-to-one map between polynomial recurrence relations and orthogonality measures, this directly implies that
\begin{equation}
    \mu^{(k)}=\mu^{(k')}, \,\,\,\,\text{for } \forall k,k'>k_c\sim O(S). \nonumber
\end{equation}

After substituting in eq.~(\ref{spectralflow}) the asymptotic formula (\ref{Qlimit}) of Theorem \ref{thm1}, we immediately see that this fixed point distribution is the Wigner semicircle:
\begin{equation}
   \lim_{\underset{k/S=r\gg 1} {k, S\to \infty}} \frac{d\mu^{(k)}(\omega)}{d\omega} = \frac{1}{4b^2} \sqrt{4b^2-\omega^2} \, . \label{semicircle}
\end{equation}
The Lanczos plateau is therefore equivalent to GUE random matrix theory controlling the dynamics of the ``environment'', namely the large complexity elements of the operator algebra \cite{Viswanath1990,Viswanath1991}. This random matrix theory at large complexities is \emph{universal}, in the sense that it is a fixed point of the complexity RG transformation (\ref{RGmap}) and (\ref{spectralflow}).

The results (\ref{Plimit})-(\ref{PQlimit}) can also be used in the expression for the RG transformation (\ref{RGmap}) to obtain the K-basis of the environment at the fixed point:
\begin{equation}
   \lim_{\underset{k/S=r\gg 1}{k, S\to \infty}} P_n^{(k)}(\omega)= \frac{1}{2\sqrt{\omega^2-4b^2}}\left( \left(\omega +\sqrt{\omega^2-4b^2}\right)^{n+1} -\left(\omega -\sqrt{\omega^2-4b^2}\right)^{n+1}\right) \, . \label{chebyshev}
\end{equation}
which are the \emph{Chebyshev polynomials of the second kind} and are indeed orthogonal with respect to the Wigner measure (\ref{semicircle}). 

The environment operator resolvent, therefore, admits the universal form:
\begin{equation}
    R_k(\omega)= \frac{1}{4b^2}(\omega - \sqrt{4b^2 -\omega^2})
\end{equation}
This can be used in (\ref{continuedfrac2}) in order to ``integrate out'' the contribution of the high-complexity Krylov elements and obtain the spectrum for the \emph{effective} Liouvillian ${\cal L}_{sys}^k$ in the limit $k/S=r\gg 1$, $k,S \to \infty$. 

\subsection{RMT universality, linear complexity growth and chaos}
What we discovered in this Section is that the spectrum of the Liouvillian ${\cal L}$  within any fixed average energy sector is described by a random matrix theory, when we integrate out large K-complexities as prescribed above. The potential of this effective RMT depends on the measure $\frac{d\mu_E(\omega)}{d\omega}$, given by (\ref{themeasure}), which combines the information about the spectral density $\rho(E,\omega)$ for energy eigenvalue pairs with the operator matrix elements $|O(E,\omega)|^2$, in a way that changes non-trivially as we vary the K-complexity RG scale. Since the physical meaning of this RMT and the large complexity universality found in chaotic systems may not be immediately clear, it is useful to elaborate on these issues here. 

\paragraph{The physics of the complexity RMT} By throwing away the large complexity dynamical data we lose information about the underlying spectral density of the system, including and especially the precise eigenvalues of ${\cal L}$. What remains in the low-complexity data of the ascent is the operator-weighted spectral density. 
The longer we can observe the time evolution of the operator, the finer the resolution of the spectral density we may achieve, and the more of the Lanczos ascent we are sensitive to. 

Our complexity RG flow after the ascent and into the plateau can then be thought of as a ``sampling'' protocol: As we increase the complexity scale in the plateau, we flow to a universal RMT ensemble that we use to sample the operator-weighted spectral density. The eigenvalues of the universal RMT are re-weighted in the sampling so that the spectral moments $m^E_{n}$ of the RMT ensemble measure match those of the exact operator-weighted spectral measure.

This ``sampling'' probability is determined by two effects. On the one hand, the operator wavefunction and the spectral density obviously assign a weight on the different frequencies driving evolution. On the other hand, there is a penalty when a pair of Liouvillian frequencies get too close, an effect encapsulated in the level repulsion of our RMT. The reason for the latter is perhaps less apparent but it follows from the construction of the Krylov basis and, in particular, the orthogonal polynomial smearing functions of the K-basis: the roots of those polynomials, which are closely related to the coarse-grained Liouvillian frequencies, ``repel'' each other with the same logarithmic repulsion potential of random matrix theory. The combination of these two effects underlies the random matrix physics of the complexity coarse-grained spectrum of the Liouvillian. 

\paragraph{The physics of universal linear complexity growth}  At early, pre-scrambling timescales, the RMT potential that controls the ``sampling rate'' at different parts of the spectrum, encodes coarse features of the operator-weighted spectral measure at frequencies small compared to the bandwidth of the given average energy sector. As we increase the complexity RG scale, we probe finer details of the measure while also becoming more sensitive to the finite $\omega$ bandwidth. The transition to the Lanczos plateau at complexities $O(S)$ corresponds, as we showed, to the potential of this RMT approaching an infinite well: Beyond this point, new eigenvalues are being sampled uniformly from within the given fixed $E$ window, i.e. $\omega \in [-2E, 2E]$, with the only effect shaping the resulting effective spectrum being the aforementioned Krylov polynomial ``root'' repulsion, i.e. the level repulsion of our complexity RMT. This coincides with the onset of the linear increase of K-complexity.

Linear K-complexity growth for $ \beta \log S \lesssim t\ll e^S$ is, therefore, associated to a complexity regime in which no further knowledge about the microscopic operator-weighted spectral measure is gained by accessing new, higher complexity operators. The sampling rate of new frequencies becomes universal and independent of the system of interest and the coarse-grained spectrum of ${\cal L}$ receives no updates as we increase the complexity RG scale. This, of course, ends when we gain access to an $O(1)$ fraction of the entire system, when parts of the exact spectrum get detected, typically in regions of low density first, as explained in the previous subsections. 

What is special about chaotic systems is the parametrically large size of this Lanczos plateau, $O(S)\lesssim n\ll O(e^S)$ during which no further spectral information is accessed by increasing the timescale of our observations/frequency resolution: The large scale features of the operator-weighted measure are detected quickly and efficiently early on, getting imprinted on the initial $O(S)$ piece of the Lanczos sequence, while the true spectral density and precise eigenvalue locations reveal themselves only after accessing an $O(e^S)$ part of the sequence. The intermediate ``desert'' is universal and responsible for the linear complexity growth ---a universality that according to our suggestion in Section \ref{sec:holographic-k-comp} may be related to the smoothness of the black hole interior.

\section{Discussion and musings}\label{sec:discussion}
In this paper we explored a notion of Krylov complexity which refines the one appearing in previous work \cite{Parker:2018yvk,Barbon:2019wsy,Rabinovici:2020ryf,Jian:2020qpp}.
This refined notion naturally generalizes to systems with an infinite spectrum. Using our  ``microcanonical'' definition we confirmed the existence of universal regimes in chaotic theories, characterized by the behaviour of the Lanczos sequence and the resulting time-dependence of the K-complexity: An early exponential growth characteristic of scrambling, followed by a parametrically long linear increase. Importantly, the ensemble dual to JT gravity was shown to exhibit the same universal behavior.

The linear K-complexity growth regime was then shown to reflect a universal dynamical property at post-scrambling timescales: Within every fixed average energy window, the Liouvillian generating the adjoint action of the Hamiltonian on $O(t)$ for $\log S \ll t\ll e^S$ behaves like a Hermitian random matrix with a \emph{uniform matrix probability distribution} and a fixed bandwidth, what we referred to as the ``infinite well'' RMT ensemble (\ref{infinitewell1}). 

We arrived at this result by developing a \emph{complexity renormalization group} framework that enabled us to integrate out the large K-complexity part of the Krylov subspace to obtain an effective description of the dynamics at different complexity scales. This formalism relates the data controlling the K-complexity growth to the matrix probability distribution for the effective Liouvillian. The validity of the universal RMT is spoiled by finite size effects that become important at timescales exponentially large in the entropy, albeit in a controlled way: the RMT level density progressively morphs into the microscopic one starting from spectral regions with lowest density; this effect characterizes the non-perturbative descent of the Lanczos coefficients that drives the ultimate saturation of K-complexity at its maximum value. The scope of this RMT approximation to chaotic operator dynamics in currently not fully understood. Whether it correctly reproduces higher moments of the time-evolved two-point function or continues to hold in higher-point correlators are important questions we leave for future work. The fact that the spectrum of ${\cal L}$ is independent of the particular operator we evolve in the regime $\log S \ll t\ll e^S$ is a promising sign of a universality present in all sufficiently simple correlation functions, though the validity of this claim remains to be investigated.

The precise formula for K-complexity used in this paper opens the door for an investigation of its holographic dual. A detailed analysis of the latter is of great interest but was beyond the scope of the present work. Nevertheless, the key observation which motivated this work in the first place, is that the different universal regimes of K-complexity growth and their associated time-scales (which are shared by the ensemble dual of JT gravity and are expected to persist in higher dimensions) are in direct correspondence with the volume increase of the maximal volume slices as a function of boundary time, a proposed candidate dual to circuit complexity \cite{Susskind:2014moa}.\footnote{Identification of the precise bulk dual of K-complexity may require ideas about infalling observers which have been explored previously in AdS/CFT in the interior reconstruction context \cite{Papadodimas:2012aq,Papadodimas:2013jku,Penington:2019kki,Jafferis:2020ora}.}

The early part of this universal volume growth corresponds to physics outside the black hole and reflects the near horizon Poincar\'e symmetry: The infalling particle's momentum grows exponentially near the horizon as a result of this local symmetry which then gets imprinted on the time-dependence of the volume due to backreaction. Quantum mechanically this is understood to describe the approach to thermalization and in particular maximal scrambling, a manifestation of which we see in the K-complexity formalism. The later linear volume increase probes the physics of the black hole interior and reflects the spatial translation symmetry $t_{sch}\to t_{sch}+\epsilon$ of the AdS-Schwarzschild metric behind the horizon. The infinite well RMT description of operator evolution underlying the linear K-complexity growth may thus provide the holographic counterpart of the spatial homogeneity of a smooth black hole interior, providing another example of a symmetry of the bulk background that manifests a chaotic property of its quantum dual, or as \cite{Lin:2019qwu} poetically put it: an example of order from chaos.

Perhaps the most ambitious application of the ideas explored in this work is the detection of particle collision in the interior of the black hole, for instance the scattering in the interior of a two-sided AdS wormhole of two particles, each falling in from a different exterior region.\footnote{Similar issues have recently been discussed in the context of circuit complexity \cite{Zhao:2020gxq,Haehl:2021prg,Haehl:2021sib,Haehl:2021dto}.}
This involves repeating the Krylov construction about the time evolved thermofield double state $e^{-i(H_R+H_L)t}|TFD\rangle$ with a pair of operators insertions in the two asymptotic boundaries at a sufficiently early time for their collision to take place before the maximal volume slice anchored at $t_L=t_R=t$. We may then hope to obtain a characteristic signature of this collision in the evolution of K-complexity that reflects the corresponding effect on the maximal volume slices. We hope to report on this interesting possibility in a future work.

\subsection*{Acknowledgments}

We thank Vijay Balasubramanian, Jos\'e Barb\'on, Bartek Czech, Jan de Boer, Jorge Garza-Vargas, Felix Haehl, Zohar Komargodski, Archit Kulkarni, Charles Marteau, M\'{a}rk Mezei, Eliezer Rabinovici, Julian Sonner, Leonard Susskind, Mark Van Raamsdonk, and Ying Zhao for helpful discussions.
AK and LL are supported by the Simons Foundation via the It from Qubit Collaboration. MR is supported by a Discovery grant from NSERC.

\appendix

\section{The Lanczos dictionary}\label{sec:lanczos-dictionary}

In this Appendix we elaborate on the broad phenomenology of Lanczos sequences and relate features in the Lanczos sequence to features of the corresponding spectral density and real-time two-point function. We hope this will help the reader build intuition for understanding the meaning of Lanczos data. Our discussion will rely heavily on the collection of useful results and examples in the very-readable textbook \cite{Viswanath1994TheRM}, with a few additions. We refer the reader to this textbook as a reference for material in this section, although most results were first derived elsewhere in the mathematics literature.  

For the sake of simplicity, we will consider Lanczos sequences for continuum spectral measures $\mu$, which we associate with the continuum operator-weighted spectral measures discussed above. Nevertheless, everything we discuss here applies equally to finite-dimensional, discrete measures: for $\mu$ discrete, but sufficiently dense, the early features of the Lanczos sequence for $n\ll O(e^S)$ reflect only the smooth continuum features of the measure $\mu$ and are insensitive to the location of eigenvalues. (We save a discussion of late-time features of the Lanczos sequence and relationship to the location of these eigenvalues for Section \ref{sec:finitesize}.) 

For a given spectral measure $\mu$, we will explain the relationship between features of the measure, the Lanczos sequence, and the real-time two-point function:
\begin{equation}
    \mu(\omega)\, \leftrightarrow \,
    \lbrace b_n \rbrace \,\leftrightarrow\, G(t) = \int d\mu(\omega) e^{i \omega t} \, .
\end{equation}
We summarize the relationships between these quantities in Table \ref{tab:lanczos-table}, which we will now proceed to explain in slightly more detail. 

\begin{table}[]
\centering
\def\arraystretch{1.75}
\begin{tabular}{lcc}
\hline
 \multicolumn{1}{c}{$\mu(\omega)$} & \multicolumn{1}{c}{$b_n\underset{n \rightarrow \infty}{\sim}$} & \multicolumn{1}{c}{$G(t)$} \\ \hline \hline
 \multicolumn{3}{l}{\cellcolor[HTML]{EFEFEF} Lanczos Ascent} \\ \hline
 $\exp{\left( - |\omega/\omega_0|^\gamma \right)}$ &   $\omega_0 \, n^{1/\gamma} $ & ``Slope'' \\
  \multicolumn{3}{l}{\cellcolor[HTML]{EFEFEF} Lanczos Plateau} \\ \hline
$\omega \in \left[-\omega_0,\omega_0\right]$ with \\ spectral edge $\sim  (\omega_0^2 - \omega^2)^\beta$  & $\frac{\omega_0}{2} + \frac{f(\beta)}{n^2}$ & ``Noise''  \\
  \multicolumn{3}{l}{\cellcolor[HTML]{EFEFEF} Alternating Sequences} \\ \hline
$\omega \in \left[-\omega_0,\omega_0\right]$ with \\ gap $|\omega|^\alpha$ near $\omega=0$ & $\frac{\omega_0}{2} + (-1)^n \frac{\alpha}{2n}$ & ``Ramp \& Plateau'' \\
\hline
\end{tabular}
\caption{\small{We tabulate some characteristic features of the spectral measure and the corresponding feature of the Lanczos sequence. We also identify for each spectral/Lanczos feature the corresponding characteristic behaviour of the real-time two point function, labelled by their commonly-used name. The ramp and plateau are famously universal features which appear in the JT gravity spectral form factor at leading semiclassical order in $e^{-S_0}$ \cite{Saad:2019lba}, and which are also manifested in any simple operator two-point function.  In the spectral measure of JT gravity, they arise from the sine kernel and contact terms in the density pair correlation function of the dual Hermitian matrix model.  The noise may be less familiar, as it only appears at leading order in the microcanonical two-point function.  A two-point function computed with a spectral measure that has compact support will oscillate with some frequency; this effect is invisible at least at leading order in the canonical ensemble of JT because the JT spectrum is double-scaled and thus non-compact. 
}}
\label{tab:lanczos-table}
\end{table}

\paragraph{Spectral decay and the growth of Lanczos coefficients}

Higher moments $m_n(\mu)$ of a spectral measure $\mu$ will increasingly concentrate on the largest frequencies supported by the measure. 
How quickly the moments $m_n$ grow with $n$, and correspondingly how quickly the Lanczos coefficients grow with $n$, depends on how the measure falls off in frequency: a measure that falls off rapidly in frequency will lead to moments that grow slowly until the polynomial in $n$ growth overwhelms the suppression of the measure. As a simple example, the discrete measure 
\begin{equation}
    \mu_2 = \delta(\omega-\omega_1) + e^{-K} \delta(\omega-\omega_2)  \quad, \quad \omega_2 > \omega_1
\end{equation}
has moments that grow like $m_n \sim \omega_1^n$ until $e^{-K}\omega_2^n\approx 1$ and then grow like $m_n \sim \omega_2^n$. 

For a measure with non-compact support and asymptotic fall-off
\begin{equation}
    \mu(\omega) \underset{\omega\rightarrow \infty}{\sim} \exp{\left( - |\omega/\omega_0|^\gamma \right)}
\end{equation}
one can show that the Lanczos coefficients grow asymptotically as
\begin{equation}
    b_n \underset{n \rightarrow \infty}{\sim} \omega_0 \, n^{1/\gamma} \, .
    \label{eq:lanczos-growth}
\end{equation}
In particular, then, linear growth in the Lanczos coefficients corresponds to a decay of the spectral measure of the form $\mu \sim e^{-\alpha |\omega|}$. And a spectral measure that falls off faster than the above will lead to a slower power-law growth of the Lanczos coefficients. 

\begin{figure}
     \centering
     \begin{subfigure}[b]{0.32\textwidth}
         \centering
         \includegraphics[width=\textwidth]{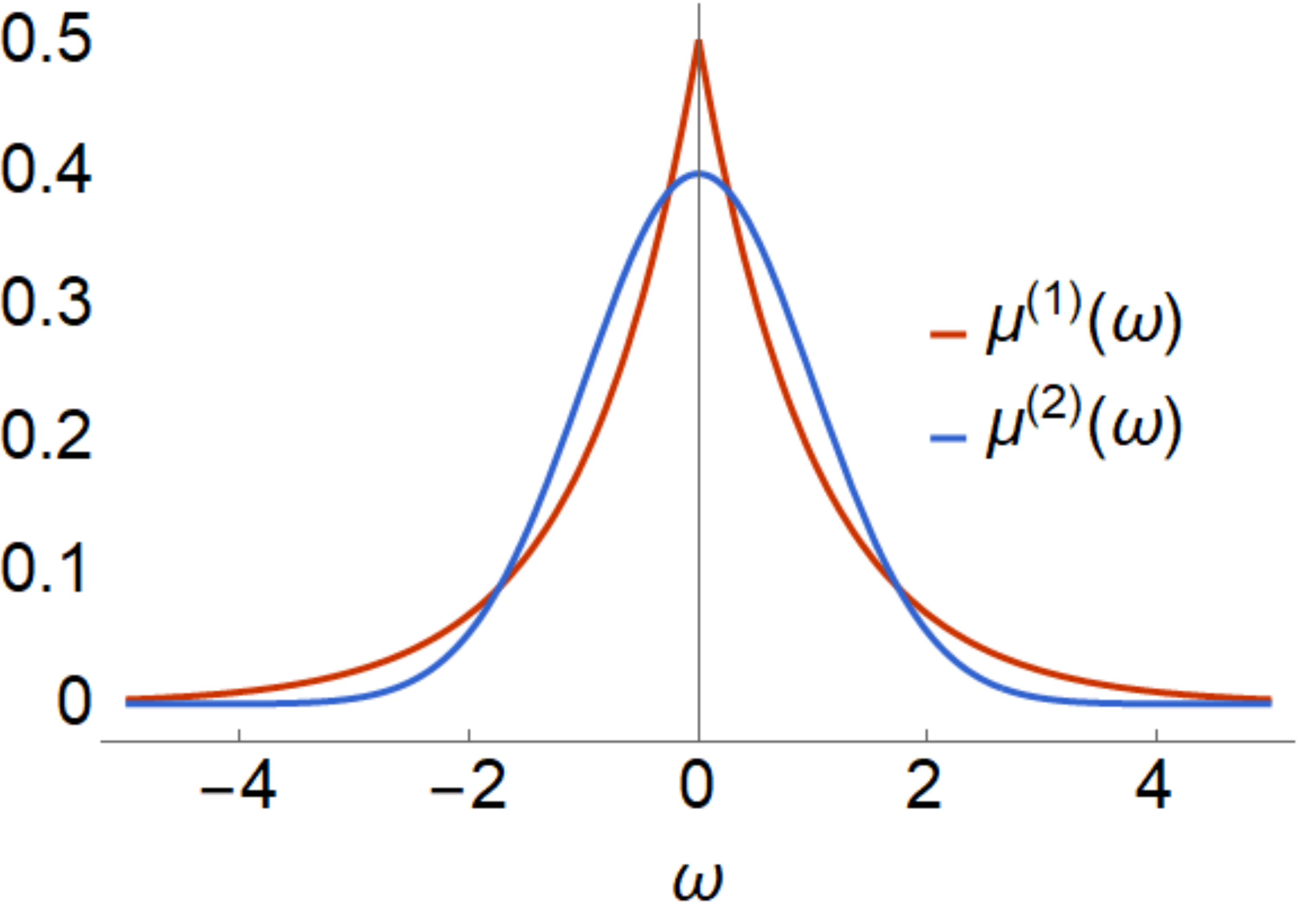}
         \caption{}
         \label{fig:lanczos-growth-w}
     \end{subfigure}
     \hfill
     \begin{subfigure}[b]{0.32\textwidth}
         \centering
         \includegraphics[width=\textwidth]{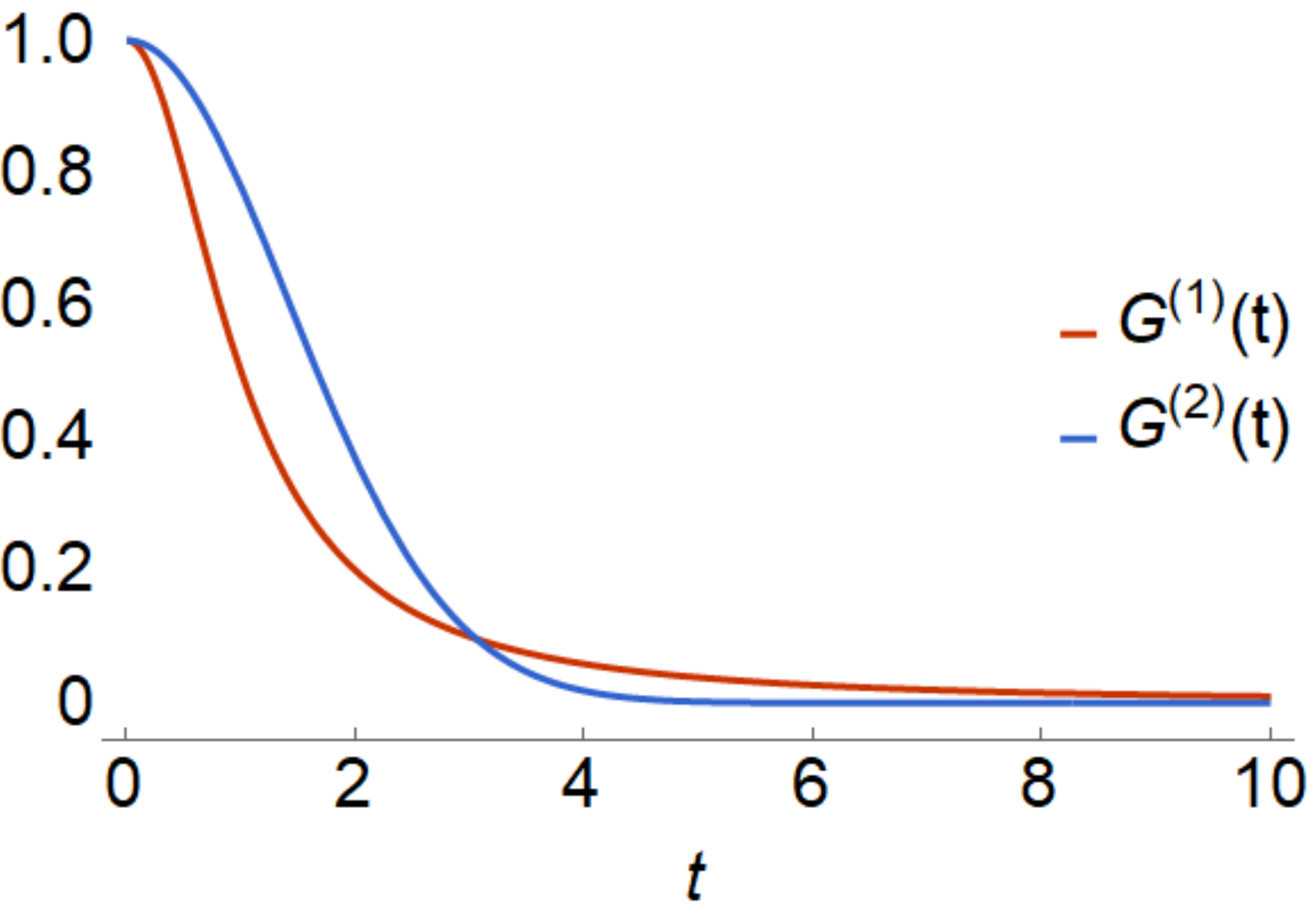}
         \caption{}
         \label{fig:lanczos-growth-t}
     \end{subfigure}
     \hfill
     \begin{subfigure}[b]{0.32\textwidth}
         \centering
         \includegraphics[width=\textwidth]{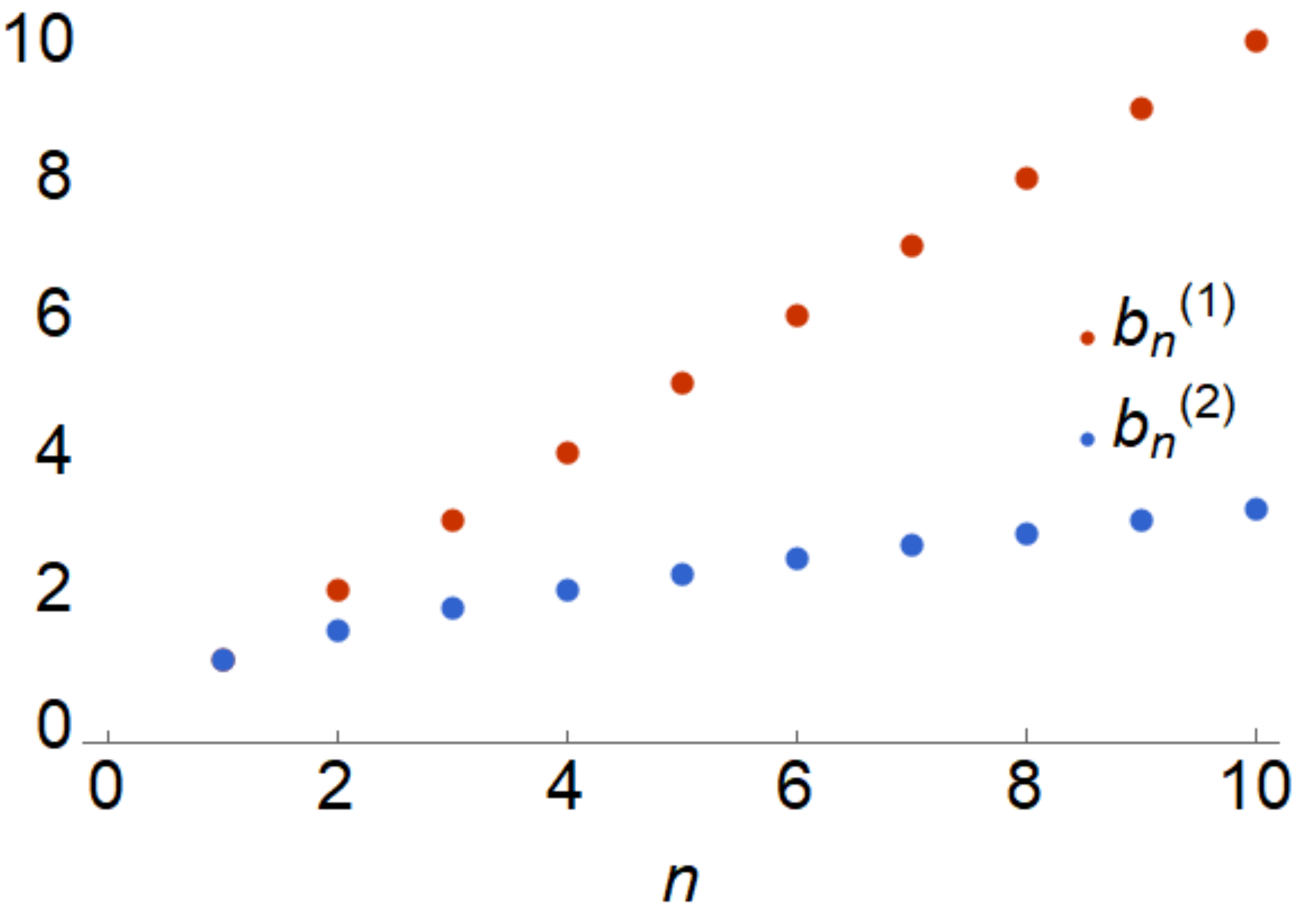}
         \caption{}
         \label{fig:lanczos-growth-b}
     \end{subfigure}
        \caption{\small{The \textbf{(a)} spectral measures, \textbf{(b)} real-time two-point functions, and \textbf{(c)} Lanczos sequences for an exponential (red) and Gaussian (blue) decay shape of the spectral measure \eqref{eq:unbounded-measures}.  Both of these have non-compact support and their Lanczos sequences grow indefinitely, though at different asymptotic rates.}}
        \label{fig:lanczos-growth}
\end{figure}

In Figure~\ref{fig:lanczos-growth-w}, we illustrate the behaviour of two of the simplest prototypical measures
\begin{equation}
    \mu^{(1)} = \frac{1}{2}e^{- |\omega|} \quad , \quad \mu^{(2)} = \frac{1}{\sqrt{2\pi}}e^{- \frac{1}{2}\omega^2} \, , \label{eq:unbounded-measures} 
\end{equation}
with corresponding Lanczos sequences (Figure~\ref{fig:lanczos-growth-b})
\begin{equation}
    b_n^{(1)} = {n} \quad , \quad b_n^{(2)} = n^{1/2}  \, ,
\end{equation}
and two-point functions (Figure~\ref{fig:lanczos-growth-t})
\begin{equation}
    G^{(1)}(t) = \frac{1}{1+t^2} \quad , \quad G^{(2)}(t) = e^{-\frac{1}{2}t^2} \, .
\end{equation}

Note that, even when the measure has compact support, or has a particular fixed decay rate only over some long regime, the Lanczos coefficients will grow according \eqref{eq:lanczos-growth} for a range of $n$ where the moments are dominated by the characteristically decaying region of the measure. 

\paragraph{Bounded spectral support and the Lanczos plateau}

When a spectral measure has bounded support, $|\omega| \leq \omega_0$, then the moments are necessarily bounded
\begin{equation}
    m_n \leq \omega_{0}^n
\end{equation}
corresponding to an asymptotic bound on Lanczos coefficients
\begin{equation}
    b_n \underset{n \rightarrow \infty}{\leq} \frac{\omega_0}{2} \, .
\end{equation}
In fact, as the moments get large, they increasingly concentrate around $\omega_0$ and this bound becomes the asymptotic plateau of the $b_n$. 

\begin{figure}
     \centering
     \begin{subfigure}[b]{0.32\textwidth}
         \centering
         \includegraphics[width=\textwidth]{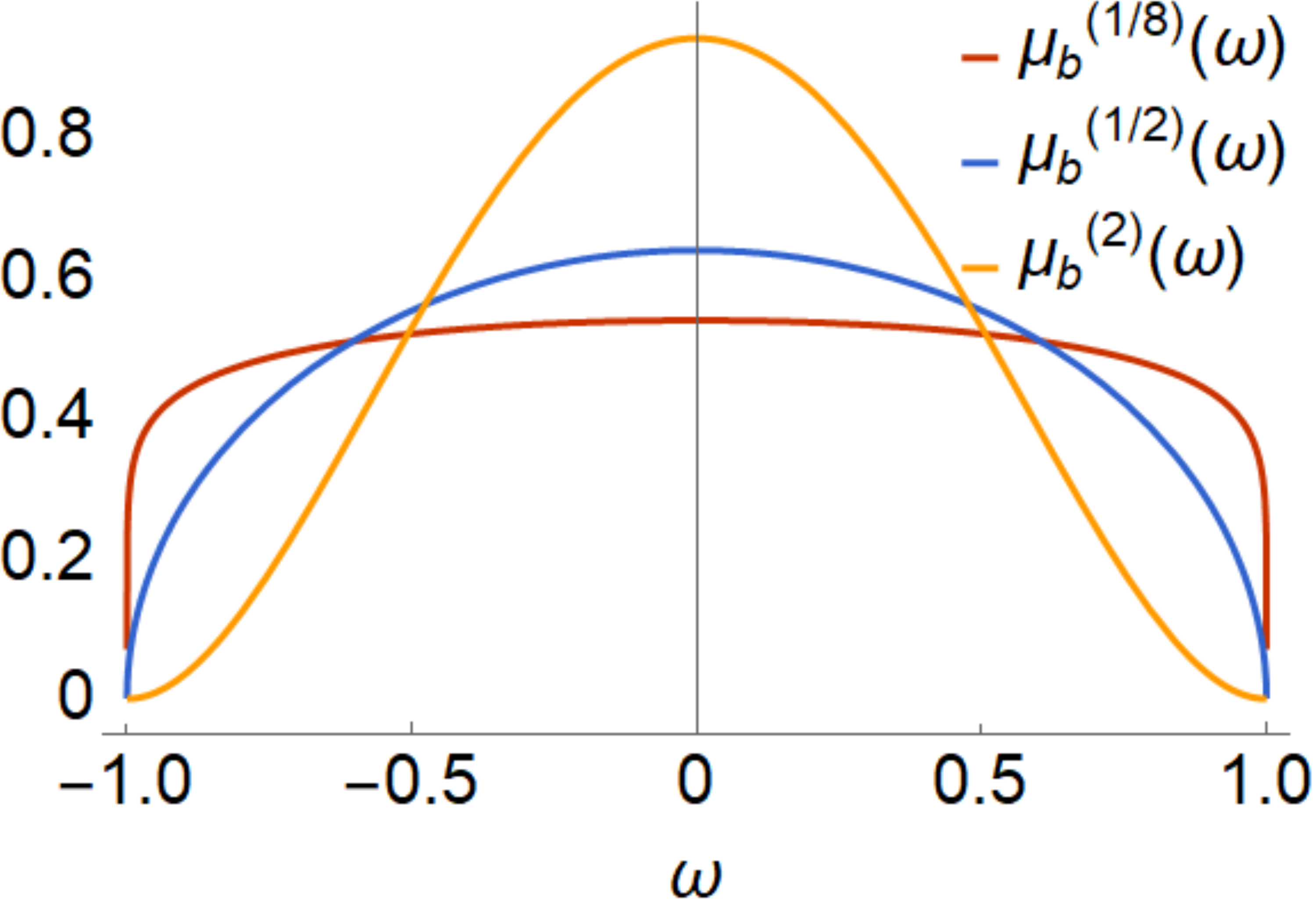}
         \caption{}
         \label{fig:lanczos-approach-w}
     \end{subfigure}
     \hfill
     \begin{subfigure}[b]{0.32\textwidth}
         \centering
         \includegraphics[width=\textwidth]{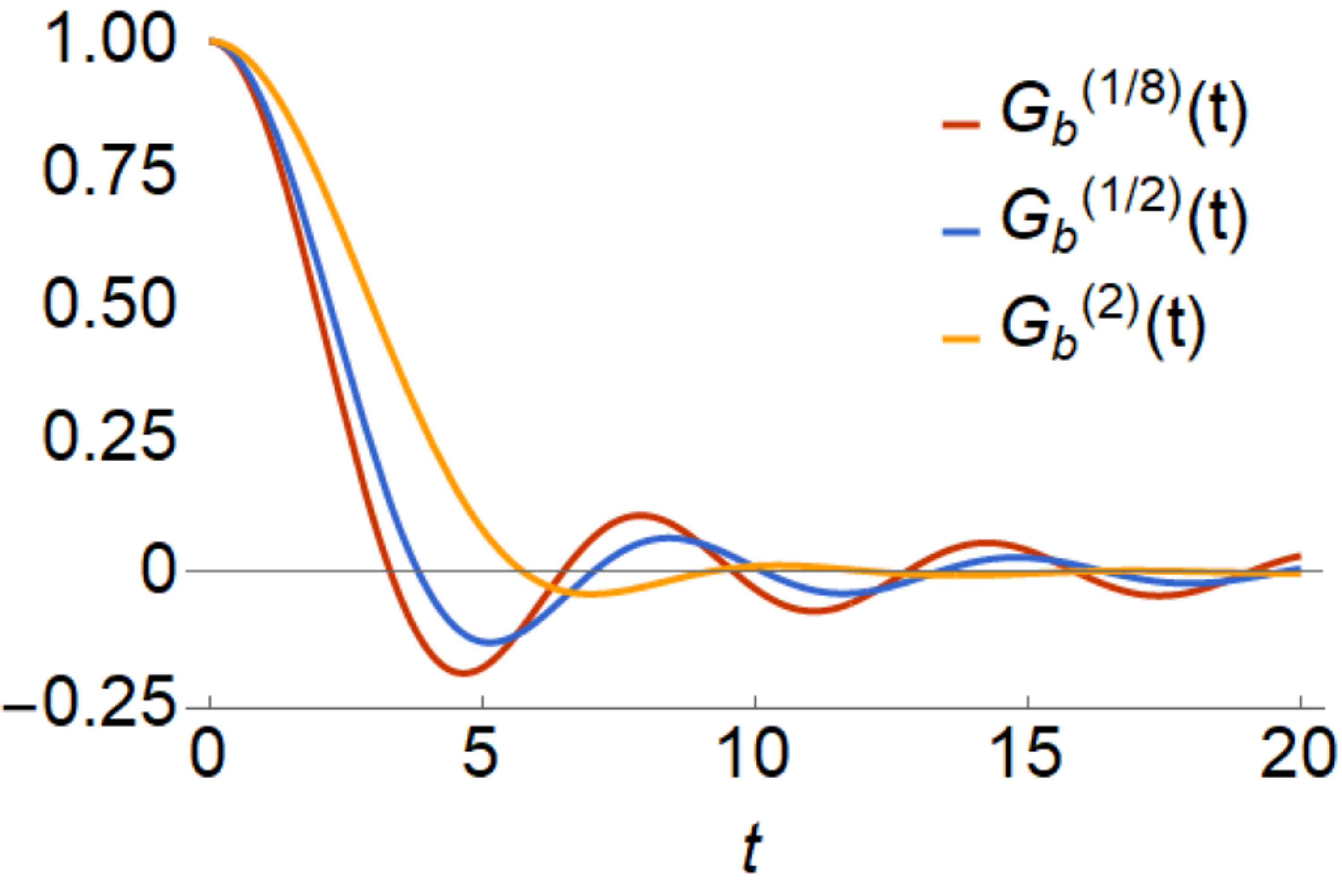}
         \caption{}
         \label{fig:lanczos-approach-t}
     \end{subfigure}
     \hfill
     \begin{subfigure}[b]{0.32\textwidth}
         \centering
         \includegraphics[width=\textwidth]{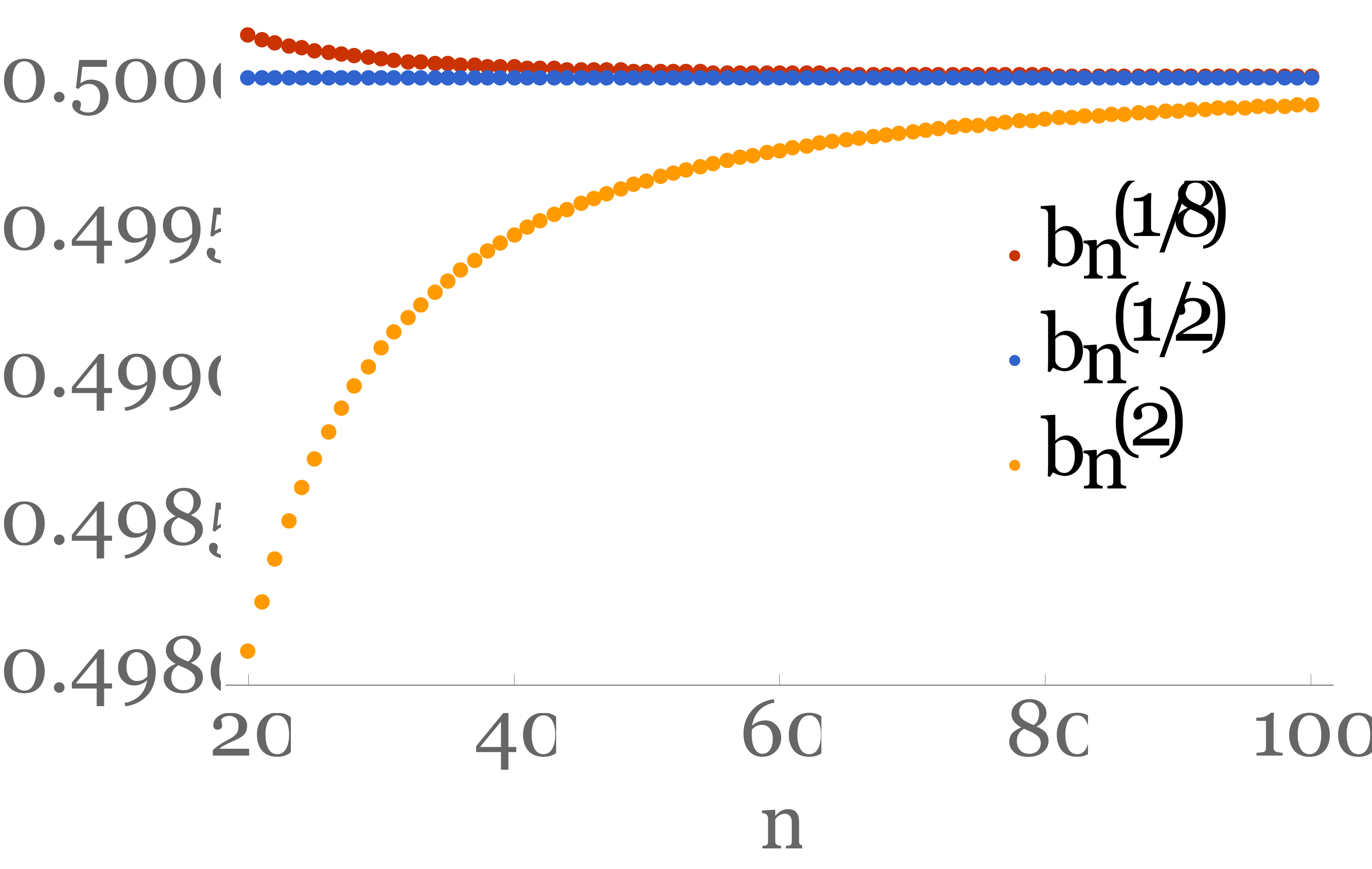}
         \caption{}
         \label{fig:lanczos-approach-b}
     \end{subfigure}
        \caption{\small{The \textbf{(a)} spectral measures, \textbf{(b)} real-time two-point functions, and \textbf{(c)} Lanczos sequences for different approaches to the spectral edge:  $\mu_b^{(1/8)}$ (red), $\mu_b^{(1/2)}$ (blue), and $\mu_b^{(2)}$ (orange). The power with which the spectrum \eqref{eq:one-parameter-ansatz} vanishes at the edge changes the asymptotic $1/n^2$ approach to the plateau.}}
        \label{fig:lanczos-approach}
\end{figure}

The prototypical example of such a measure with bounded support is
\begin{equation}
    \mu_b^{(1/2)} = \frac{2}{\pi \omega_0^2}\sqrt{\omega_0^2 - \omega^2} \quad , \quad \omega \in [-\omega_0,\omega_0]
\end{equation}
for which the Lanczos coefficients reach their asymptotic limit immediately
\begin{equation}
    b^{(1/2)}_n \equiv \frac{\omega_0}{2} \, ,
\end{equation}
and the two-point function is given by
\begin{equation}
    G^{(1/2)}_b(t) = \frac{2 J_1(\omega_0 t)}{\omega_0 t} \, .
\end{equation}
More generally, the fall-off of the spectral measure as one approaches the upper-bound determines the asymptotic approach to the plateau. 
For example, we can generalize the above measure to have a different fall-off near the spectral edge,
\begin{equation}
    \mu_b^{(\beta)}(\omega) = N_\beta (\omega_0^2 - \omega^2)^\beta \quad , \quad \omega \in [-\omega_0,\omega_0] \quad , \quad  N_\beta = \frac{\Gamma(\beta+3/2)}{\sqrt{\pi}\omega_0^{2\beta+1}\Gamma(\beta+1)}
    \, , \label{eq:one-parameter-ansatz}
\end{equation}
and still calculate exactly the Lanczos coefficients
\begin{equation}
    b_n^{(\beta)} = \omega_0\sqrt{\frac{n(n+ 2\beta)}{(2n+2\beta-1)(2n+2\beta+1)}} \, .
\end{equation}
It's then easy to see that the Lanczos coefficients asymptotically approach the plateau like
\begin{equation}
     b_n^{(\beta)} - \frac{\omega_0}{2} \underset{n \rightarrow \infty}{\sim}  \frac{1-4\beta^2}{8n^2} + \ldots
\end{equation}
where the edge of the spectrum is encoded in the $1/n^2$ approach. The corresponding two-point function is
\begin{equation}
   G_b^{(\beta)}(t) =  {_0 F_1}\left(3/2+\beta,-(t\omega_0/2)^2\right) \, .
\end{equation}
We plot a few examples of the approach to the plateau in Figure~\ref{fig:lanczos-approach}. 

\paragraph{Spectral gap and alternating Lanczos sequences}

When the support of the spectral measure has two connected components, the orthogonal polynomials of odd and even degree will probe the density differently, leading to an alternating structure to the Lanczos coefficients. This structure persists even when there is no true gap, but merely a power-law suppression near $\omega=0$. 

\begin{figure}
     \centering
     \begin{subfigure}[b]{0.32\textwidth}
         \centering
         \includegraphics[width=\textwidth]{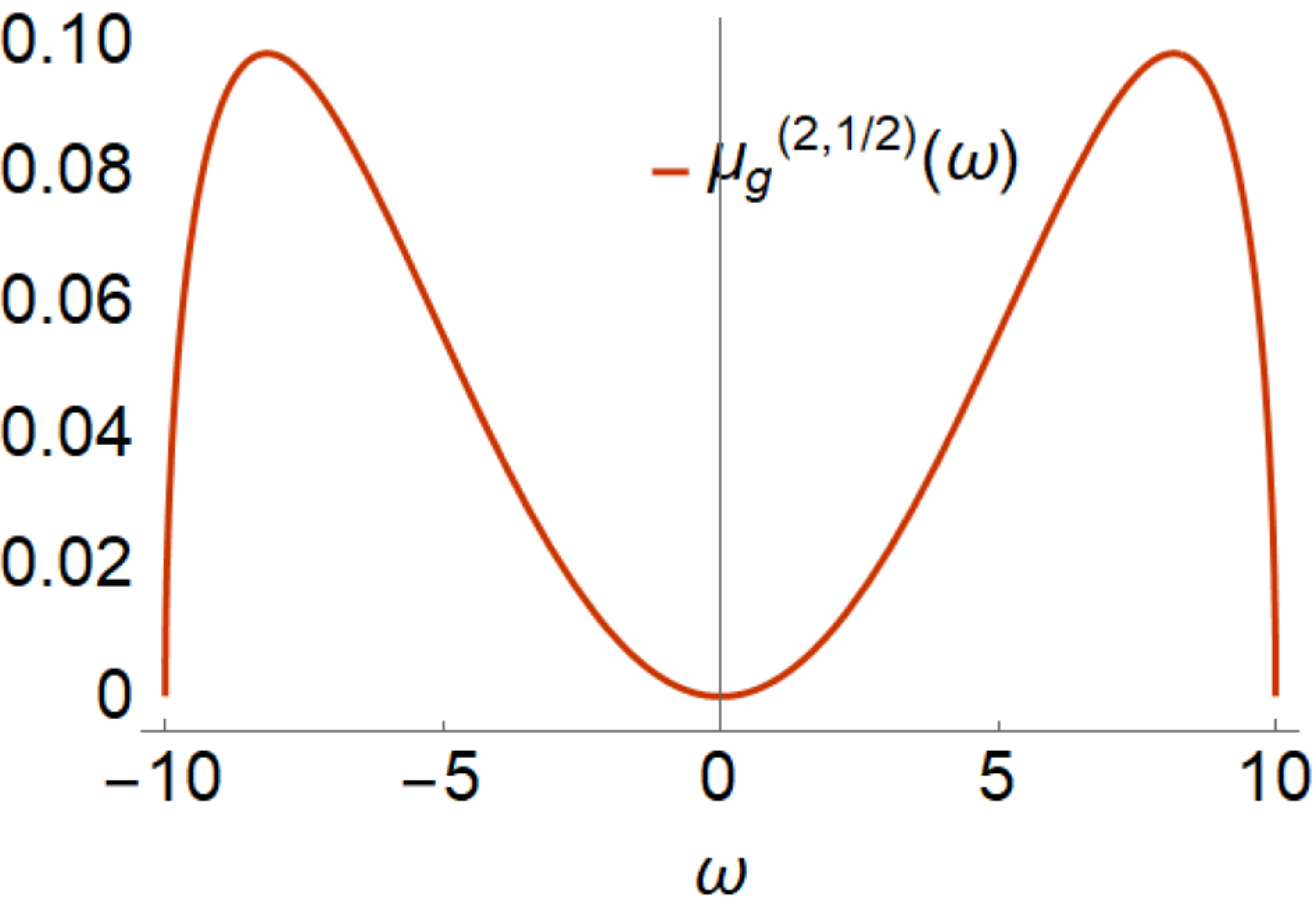}
         \caption{}
         \label{fig:lanczos-gap-w}
     \end{subfigure}
     \hfill
     \begin{subfigure}[b]{0.32\textwidth}
         \centering
         \includegraphics[width=\textwidth]{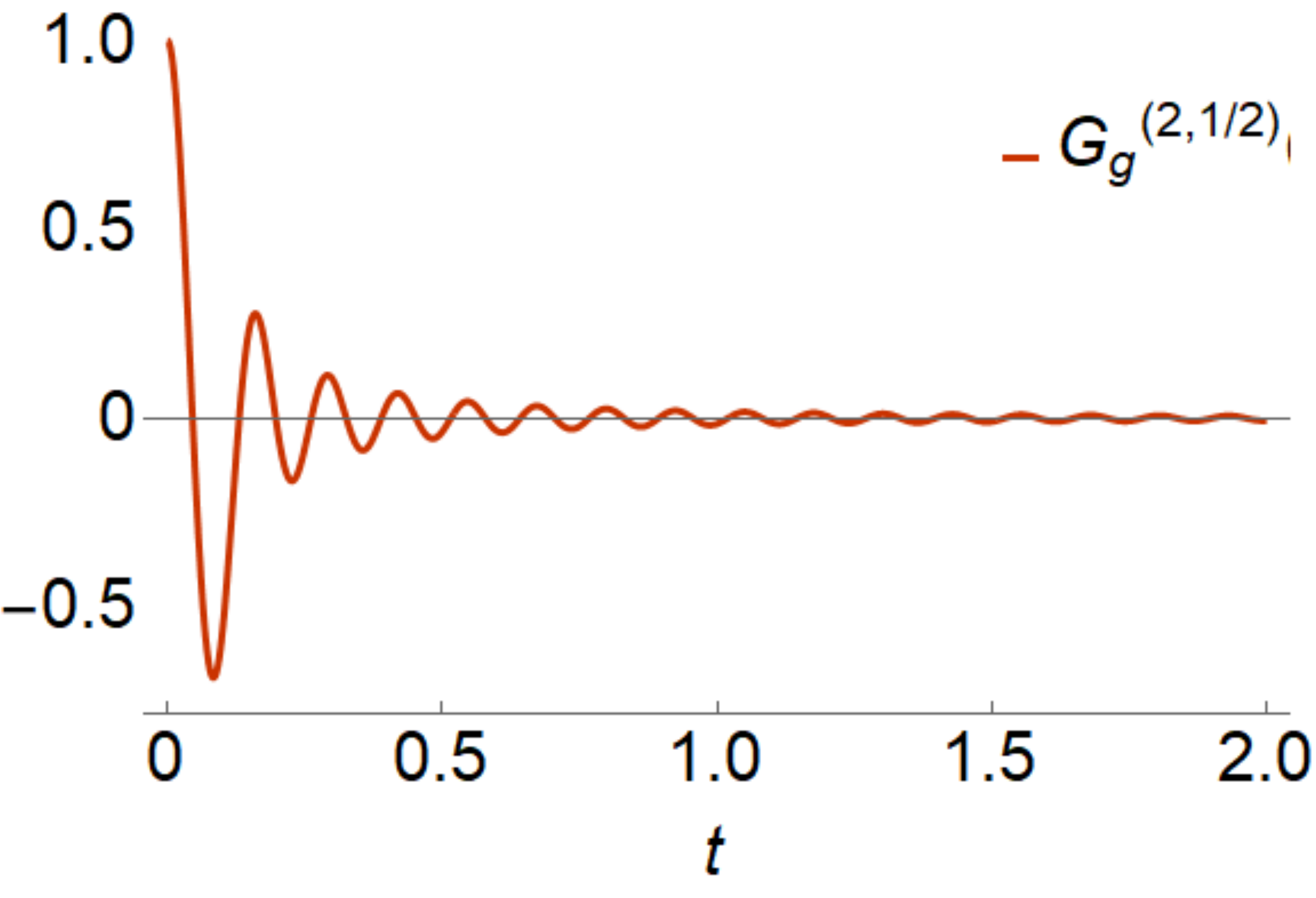}
         \caption{}
         \label{fig:lanczos-gap-t}
     \end{subfigure}
     \hfill
     \begin{subfigure}[b]{0.32\textwidth}
         \centering
         \includegraphics[width=\textwidth]{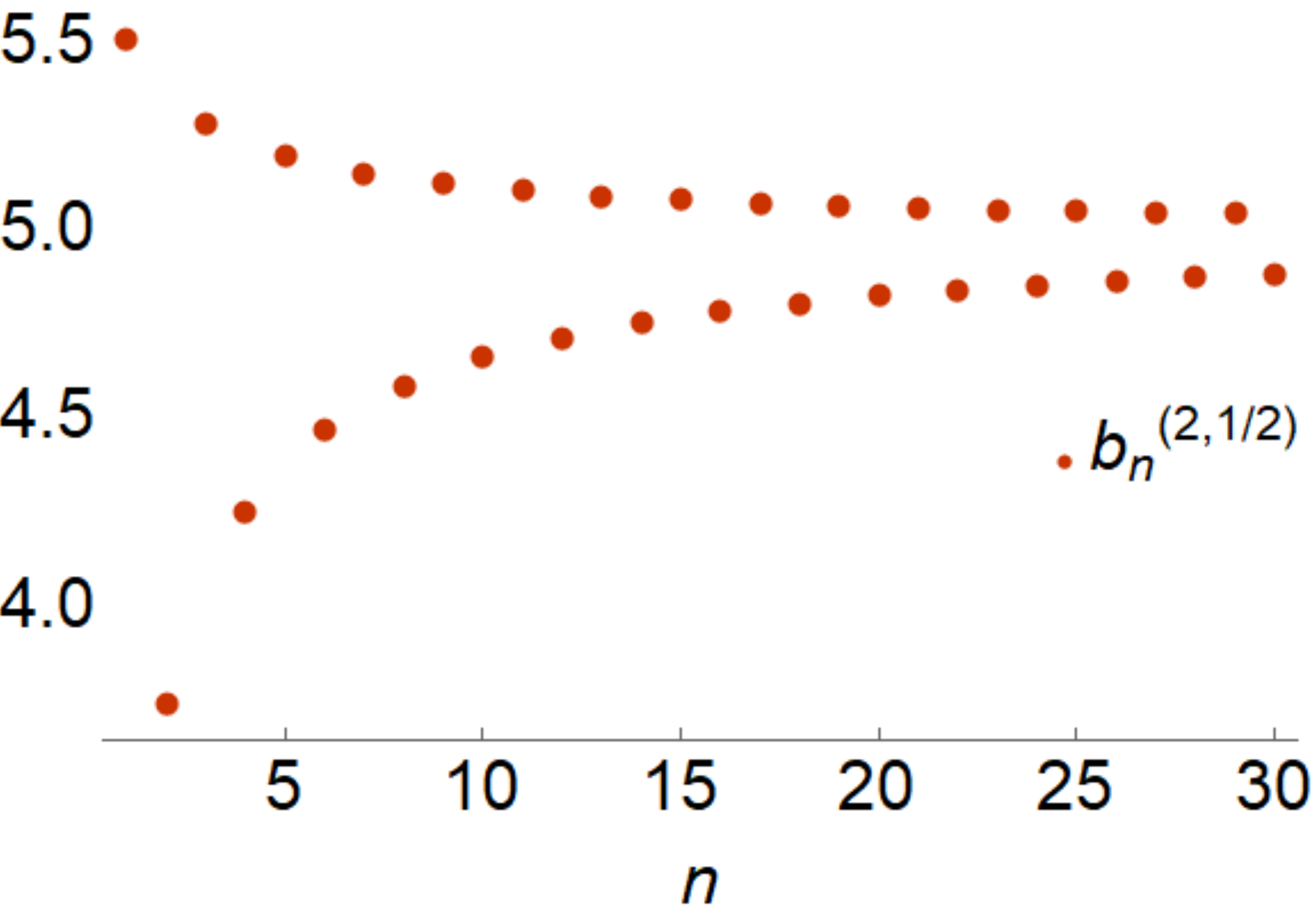}
         \caption{}
         \label{fig:lanczos-gap-b}
     \end{subfigure}
        \caption{\small{The \textbf{(a)} spectral measure, \textbf{(b)} real-time two-point function, and \textbf{(c)} Lanczos sequence for the $\left(2,\frac{1}{2} \right)$ instance of the $(\alpha,\beta)$ spectral measure family \eqref{eq:two-parameter-ansatz}.  An infinitesimal gap in the spectral measure leads to an alternating Lanczos sequence.}}
        \label{fig:lanczos-gap}
\end{figure}

A simple, solvable model is the measure
\begin{equation}
       \mu_g^{(\alpha,\beta)}(\omega) = N_{\alpha,\beta} |\omega|^\alpha (\omega_0^2 - \omega^2)^\beta \quad , \quad \omega \in [-\omega_0,\omega_0] \quad , \quad  N_{\alpha,\beta} = \frac{\omega_0^{-2\beta-\alpha-1}}{B\left(\beta+1,\frac{\alpha+1}{2}\right)} \, , \label{eq:two-parameter-ansatz}
\end{equation}
which has Lanczos coefficients that now approach the plateau with an alternating $1/n$ term:
\begin{equation}
    b^{(\alpha,\beta)} - \frac{\omega_0}{2} \underset{n \rightarrow \infty}{\sim} (-1)^n \frac{\alpha}{2n} + \ldots \, .
\end{equation}
The corresponding two-point function is 
\begin{equation}
       G_g^{(\alpha,\beta)}(t) =  {_1 F_2}\left(1/2+\alpha/2;1/2,3/2+\alpha/2+\beta,-(t\omega_0/2)^2\right) \, .
\end{equation}
We plot the particularly interesting example of $\alpha=2,\beta=1/2$ in Figure~\ref{fig:lanczos-gap}, where we see the gap in the spectrum gives rise to the `ramp' in the-two point function, and manifests in the alternating approach to the Lanczos plateau. 

\begin{figure}
     \centering
     \begin{subfigure}[b]{0.32\textwidth}
         \centering
         \includegraphics[width=\textwidth]{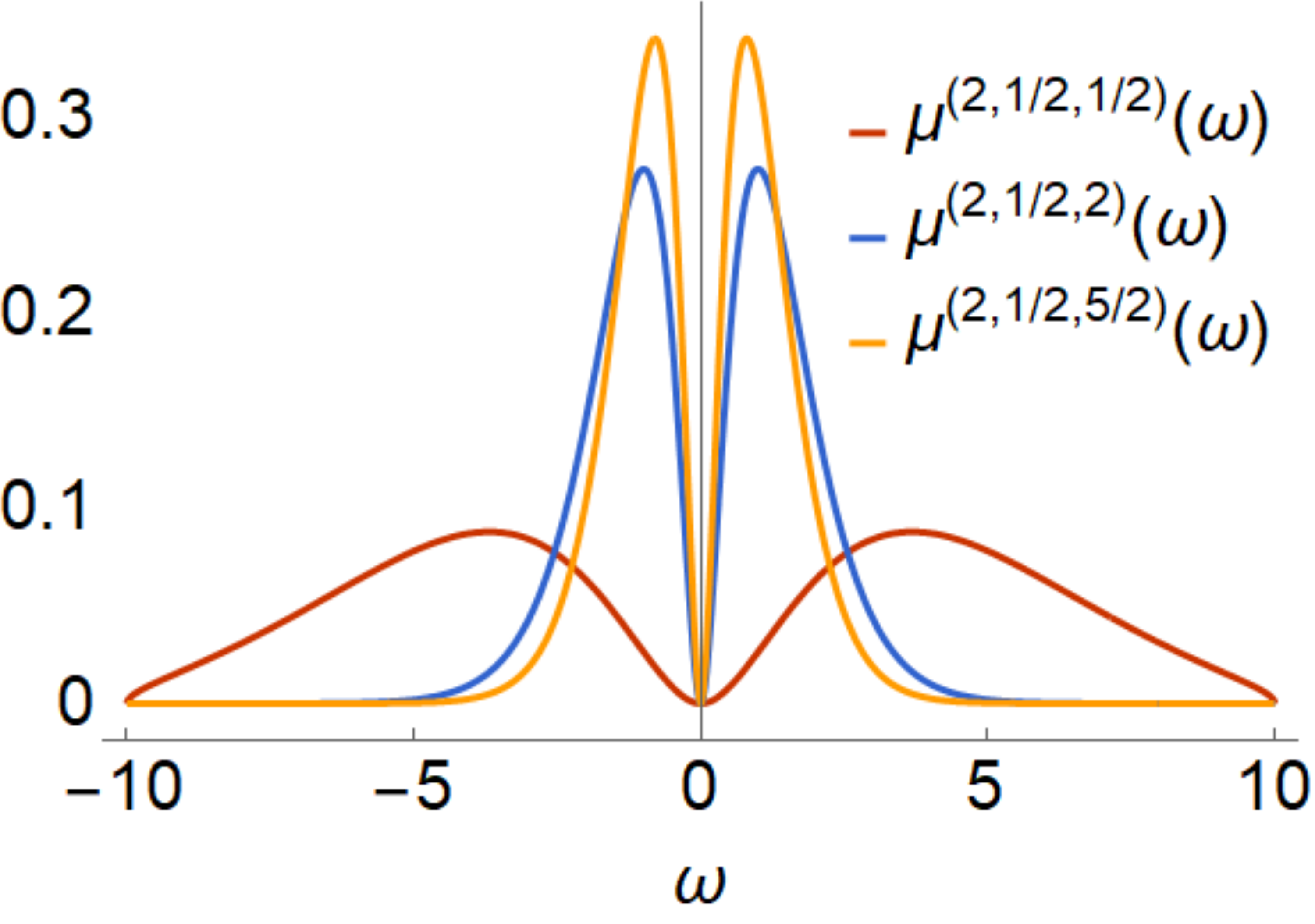}
         \caption{}
         \label{fig:lanczos-combined-w}
     \end{subfigure}
     \hfill
     \begin{subfigure}[b]{0.32\textwidth}
         \centering
         \includegraphics[width=\textwidth]{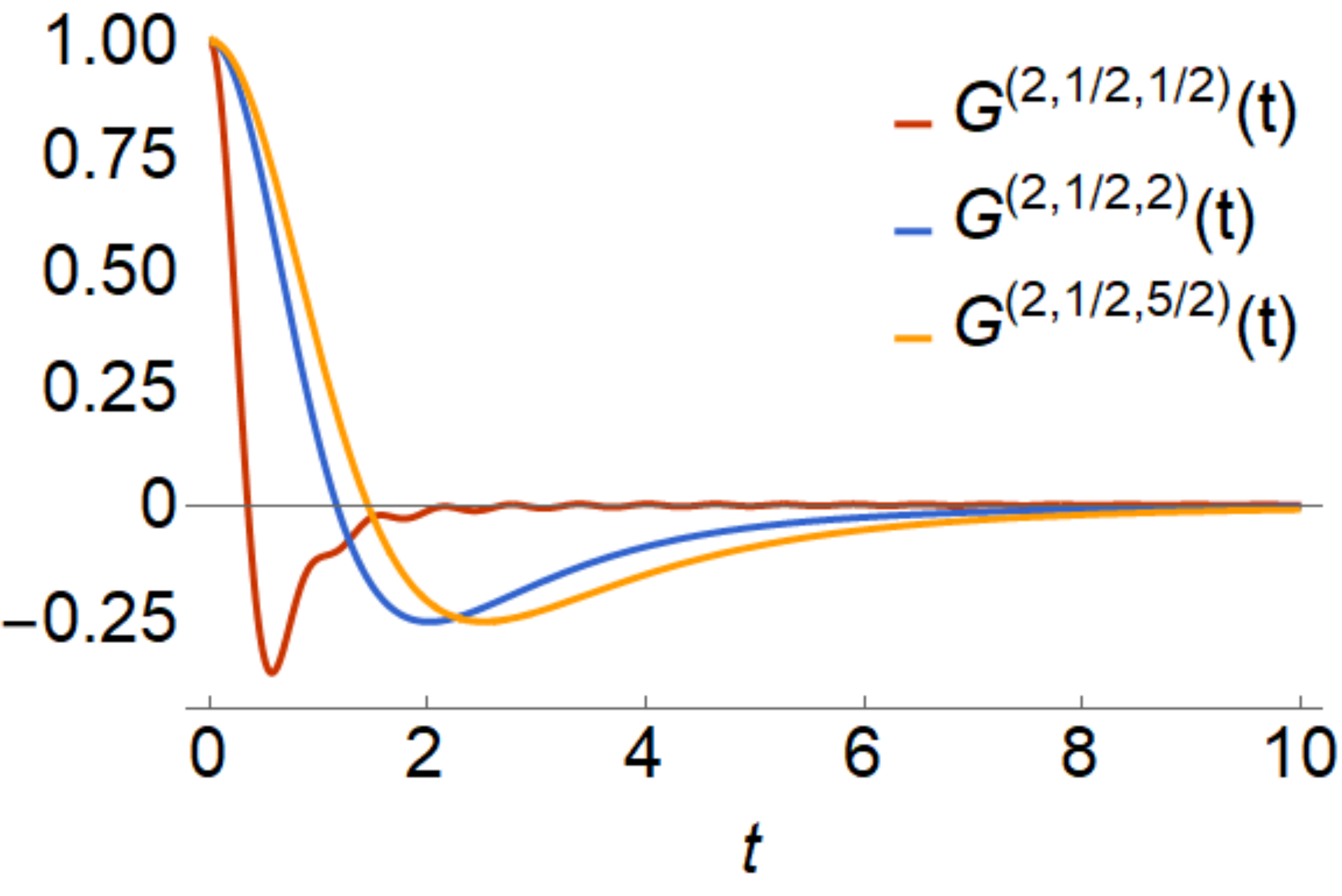}
         \caption{}
         \label{fig:lanczos-combined-t}
     \end{subfigure}
     \hfill
     \begin{subfigure}[b]{0.32\textwidth}
         \centering
         \includegraphics[width=\textwidth]{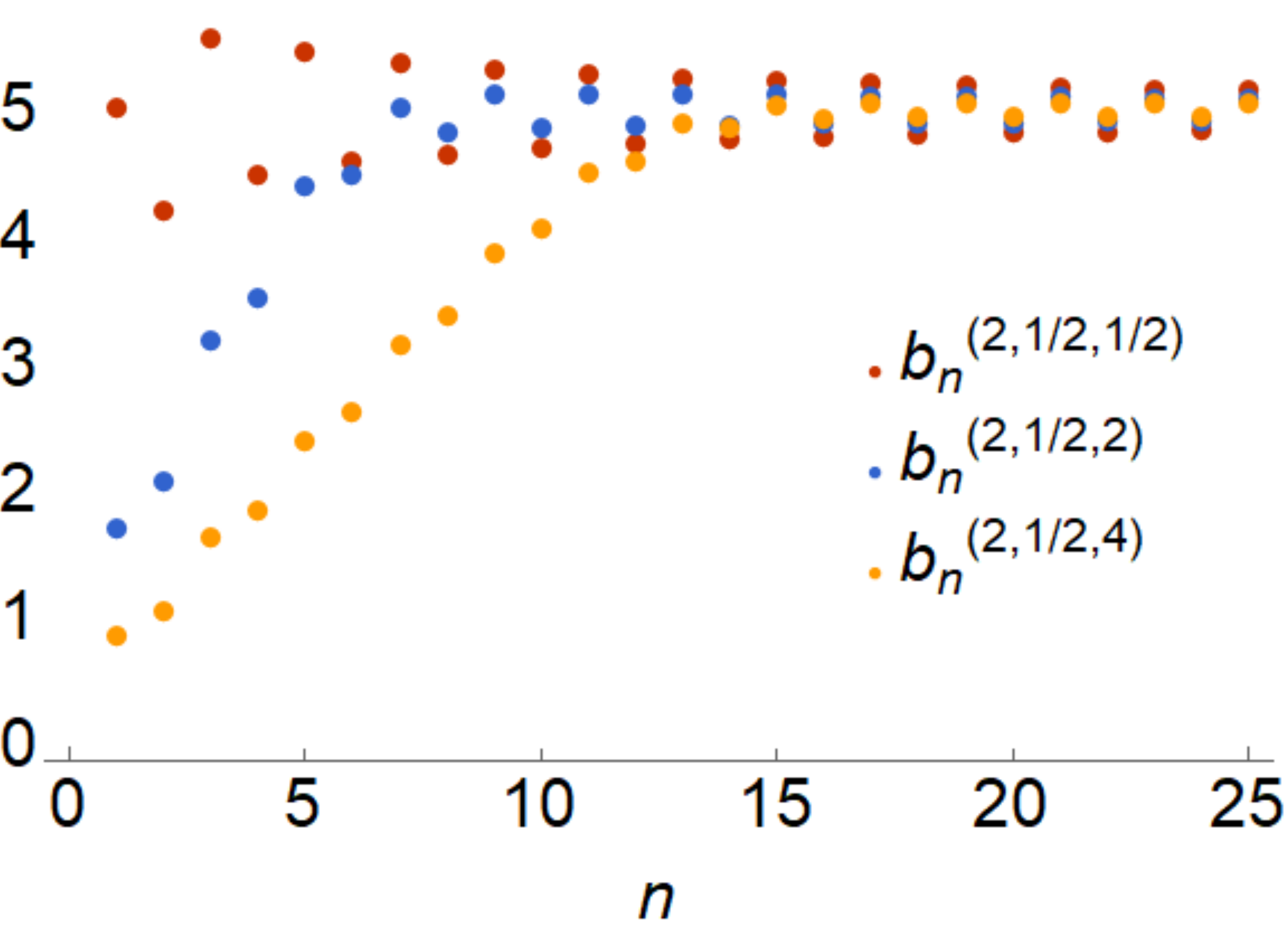}
         \caption{}
         \label{fig:lanczos-combined-b}
     \end{subfigure}
        \caption{\small{The \textbf{(a)} spectral measures, \textbf{(b)} real-time two-point functions, and \textbf{(c)} Lanczos sequences for the $\left(2,\frac{1}{2},\frac{1}{2} \right)$ (red), $\left(2,\frac{1}{2},2 \right)$ (blue),  $\left(2,\frac{1}{2},\frac{5}{2} \right)$ (orange, (a) and (b)), and $\left(2,\frac{1}{2},4 \right)$ (orange, (c)) instances of the $(\alpha,\beta,\gamma)$ spectral measure family \eqref{eq:three-parameter-ansatz}.  As the exponential decay coefficient $\gamma$ increases, a clear Lanczos ascent emerges.}}
        \label{fig:lanczos-combined}
\end{figure}

Lastly, we can combine all of the above features together in a simple phenomenologically interesting toy model with measure
\begin{equation}
       \mu^{(\alpha,\beta,\gamma)}(\omega) = N_{\alpha,\beta,\gamma} |\omega|^\alpha (\omega_0^2 - \omega^2)^\beta e^{-\gamma | \omega |} \quad , \quad \omega \in [-\omega_0,\omega_0] \, , \label{eq:three-parameter-ansatz}
\end{equation}
with two-point function
\begin{align}
G^{(\alpha,\beta,\gamma)}(t) =\omega_0 \Gamma \left(\frac{\alpha }{2}+1\right) \bigg[ &(\gamma +i t)
   \, _1\tilde{F}_2\left(\frac{\alpha +2}{2};\frac{3}{2},\frac{\alpha
   }{2}+\beta +2;-\frac{1}{4} \omega_0^2 (t-i \gamma
   )^2\right) \nonumber \\
   & +(\gamma -i t) \, _1\tilde{F}_2\left(\frac{\alpha
   +2}{2};\frac{3}{2},\frac{\alpha }{2}+\beta +2;-\frac{1}{4}
   \omega_0^2 (t+i \gamma )^2\right)\bigg] \\
    -2 \Gamma
   \left(\frac{\alpha +1}{2}\right) \bigg[ & \,
   _1\tilde{F}_2\left(\frac{\alpha +1}{2};\frac{1}{2},\frac{\alpha
   +3}{2}+\beta ;-\frac{1}{4} \omega_0^2 (t-i \gamma )^2\right) \nonumber\\ 
   & +\,
   _1\tilde{F}_2\left(\frac{\alpha +1}{2};\frac{1}{2},\frac{\alpha
   +3}{2}+\beta ;-\frac{1}{4} \omega_0^2 (t+i \gamma
   )^2\right)\bigg] \nonumber \, .
\end{align}
We plot this example in Figure~\ref{fig:lanczos-combined}. 

\section{Review of random matrices} 
\label{app:rmtreview}
It is useful to first recall some basic concepts and classic results from the theory of random matrices. Suppose that our matrix $\mathcal{L}$ of size $D \times D = e^{2N} \times e^{2N}$ is chosen randomly from some matrix ensemble with measure:
\begin{equation}
    \Pi_D(\mathcal{L}) d\mathcal{L} = Z^{-1}_D e^{-D\, Tr[V(\mathcal{L})]} d\mathcal{L} \, , \label{matrixprop1}
\end{equation}
where $d\mathcal{L}$ is the Haar measure on $D\times D$ matrices and $Z_D$ is the associated partition function. After diagonalizing $\mathcal{L}= U \Omega U^\dagger$ with $\Omega= diag\{ \omega_1, \omega_2, \dots, \omega_D\} $ we can integrate out the unitary basis choice $U$ and obtain a probability distribution for the eigenvalues of $\mathcal{L}$:
\begin{equation}
    \rho_D(\omega_i ) \prod_{i=1}^D d\omega_i = Z^{-1}_D e^{-D\, \sum_{i=1}^{D}V(\omega_i)} \det\left[\Delta_D(\omega_i) \right] \,\prod_{i=1}^{D}d\omega_i \, , \label{matrixprop2}
\end{equation}
where $\Delta_D(\omega_i)$ is the well-known Vandermonde matrix
\begin{equation}
    \Delta_D(\omega_i)= \begin{pmatrix}
    1 & 1 &\cdots & 1\\
    \omega_1 & \omega_2 &\cdots &\omega_D \\
    \omega_1^2 & \omega_2^2 & \cdots &\omega_D^2 \\
    \vdots & & & \\
    \omega_1^{D-1} & \omega_2^{D-1} &\cdots & \omega_D^{D-1}
    \end{pmatrix} \,  . \label{vandermonde}
\end{equation}
Its determinant in (\ref{matrixprop2}) can be absorbed in a redefinition $\sum_i V(\omega_i) \to \sum_i V(\omega_i)- 2\sum_{i,j}\log (\omega_i- \omega_j)$. This logarithmic contribution to the potential $V$ controlling the eigenvalue statistics describes a universal ``repulsion'' between eigenvalues of a random matrix. 

\paragraph{Average spectral density and equilibrium measure} The average spectral density of this matrix ensemble at finite dimension $D=e^{2N}$ can be obtained from (\ref{matrixprop2}) simply by integrating out all but one eigenvalue:
\begin{equation}
    \rho_D( \omega_1) = Z^{-1}_D \int \prod_{i=2}^D d\omega_i \, e^{-D\, \sum_{i=1}^{D}V(\omega_i)} \det\left[\Delta_D(\omega_i) \right] \, . \label{spectraldensity}
\end{equation}
The key step in computing this spectral density as well as in connecting random matrix technology with the results of the previous Section, is to observe that the Vandermonde determinant is invariant under replacing the $\omega_i^j$ entries in the matrix (\ref{vandermonde}) with $\pi_j(\omega_i)$ where $\pi_j$ are a family of orthogonal polynomials and $j$ denotes, as before, the polynomial degree. If those polynomials are chosen to be orthogonal with respect to the measure $d\mu_\pi (\omega)= e^{-D V(\omega)} d\omega $ then by virtue of ... eq. (\ref{spectraldensity}) simplifies to:
\begin{equation}
    \rho_D(\omega) = \frac{1}{D}e^{-V(\omega) } \sum_{i=0}^{D-1} \pi_i^2 (\omega) \, . \label{spectraldensity2}
\end{equation}
The continuous approximation 
\begin{equation}
    \rho(\omega) = \lim_{D\to \infty}\rho_D(\omega) \, , \label{equilibriummeasure}
\end{equation}
to the spectral density (\ref{spectraldensity2}) is called the \emph{equilibrium measure} and it extremizes the ``energy functional'':
\begin{equation}
  I[\rho]= \int d\omega_1 d\omega_2 \,\rho(\omega_1) \rho(\omega_2) \,\left( V(\omega) - 2\log (\omega_2 -\omega_1) \right)  \, .
\end{equation}
Two classic examples of matrix ensembles are the Gaussian potential $V(\omega)=\omega^2$ whose equilibrium measure is the famous Wigner semi-circle distribution of eigenvalues $\rho(\omega)= \sqrt{1-\omega^2}$ and the ``infinite well'' potential 
\begin{equation}
 V(\omega)=\begin{cases}
 0,\,\, \text{for } \omega \in [-1,1]\\
 \infty\,\, \text{for }\omega \notin [-1,1] \, .
 \end{cases} \label{infinitewell}
\end{equation} 
with equilibrium measure $\rho(\omega)= \frac{1}{\sqrt{1-\omega^2}}$.

\paragraph{A bound on spectral density due to eigenvalue repulsion} An important mathematical fact is that for ensembles of matrices with eigenvalues concentrated in a single connected interval, taken here to be $[-1,1]$, and for sufficiently attractive potentials $(\lim_{\omega\to \infty} \frac{V(\omega)}{\log (1+\omega^2)} =\infty )$ all equilibrium measures are bounded by the inverse semicircle distribution:
\begin{equation}
    \rho(\omega)\leq \frac{c}{\sqrt{1-\omega^2}} \label{measurebound}
\end{equation}
where $c\in \mathbb{R}^+$. The content of the inequality (\ref{measurebound}) is an upper bound on the possible growth of the spectral density as we approach the edge of the spectrum $|\omega|\to 1$ and this is a consequence of the universal eigenvalue repulsion in RMT. Intuitively, (\ref{measurebound}) states that the specific choice of matrix potential $V$ (among the class allowed by the assumptions of the theorem) can only decrease the accumulation of eigenvalues near $|\omega|=1$, while the densest concentration at the edge is the one constrained only by the logarithmic eigenvalue repulsion term. The latter is, in turn, the only effect controlling the spectral density for the infinite well potential (\ref{infinitewell}), hence the bound (\ref{measurebound}) follows.

\section{An important orthogonal polynomial theorem}
\label{app:asymptotics}
A number of key results of this paper rely on a beautiful theorem about asymptotics of orthogonal polynomials when the Lanczos coefficients approach a plateau \cite{VANASSCHE1991237}. For organizational convenience, we opt to state it here as a mathematical fact, disconnected from physical context, and simply employ it in the appropriate points of our analysis.

First, we introduce the \emph{Stieltjes transform} $Q_n$ of the orthogonal polynomial sequence $P_n$, also known as the \emph{functions of the second kind}, defined as:
\begin{equation}
    Q_n(\omega) = \int d\mu(x) \frac{P_n(x)}{\omega-x} =(O_0|\frac{1}{\omega- {\cal L}} |O_n) \, . \label{stieltjes}
\end{equation}
These functions clearly satisfy the same recurrence relation as the orthogonal polynomials $P_n$ albeit with different initial conditions. 

The theorem of interest, then, establishes the following facts about $P_n$ and $Q_n$ in the $n\to \infty$ limit:
\begin{theorem}
Consider a family of polynomials $P_n$, orthogonal with respect to a measure $\mu$, whose recurrence relation coefficients asymptote to a constant $b_n\to b$ as $n\to \infty$. Then the following properties hold:
 \begin{align}
\lim_{n\to \infty}\frac{P_{n+1}(\omega)}{P_n(\omega)}&=\frac{1}{2b}\left( \omega +\sqrt{\omega^2-4b^2} \right) \, , \label{Plimit}\\
 \lim_{n\to \infty}\frac{Q_{n+1}(\omega)}{Q_n(\omega)}&=\frac{1}{2b}\left( \omega -\sqrt{\omega^2-4b^2} \right) \, , \label{Qlimit}\\
\lim_{n\to\infty}P_n(\omega) Q_n(\omega) &= \frac{1}{\sqrt{\omega^2-4b^2}} \, . \label{PQlimit}
\end{align}
It also follows that for every continuous function $f:supp[\mu] \to \mathbb{R}$ that is bounded outside the support of $\mu$ we have
\begin{align}
\lim_{n\to \infty} \int f(\omega) P_n^2(\omega) d\mu(\omega) =\lim_{n\to \infty} (O_n| f(\mathcal{L})|O_n) =\frac{1}{\pi}\int \frac{f(\omega)}{\sqrt{4b^2-\omega^2}}d\omega \, . \label{P2measure} 
\end{align} 
    \label{thm1}
\end{theorem}

\section{Canonical K-complexity in JT gravity}\label{app:canonical-k-jt}

A more na\"ive way to compute K-complexity in JT gravity is to take the entire two-point function $\overline{G_\beta(t)}$ and directly compute moments, without first passing to a microcanonical picture.
As we discussed, the microcanonical picture is where we really ought to think about questions of K-complexity for black holes, but there is nothing in principle that stops us from considering the canonical picture immediately.
We include the calculation of the Lanczos sequence in this scenario here, for completeness.

We can extract the approximate growth rate of the moments $m_{2n}$ by using a saddle point approximation, just as in the microcanonical picture.
In this version of K-complexity, the moments are given by
\begin{equation}
    m_{2n} = \left[ \frac{\partial^{2n}}{\partial t^{2n}} \overline{G_\beta(t)} \right]_{t=0} ,
\end{equation}
where $\overline{G_\beta(t)}$ is defined in \eqref{eq:jt-thermal-2pt}.
The exact saddle point equations for the above moments are difficult to solve analytically, but a cursory numerical exploration shows that there are essentially two relevant saddle points (with equal height, exchanged by $E_1 \leftrightarrow E_2$) whose locations depend nontrivially on both $n$ and $\beta$, while the dependence on $S_0$ and $\Delta$ is a bit more mild.
The location of the saddle point with $E_1 > E_2$ is roughly
\begin{equation}
    E_1 \approx \frac{n}{\beta} , \quad E_2 \approx 0 ,
\end{equation}
and this leads to a moment approximation
\begin{equation}
    m_{2n} \approx \exp \left( 2n \log n + O(n) \right) ,
\end{equation}
which is the universal growth rate discussed in \cite{Parker:2018yvk}.
The intuition for this growth in the JT context is that the saddle points dominating the moment growth quickly move away from the $E_1=E_2$ locus, and so the subleading corrections to the density pair correlation function $\rho(E_1,E_2)$ quickly become unimportant.
Once the leading term is the only relevant contribution, the ensemble-averaged thermal two-point function follows an initial exponential decay $e^{-t^2}$ which transitions to a universal $t^{-3}$ decay, similarly to the spectral form factor.
This pattern of decay, if left unchecked, generally leads to linear growth of the Lanczos sequence
\begin{equation}
    b_n \approx \alpha n .
\end{equation}
This matches the result obtained in \cite{Dymarsky:2021bjq}, where K-complexity was computed for a two-point function in a 2d CFT in the thermal state on a line.

We have computed the canonical Lanczos sequence using the average thermal two-point function $\overline{G_\beta(t)}$.
Though exact results were hard to obtain, our approximate understanding seems to suggest that even the complete semiclassical two-point function, which includes contributions from both level repulsion and discrete eigenvalue effects in the matrix model, is not enough to see a plateau in the Lanczos sequence.
This is in stark contrast to the black hole information paradox, where level repulsion effects in the matrix model have a geometric interpretation in gravity and end up rectifying the unitary Page curve directly in the canonical ensemble.
If it is possible to see a plateau in the Lanczos sequence in semiclassical gravity using this definition of K-complexity, this effect requires much more detailed information about the ensemble of boundary theories than the sort of information which leads to a resolution of the information paradox.
In fact, a hint of the more detailed information which may be necessary has already been studied by Stanford in an effort to find Euclidean wormhole saddles in JT gravity plus matter which require a more direct ensemble interpretation \cite{Stanford:2020wkf}.

This perspective reveals a potential issue with our na\"ive calculation.
It might be that the effects which lead to a Lanczos plateau in the direct canonical picture are revealed only in a \textit{quenched} calculation of the Lanczos sequence, where we do not take the ensemble average until after we have run the Lanczos algorithm to convert the thermal two-point function into the Lanczos sequence.
This quenched sequence $\{\overline{b_n}\}$ will involve a replica trick computation that incorporates correlation functions of much higher order on multiple boundaries, and the simplest such object is the double trumpet topology with four boundary operator insertions studied in \cite{Stanford:2020wkf}.
It is not obvious that the quenched calculation will yield an effective moment sequence 
\begin{equation}
    \{\overline{b_n}\} \leftrightarrow \{m_{2n}^{\text{eff}}\} ,
\end{equation}
that can be written as an integral over energies, but we may be able to find an effective description of that sort by modifying the operator wavefunction $|\bra{E_1} O_\Delta \ket{E_2}|^2$ (perhaps inspired by universal operator dynamics in CFTs \cite{Collier:2019weq}).
The general problem of computing a quenched Lanczos sequence for an operator defined by a smooth function\footnote{By this we mean that, as in JT gravity, the operator matrix elements are defined in any given member of the ensemble by evaluation of a smooth function of energies on the spectrum of the particular Hamiltonian in that member.  For a free scalar of dimension $\Delta$ in JT gravity, this function is \eqref{eq:JT-scalar-matrix-elts}.} in a double-scaled random matrix ensemble seems relatively unstudied in both the Lanczos algorithm and random matrix theory literatures.

The quenched calculation we described above presents many technical challenges, and would require a much higher degree of control in JT gravity than has previously been achieved in e.g. calculations of entanglement entropy.
It would be very interesting to make progress on such a calculation, perhaps using the detailed correlation function results of \cite{Blommaert:2020seb}.
However, due to the arguments presented in the main body of the paper concerning the interpretation of $\mathcal{E}$ as a conserved charge for $\mathcal{L}$, it is not clear that such an effort would succeed.

\bibliographystyle{jhep}
\bibliography{refs}

\end{document}